\documentclass[fleqn,usenatbib,useAMS]{mnras}

\usepackage[dvipdfmx]{graphicx}	
\usepackage{amsmath}	
\usepackage{amssymb}	
\usepackage{multicol}        
\usepackage{bm}		
\usepackage{pdflscape}	
\usepackage{ulem}
\usepackage{hjkw}
\usepackage[subrefformat=parens]{subcaption}

\usepackage[T1]{fontenc}
\usepackage{ae,aecompl}

\usepackage{newtxtext,newtxmath}

\title[Pop. III BBH mergers with IMBHs]{On the population III binary black hole mergers with intermediate mass black holes:
dependence on common envelope parameter}

\author[Kotaro Hijikawa]{
Kotaro Hijikawa,$^{1}$\thanks{E-mail: hijikawa@astron.s.u-tokyo.ac.jp}
Tomoya Kinugawa,$^{2}$ Ataru Tanikawa,$^{3}$ Takashi Yoshida$^{4,1}$ and
Hideyuki Umeda$^{1}$
\\
$^{1}$Department of Astronomy, Graduate School of Science, 
The University of Tokyo, 7-3-1 Hongo, Bunkyo-ku, 113-0033 Tokyo, Japan\\
$^{2}$Institute for Cosmic Ray Research, The University of Tokyo,
Kashiwa, Chiba, Japan\\
$^{3}$Department of Earth Science and Astronomy, College of
Arts and Sciences, The University of Tokyo, Meguro-ku, Tokyo, Japan\\
$^{4}$Yukawa Institute for Theoretical Physics, Kyoto University, Kitashirakawa Oiwakecho, Sakyo-ku, Kyoto 606-8502, Japan}

\date{Accepted XXX. Received YYY; in original form ZZZ}

\pubyear{2022}

\begin{document}
\label{firstpage}
\pagerange{\pageref{firstpage}--\pageref{lastpage}}
\maketitle

\begin{abstract}
The current gravitational wave (GW) detectors have successfully observed many binary 
compact objects, and the third generation ground-based GW detectors such as Einstein telescope
and space-borne detectors such as LISA will start their GW observation in a decade.
Ahead of the arrival of this new era,
we perform a binary population synthesis calculation for
very massive ($\sim$ 100--1000 $\msun$) population (Pop.) III stars, 
derive the various property of binary black hole (BBH) mergers with 
intermediate mass black holes (IMBHs) and investigate the dependence on
common envelope parameter $\alpha\lambda$ which is still not a well understood parameter.
We find that the maximum mass of primary BH mass is larger
for smaller value of common envelope parameter.
In this study, we adopt double power law initial mass function (IMF) for Pop. III
stars, and put some constraints on Pop. III IMF by comparing our obtained merger rate
density at the local Universe with that derived from gravitational wave (GW) observation.
We compute the detection rate and show that the third generation ground-based
GW detector, Einstein telescope, have a potential to detect
$\sim$ 10--1000 BBHs with IMBHs per year.
We also find that we may be able to obtain the insight into $\alpha\lambda$
if a BBH with total mass $\gtrsim500\msun$ are detected by advanced LIGO (O4) or LISA.
\end{abstract}





\section{Introduction}
Many binary black hole (BBH) mergers have been discovered by the gravitational
wave (GW) detectors, advanced LIGO and Virgo.
The catalog, GWTC-3, was released last year by the LIGO Scientific Collaboration,
the Virgo Collaboration and the KAGRA Collaboration \citep{gwtc3}, and 
the number of discovered BBH mergers is approaching 100.
Detected BBHs have relatively high mass ($\sim$ 20--50 $\msun$) compared to Galactic 
BHs, and have low spin ($\sim$ 0).
Origin of such discovered BBH mergers have been suggested by many authors.
For example, field population (Pop.) I or II isolated binary stars 
\citep[e.g.,][]{bethe98, marchant16, mandel16, stevenson17, eldridge17, eldridge19, 
mapelli17, mapelli19, mapelli21, kruckow18, belczynski20, bavera21, riley21}, Pop.
III isolated binary stars \citep[e.g.,][]{kinugawa14, kinugawa16, kinugawa20, kinugawa21, 
hartwig16, belczynski17, tanikawa21, tanikawa22}, a dynamical formation in stellar clusters
\citep[e.g.][]{portegies00, oleary06, sadowski08, banerjee10, banerjee17, banerjee18,
banerjee18b, banerjee21, banerjee21b, downing10, downing11, tanikawa13, bae14, ziosi14, 
mapelli16, rodriguez16, rodriguez18, rodriguez19, fujii17, park17, hong18, hong20,
samsing18, samsing18b, dicarlo19, dicarlo20, kumamoto19, kumamoto20, kumamoto21,
anagnostou20, kremer20, fragione21, trani21, weatherford21}, a Pop. III dynamical 
formation \citep[e.g.][]{liu20, liu21, liu21b, wang22}, a formation in the galactic nuclei
\citep[e.g.][]{oleary09, antonini12, antonini16, stone17, leigh18, mckernan18, rasskazov19,
yang19, arcasedda20, tagawa20, tagawa20b, tagawa21, tagawa21b}, or triple or multiple 
star systems \citep[e.g.][]{antonini17, silsbee17, rodriguez18b, fragione19, fragione20,
hamers21, trani22, vynatheya22}.

In addition to the typical population, a more massive BBH, GW190521, with total
mass $\sim150\msun$ is also discovered \citep{abbott20}.
Since its primary BH mass is about 90 $\msun$ and it lies in 
the pair-instability mass gap \footnote[1]{Note that GW190521 may be a pair of two 
BHs above and below the pair-instability mass gap \citep{fishbach20, nitz21}.}, 
it is considered to be difficult to form such BBHs through a general isolated binary formation
channel, and the hierarchical merger process in the dynamical formation scenario
can explain the existence of $\sim90\msun$ BH.
However, there are some isolated binary scenarios to explain the formation of GW190521.
For example, by reducing ${}^{12}\mathrm{C}(\alpha,\gamma)^{16}\mathrm{O}$ reaction 
rate within its uncertainty, the pair-instability mass gap shifts toward the heavier
side, and GW190521-like BBHs can be formed through isolated binary formation channel
\citep{farmer19, farmer20, belczynski20b}.
Furthermore, a massive extremely metal-poor star or a Pop. III star may be able to
become a BH with 65--100 $\msun$, and thus the pair-instability mass gap may be filled,
and GW190521-like BBHs can be formed through isolated binary formation channel
\citep{umeda20, kinugawa21a, tanikawa21a, tanikawa21, farrell21, vink21}.
From the discovery of GW190521 and studies on this observation, the equivalent objects
or more massive BBH mergers, i.e., the mergers of an intermediate mass black hole
(IMBH; $\gtrsim100\msun$) with a stellar mass BH or an IMBH are expected to be
detected by the future GW observations.

So far, a number of stellar mass BHs ($\lesssim100\msun$) have been found by GW observations
or X-ray binary observations. 
On the other hand, few IMBHs have yet been discovered \citep[one can find more information
about the searchs for IMBHs in the review, ][]{greene20}.
The most reliable one is the merger remnant of the BBH merger, GW190521 \citep{abbott20}.
The X-ray sources HLX-1 is also considered to be an IMBH \citep{farrell09}.
The presence of the supermassive BHs (SMBHs; $\gtrsim10^6\msun$) at the center of galaxies
has been confirmed, and recently the Event Horizon Telescope successfully got images
of SMBHs at the center of M87 \citep{eht19, eht19a, eht19b, eht19c, eht19d, eht19e}
and the Milky Way \citep{eht22, eht22a, eht22b, eht22c, eht22d, eht22e}.
The formation and evolution of SMBHs are still not well understood.
According to some models, the seed of SMBHs has repeatedly merged with other BHs
multiple times and increases its mass. 
Since there are few IMBHs that have been found, it is difficult to constrain the
formation theory and to know where the SMBHs came from.
Therefore, it is important to know what mass of IMBHs existed at each redshift
in order to impose restrictions on the formation model of the SMBH.
As well as the probe for SMBH, IMBHs may be useful for other problems.
For example, the findings of IMBHs just above the pair-instability mass gap can be
used as the ``standard silen'' for the cosmology \citep{farr19, farmer19, farmer20}.
From the GW observation, the luminosity distance and the redshifted mass can be
obtained.
If we have a knowledge of the upper end of the pair-instability mass gap, then 
we can get the redshift of the sources from the true and redshifted mass.
Furthermore, the pair of stellar mass BH and IMBH, or the intermediate mass ratio
inspiral (also know as IMRI) will allow us to test the gravitational theory
\citep[e.g., ][]{sopuerta09,amaro-seoane18}.

In the future, the third generation ground-based GW detector, Einstein telescope
\citep{punturo10, sathyaprakash2012} or Cosmic Explorer \citep{reitze19}, and 
space-borne GW detector, LISA \citep{amaro-seoane2017}, TianQin \citep{luo16, wang19},
TianGO \citep{kuns20}, or DECIGO \citep{Seto01} will start the construction in ten years.
Since their sensitivity will improve and their horizon redshift will be $z\sim$ 10--1000
for BBH mergers with total mass $M_\mathrm{t}\sim$ 10--10 000 $\msun$,
the inspiral and merger of more massive BBHs, or binary IMBHs, located in the earlier
Universe are expected to be discovered.
Previously, the GWs from IMBH-IMBH mergers during the runaway merging process were
investigated \citep[e.g.,][]{matsubayashi04,shinkai17}.
As to the isolated binary, the BBH mergers above the pair-instability mass gap originated
from isolated Pop. I/II binaries were investigated by \citet{mangiagli19}.
Pop. III star formation simulations have been performed \citep[e.g.,][]{hirano14,
hirano15, susa14}, and it is claimed that the typical Pop. III stellar mass is 10--100 $\msun$.
Furthermore, very massive ($\sim$ 100--1000 $\msun$) Pop. III stars might be formed
\citep[e.g.,][]{hirano15}. A Pop. III star with 100--200 $\msun$ finally causes the
pair-instability supernova, and leaves no remnant.
On the other hand, more massive Pop. III star than this directly collapses and finally
becomes an IMBH.
Therefore, in our previous paper \citep{hijikawa21}, we focused on the very massive 
($\sim$ 100--1000 $\msun$) Pop. III stars, performed binary population synthesis
calculation, derived the merger rate density of BBHs with IMBHs assuming
single power law initial mass function (IMF), and compared this with
the observation.
Thus far, no BBH with IMBHs have been discovered, and hence the upper limit on the 
merger rate density is obtained.
We compared our results with this upper limit, and found that the power of single power law
IMF needs to be lower than 2.8.
However, this is inconsistent with the results of Pop. III star formation simulation claiming
a relatively top heavy IMF where $\lesssim100\msun$.

In this study, we adopt the double power law IMF with extensive parameter range,
perform a binary population synthesis calculation, and investigate what types of BBH mergers 
with IMBHs will be observed by current and future GW detectors
when using a reasonable IMF which meets the Pop. III star formation simulation and current
GW observation.
In addition, we investigate the dependence of the properties of BBHs with IMBHs on the common envelope parameter $\alpha\lambda$.

This paper is constructed as follows.
In Section \ref{sec: method}, we describe our method of binary population synthesis calculations.
In Section \ref{sec: results}, we show the formation channel, primary BH mass distribution,
mass ratio distribution, effective spin distribution, delay time distribution and 
the redshift evolution of merger rate density
for all sub-populations of BBH mergers with IMBHs.
We also show in detail the dependence of such distribution on the common envelope parameter $\alpha\lambda$.
We compare our obtained merger rate density with the upper limit and impose
restrictions on IMF and $\alpha\lambda$ models, and compute the detection rate and detected BBH mass 
distribution in Section \ref{sec: discussion}.
Finally, we summarize our paper in Section \ref{sec: summary}.


\section{Method} \label{sec: method}
In order to calculate the merger rate density of Pop. III BBH mergers for various Pop. III IMFs, 
we perform binary population synthesis calculations.
Our binary population synthesis code is identical with that of \citet{hijikawa21}.
In section \ref{subsec: single and binary evolution}, 
we overview the stellar and binary evolution implemented in our code.
In section \ref{subsec: initial conditions}, 
we describe IMFs and other initial conditions we adopt.
How we compute the merger rate density of Pop. III BBH mergers is described 
in Section \ref{subsec: rate calculation}.
Finally, in Section \ref{subsec: parameter sets}, we summarize our parameter sets.

\subsection{Stellar and Binary Evolution Physics}\label{subsec: single and binary evolution}
Our binary population synthesis code is based on {\tt BSE} \citep{hurley00, hurley02}
\footnote[1]{\url{http://astronomy.swin.edu.au/\~jhurley/}}.
Some additions and changes have been made to this, and these are described below.

In our code, the fitting formulae for $Z=10^{-8}\zsun$ stars were
implemented \citep{tanikawa20, tanikawa21a, tanikawa21, tanikawa22}, published as
{\tt BSEEMP}\footnote[2]{\url{https://github.com/atrtnkw/bseemp}}. 
$Z$ and $\zsun=0.02$ are the stellar metallicity and sollar metallicity, respectively.
In their studies, the evolution of $Z=10^{-8}\zsun$ stars up to 1280 $\msun$.
are calculated by the 1D stellar evolution code, HOSHI \citep{takahashi16,takahashi19,takahashi18,yoshida19}.
In our study, we treat $Z=10^{-8}\zsun$ stars as Pop. III stars ($Z=0$).
There is little difference in the stellar evolution between $Z=10^{-8}\zsun$ stars
and $Z=0$ stars for the following reason. 
As the abundance of heavy elements in a $Z=10^{-8}\zsun$ star is extremely low, 
the CNO cycle does not operate initially. 
The abundance of carbon, nitrogen and oxygen are increased by the triple alpha reaction, 
and then the CNO cycle commences. 
These evolutionary properties are same as those of $Z=0$ stars. 
Therefore, the difference in metallicity, $Z=10^{-8}\zsun$ and $Z=0$, 
has little effect on the stellar evolution. Our code has two stellar models in which
the convective overshooting is more effective and less so, L and M models
\citep{yoshida19, tanikawa20, tanikawa21a, tanikawa21, tanikawa22}.
In this paper, we use only L model.

We adopt the `rapid' model in \citet{fryer12} with the modification for 
the pulsational pair-instability and pair-instability supernova 
\citep[see equations 5--7 in][]{tanikawa21}.
We assume that the mass of a BH formed through pulsational 
pair-instability is 45 $\msun$, and the minimum helium core mass 
of direct collapse is 135 $\msun$ in the same way as \citet{belczynski16b}.
Therefore, the mass range of pair-instability mass gap is 45--135 $\msun$.

In a binary system, orbital angular momenta and spin angular momenta are exchanged by 
tidal interaction.
In our code, the tidal coefficient factor $E$ for the dynamical tide is implemented as 
\begin{equation}
   E=\left\{
   \begin{aligned}
		&10^{-0.42}\left(\frac{R_\mathrm{con}}{R}\right)^{7.5}&&\text{for H-rich stars},\\
      &10^{-0.93}\left(\frac{R_\mathrm{con}}{R}\right)^{6.7}&&\text{for He-rich stars},
   \end{aligned}
   \right.
\end{equation}
\citep{qin18}, where $R$ is the stellar radius and $R_\mathrm{con}$ is the convective core radius, respectively.
We use the values $R_\mathrm{con}=1\rsun$ and $0.5\rsun$ for 
main-sequence stars and naked helium stars, respectively 
\citep{tanikawa20,kinugawa20}.

When one of the star in the binary system fills its Roche lobe, the stellar mass overflows.
If this mass transfer is unstable, the binary system enters a common envelope
phase; if not, the stable mass transfer occurs.
In our code, the mass transfer rate of stable mass transfer is calculated as 
\begin{align}
	\dot{M}_\don = 
	-\frac{f(\mu)M_\don}{\sqrt{R_\don^3/GM_\don}}
	\left(\frac{\Delta R_\don}{R_\don}\right)^{n+3/2}d_n,
\end{align}
where
\begin{align}
	f(\mu) = 
	\frac{4\mu\sqrt{\mu}\sqrt{1-\mu}}{(\sqrt{\mu}+\sqrt{1-\mu})^4}
	\left(\frac{a}{R_\don}\right)^3,
\end{align}
$\mu=M_\don/M_\tot$, $\Delta R_\don=R_\don-R_\mathrm{L,\don}$, $G$, $M_\don$, 
$R_\don$, $n$, $d_n$, $a$, $M_\tot$, and $R_\mathrm{L,\don}$ are the 
gravitational constant, the donor mass, the donor radius, the polytropic
index of the donor star, the normalization factor depending on $n$, 
orbital separation, the total mass, and the donor Roche lobe radius,
respectively \citep[e.g.,][]{paczynski72,savonije78,edwards87}.
In our study, we assume that the mass transfer is conservative, unless the 
accretor is a compact object.

If a mass transfer becomes unstable ($\zeta_\mathrm{L}>\zeta_\mathrm{ad}$), 
the binary system enters a common envelope phase.
$\zeta_\mathrm{L}=\mathrm{d}\log R_\mathrm{L}/\mathrm{d}\log M$ is the response of the Roche lobe
radius $R_\mathrm{L}$ to the change in stellar mass $M$ and
$\zeta_\mathrm{ad}=(\mathrm{d}\log R/\mathrm{d}\log M)_\mathrm{ad}$
is the response of stellar radius $R$ to the change in stellar mass within the dynamical timescale. 
We use the same $\zeta_\mathrm{ad}$ value as \citet{kinugawa14}, and this criteria is roughly
consistent with \citet{pavlovskii17}.
In this study, we use the $\alpha\lambda$ formalism for common envelope
evolution \citep{webbink84}, and the orbital separation just after the common envelope phase $a_\mathrm{f}$ is
calculated by the following energy budget if the accretor star does not have
a clear core-envelope structure.
\begin{align}
   \alpha\left(\frac{GM_\mathrm{c,don}M_\mathrm{acc}}{2a_\mathrm{f}}-\frac{GM_\mathrm{don}M_\mathrm{acc}}{2a_\mathrm{i}}\right)=
   \frac{GM_\mathrm{don}M_\mathrm{env,don}}{\lambda R_\mathrm{don}}.
\end{align}
Here, $M_\mathrm{c,don}$ and $M_\mathrm{env,don}$
are the stellar core and envelope mass of the donor star, 
$M_\mathrm{acc}$ is the accretor stellar mass and 
$a_\mathrm{i}$ is the orbital separation just before the common envelope phase.
$\alpha$ is the efficiency parameter showing how much orbital energy is used to strip the stellar envelope.
$\lambda$ is the binding energy parameter.
The value of the common envelope parameters, $\alpha$ and $\lambda$, 
are still not well known.
Thus, in this study, we use various values for the product of the 
common envelope parameters, $\alpha\lambda$:
$\alpha\lambda=$ 0.01, 0.1, 0.5, 1.0, 2.0 and 5.0.
We call the common envelope phase during which the accretor star does not have a core-envelope structure
``single common envelope''.

When the accretor star which does not fill its Roche lobe is also a giant star, 
the orbital energy is used to strip the envelope of both stars.
We call this common envelope ``double common envelope'', and 
the orbital separation just after the common envelope phase $a_\mathrm{f}$ is calculated as
\begin{align}\label{eq: dce}
   \alpha\left(\frac{GM_\mathrm{c,don}M_\mathrm{c,acc}}{2a_\mathrm{f}}-\frac{GM_\mathrm{don}M_\mathrm{acc}}{2a_\mathrm{i}}\right)=
   &\frac{GM_\mathrm{don}M_\mathrm{env,don}}{\lambda R_\mathrm{don}}+\notag\\
   &\frac{GM_\mathrm{acc}M_\mathrm{env,acc}}{\lambda R_\mathrm{acc}}.
\end{align}
$M_\mathrm{c,acc}$, $M_\mathrm{env,acc}$ and $R_\mathrm{acc}$ are the core and envelope mass and stellar radius of accretor star.
In our calculation, when both stars fill their Roche lobes, the binary system enters a common envelope phase
even though $\zeta_\mathrm{L}>\zeta_\mathrm{ad}$ is not met.
We call this common envelope ``contact common envelope'', and 
the orbital separation just after the common envelope phase $a_\mathrm{f}$ is 
calculated as equation (\ref{eq: dce}).

\subsection{Initial Conditions} \label{subsec: initial conditions}
In this study, we adopt various IMFs and calculate the merger rate density of BBH mergers.
In Section \ref{subsec: imfs}, we briefly review the recent Pop. III star formation simulation,
and describe our IMFs for Pop. III stars.
Other initial conditions such as initial mass ratio function, initial orbital separation function and 
initial eccentricity function are shown in Section \ref{subsec: other initial conditions}.

\subsubsection{Initial Mass Functions} \label{subsec: imfs}
Pop. III stars are formed from metal free gas.
Star forming clouds are formed in the mini-halo and at the center the gas cloud
collapses due to the $\mathrm{H}_2$ and HD cooling.
The cooling efficiency is weaker than in the Pop. I/II case, and hence
the Jeans mass of the gas cloud is larger.
\memo{According to \citet{hirano15}, 
the Pop. III stellar mass
$M_\mathrm{III}\propto \dot{M}_\mathrm{Jeans}\propto \dot{M}_\mathrm{vir}\propto M_\mathrm{vir}$
is roughly proportional to the virial mass $M_\mathrm{vir}$ for the $\mathrm{H}_2$ cooling.
Here, $\dot{M}_\mathrm{Jeans}$ and $\dot{M}_\mathrm{vir}$ are the mass infall rate
measured at the Jeans scale and the halo scale, respectively.
The virial mass distribution of dark matter mini-halo derived by cosmological simulation
shows that the number of mini-halo decreases above a certain virial mass,
and this corresponds to 200--300 $\msun$ in the Pop. III stellar mass distribution
using the above relationship.}
According to Pop. III star formation simulation \citep{hirano14,hirano15,susa14}, 
the ZAMS mass distribution of Pop. III stars is roughly flat for $\lesssim100\msun$
and there is break around at $\sim100\msun$.
In this study, in order to reproduce this structure, 
we adopt a double power law IMF for the Pop. III single stars 
and primary stars of Pop. III binaries.
Our double power law IMFs $f(m)$ have three parameters as shown in the following equations.
\begin{equation}
   f(m)\propto\left\{
   \begin{aligned}
		&m^{-\gamma_1}&&(m_\mathrm{min}\le m \le m_\mathrm{crit}),\\
      &m_\mathrm{crit}^{\gamma_2-\gamma_1}\times m^{-\gamma_2}&&(m_\mathrm{crit}<m\le  m_\mathrm{max}),
   \end{aligned}
   \right.
\end{equation}
where $m$ is the stellar mass and 
$m_\mathrm{min}=$ 10 $\msun$ and $m_\mathrm{max}=$ 1500 $\msun$ are the minimum mass and maximum mass
in our calculation, respectively.
At the critical mass $m_\mathrm{crit}$, the slope of the IMF changes from $\gamma_1$ to $\gamma_2$.
In order to connect the IMF continuously at $m=m_\mathrm{crit}$, 
the coefficient $m_\mathrm{crit}^{\gamma_2-\gamma_1}$ is multiplied.
According to Pop. III star formation simulations \citep{susa14,hirano14,hirano15},
we set the critical mass $m_\mathrm{crit}$ to 100, 200 and 300 $\msun$.
The slope of lower mass region ($m<m_\mathrm{crit}$), $\gamma_1$, is set to 0 and 1
to reproduce a top heavy function in the lower mass region.
The slope of higher mass region ($m>m_\mathrm{crit}$), $\gamma_2$, 
is set to 1.5, 2.0, 2.5, 3.0, 3.5, 4.0, 4.5 and 5.0.

\subsubsection{Other Initial Conditions} \label{subsec: other initial conditions}
In order to determine the initial condition, we set the initial mass ratio, 
initial orbital separation, and initial eccentricity distribution in addition to the IMF.
The initial mass ratio distribution is flat ($\propto\const.$), and the
mass ratio range is $q_\mathrm{min}$--1, where 
$q_\mathrm{min}=10/M_1$ and $M_1$ is the mass of initially heavier star.
The initial orbital separation distribution is logarithmically flat 
distribution, and the orbital separation range is $a_\mathrm{min}$--
$10^5\rsun$, where $a_\mathrm{min}$ is determined so that the ZAMS 
radius does not exceed the Roche lobe radius.
In this study, we assume that the initial eccentricity is zero (circular
orbit).

\subsection{Rate Calculation} \label{subsec: rate calculation}
In this study, we sample $N_\mathrm{sam}=10^6$ binaries consisting of
ZAMS stars with the metallicity $10^{-8}Z_\odot$ and primary mass $\in[10,1500]\msun$
according to various initial functions and
follow the evolution of $N_\mathrm{sam}$ binaries for each parameter sets.
The detail of parameter sets are described in Section \ref{subsec: parameter sets}.
The merger rate $\mathcal{R}(z)$ at the redshift $z$ is calculated as follows.
\begin{align}
   \mathcal{R}(z(t))=\int_0^t f_\mathrm{b}\frac{\mathrm{SFR}(\tau)}{\langle m\rangle}\frac{n(t-\tau)}{N_\mathrm{sam}}~\mathrm{d}\tau.
\end{align}
$t$ is the age of the Universe at the redshift $z$ and is calculated assuming the $\Lambda$-CDM cosmology.
The Hubble constant is $H_0=67.4~\mathrm{km}~\mathrm{s}^{-1}~\mathrm{Mpc}^{-1}$ and
the matter density parameter is $\Omega_\mathrm{m0}=0.315$ \citep{planck20}.
$\mathrm{SFR}(\tau)$ is the star formation rate, and we adopt the star 
formation history in \citet{desouza11} for Pop. III stars, 
but reduce it by a factor of three \citep{inayoshi16,kinugawa20}.
We assume that the binary fraction $f_\mathrm{b}$ is 0.5.
\begin{align}
   \langle m\rangle = \int_I mf(m)~\mathrm{d}m+f_\mathrm{b}\int_I\mathrm{d}m~f(m)\int_{J(m)}mqg(q;m)~\mathrm{d}q,
\end{align}
is the mean mass of Pop. III stellar objects (single stars and binaries).
$q\in(0,1]$ is the mass ratio,
$I=[m_\mathrm{min}, m_\mathrm{max}]=[10,1500]\msun$ is the primary mass range, 
$g(q;m)=1/(1-m_\mathrm{min}/m)$ is the initial mass ratio function 
when the primary mass is $m$ and 
$J(m)=[m_\mathrm{min}/m,1]$ is the initial mass ratio range.
\begin{align}
   n(t)=\frac{\text{the number of BBHs whose delay time is in }[t,t+\delta t]}{\delta t},
\end{align}
is the delay time distribution function and we set $\delta t=0.1$ Gyr. 
As long as we use such a small value for $\delta t$, the results do not change.

For all parameter sets, we additionally sample the $N_{\mathrm{sam},J}=10^6$ binaries
in the primary initial mass range $J=[150,600]\msun\subset I$
in order to increase the number of samples efficiently and to obtain the more precise results.
In such a case, the merger rate density $\mathcal{R}(z)$ is calculated as follows.
\begin{align}
   \mathcal{R}(z(t))=\int_0^t &f_\mathrm{b}\frac{\mathrm{SFR}(\tau)}{\langle m\rangle} \left(A_J\frac{n_{I}^{(J)}(t-\tau)+n_J(t-\tau)}{N_{\mathrm{sam},I}^{(J)}+N_{\mathrm{sam},J}} + \notag\right.\\
   &(1-A_J)\left.\frac{n_I(t-\tau)-n_I^{(J)}(t-\tau)}{N_{\mathrm{sam},I}-N_{\mathrm{sam},I}^{(J)}}\right)~\mathrm{d}\tau,
\end{align}
$N_{\mathrm{sam},I}=N_\mathrm{sam}$
is the number of samples for the sampling in the primary initial mass range $I$.
$N_{\mathrm{sam},I}^{(J)}<N_{\mathrm{sam},I}$ is the number of samples whose primary initial mass is in the 
range $J\subset I$ for the sampling in the primary initial mass range $I$.
$n_I(t)=n(t)$ and $n_J(t)$ are the delay time distribution 
for the sampling in the primary initial mass range $I$ and $J$, respectively.
$n^{(J)}_I(t)<n_I(t)$ is the delay time distribution of BBHs 
whose primary initial mass is in the range $J$ for the sampling
in the primary initial mass range $I$. Finally,
\begin{align}
   A_J=\int_J f(m)~\mathrm{d}m < \int_I f(m)~\mathrm{d}m=1,
\end{align}
is calculated from an IMF.

\subsection{Parameter Sets} \label{subsec: parameter sets}
As written in Section \ref{subsec: single and binary evolution} and 
Section \ref{subsec: imfs}, we use $\alpha\lambda=$ 0.01, 0.1, 0.5, 1.0, 2.0 and 5.0 for 
common envelope evolution, and $\gamma_1=$ 0.0, 1.0, $\gamma_2=$ 1.5, 2.0, 2.5, 3.0, 3.5, 4.0, 4.5 and 5.0, and 
$m_\mathrm{crit}/\msun=$ 100, 200 and 300 for the Pop. III IMF.
Therefore, the number of IMFs in our study is $2\times8\times3=48$, and thus
the number of parameter sets in our study is $6\times48=288$.
For these sets, we perform the binary population synthesis calculation and 
commpute the merger rate density of Pop. III BBH mergers.
Finally, we summarize our parameter sets in Table \ref{table: parameter}.

\begin{table}
   \caption{Parameter sets}
   \label{table: parameter}
   \begin{center}
     \begin{tabular}{cc}\hline
      Parameter & Value\\
      \hline
      common envelope parameter, $\alpha\lambda$ & 0.01, 0.1, 0.5, 1.0, 2.0, 5.0\\
      the slope of the lower mass region, $\gamma_1$ & 0.0, 1.0\\
      the slope of the higher mass region, $\gamma_2$ & 1.5, 2.0, 2.5, $\ldots$, 4.5, 5.0\\
      critical mass, $m_\mathrm{crit}$ & 100, 200, 300 $\msun$\\
      \hline
     \end{tabular}
  \end{center}
\end{table}


\section{Results} \label{sec: results}
The BHs in the mass range 45--135 $\msun$ cannot be formed due to 
the pair-instability effect (see Section \ref{subsec: single and binary evolution}).
Therefore, the BBHs with IMBHs ($>100$ $\msun$) are classified into two types.
In this paper, we call a BH whose mass is lower than the lower edge of pair-instability mass gap (45 $\msun$)
``low mass BH'' and call a BH whose mass is higher than the upper edge of pair-instability mass gap (135 $\msun$)
``high mass BH''.
We describe the pair of ``low mass BH'' and ``high mass BH'' as ``low mass $+$ high mass'' and
describe the pair of ``high mass BH''s as ``high mass $+$ high mass''.
In Section \ref{subsec: hh} and \ref{subsec: lh}, we show the primary BH mass, mass ratio, effective spin 
, delay time and merger rate density of ``high mass $+$ high mass'' and ``low mass $+$ high mass''.
We also discuss the dependence of these distribution on common envelope parameter $\alpha\lambda$.

\subsection{``high mass $+$ high mass''} \label{subsec: hh}
\subsubsection{formation channel} \label{subsubsec: channel hh}
\begin{figure*}
   \begin{center}
      \includegraphics[width=\linewidth]{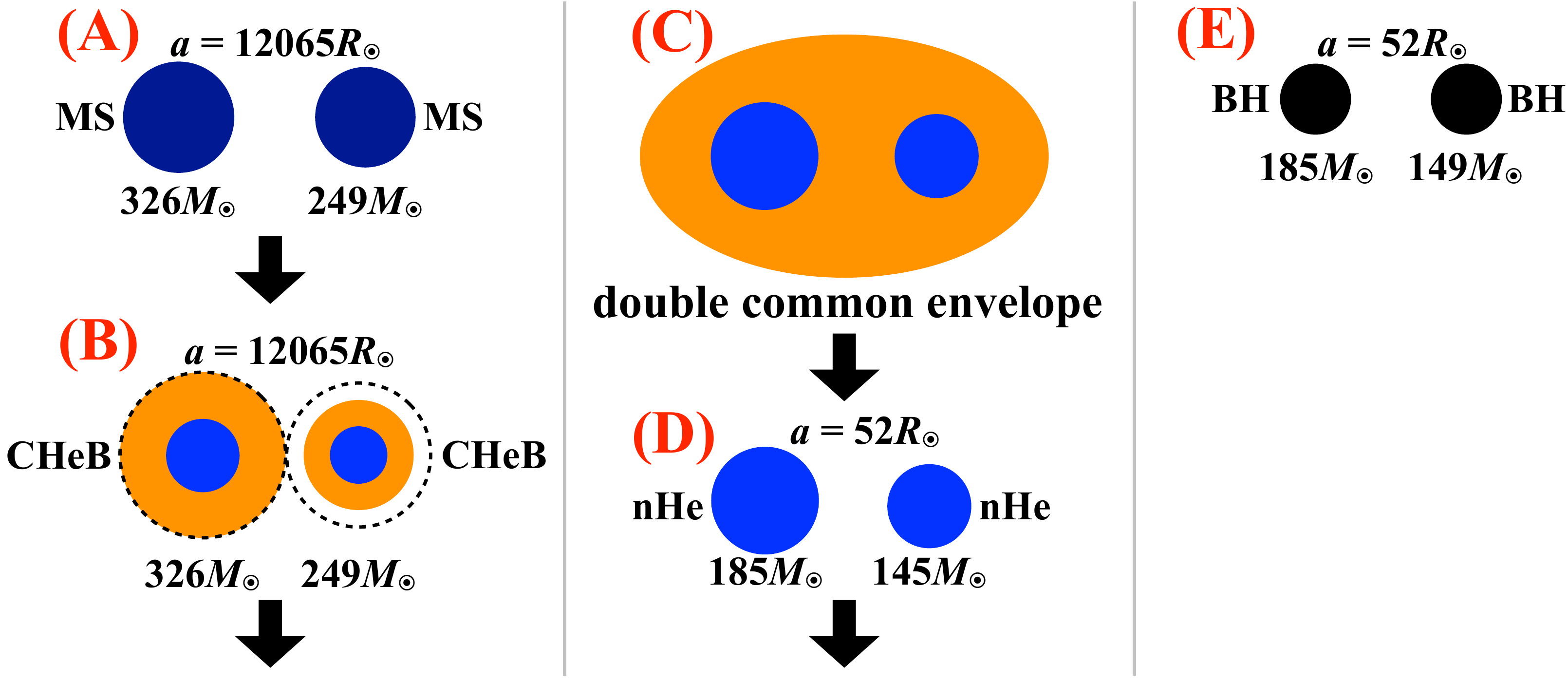}
   \end{center}
   \caption{An example of ``double common envelope channel'' when $\alpha\lambda=1.0$.
   ``CHeB'' and ``nHe'' stand for the core helium burning
   phase and naked helium star, respectively.
   The dashed circles in phase (B) indicate the Roche lobes of both stars.}
   \label{fig: double ce}
\end{figure*}
\begin{figure*}
   \begin{center}
      \includegraphics[width=\linewidth]{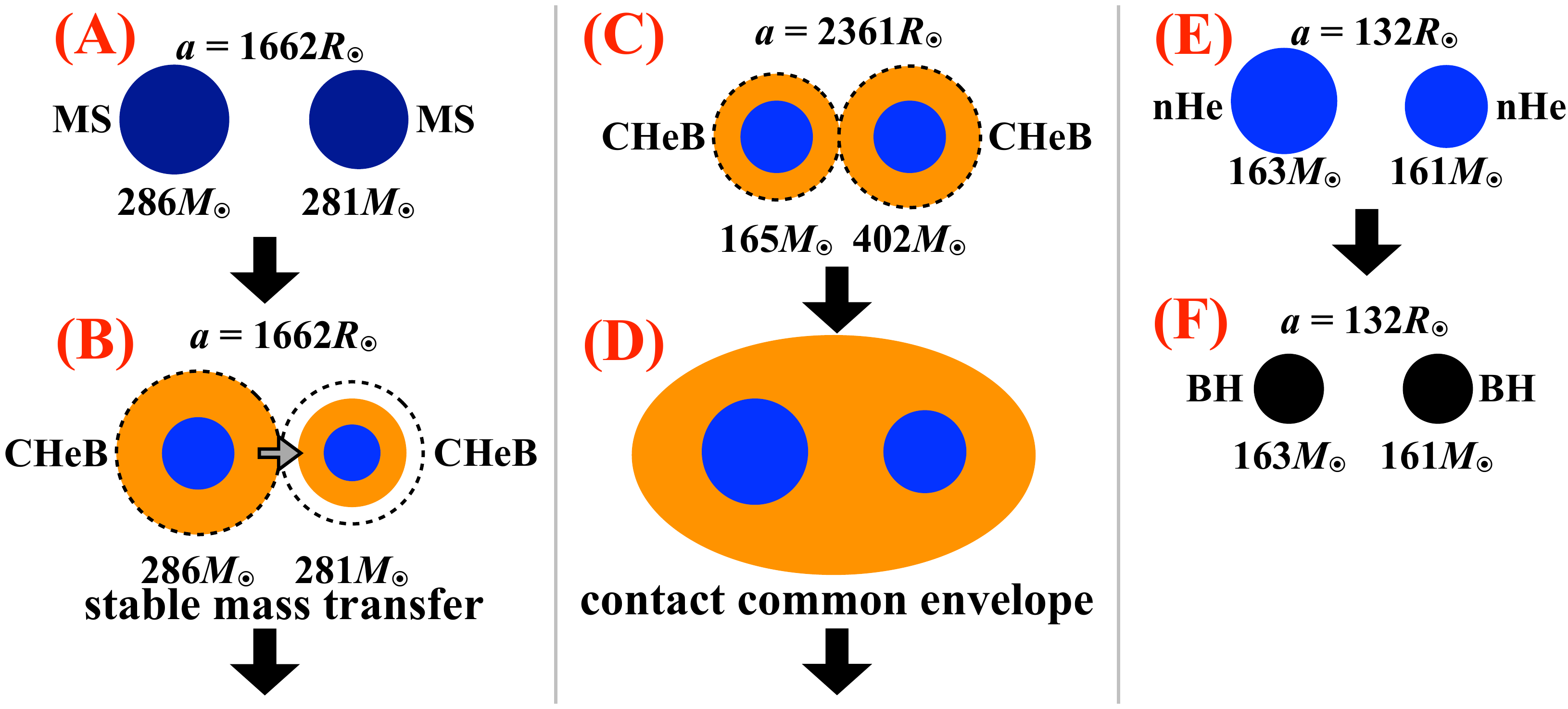}
   \end{center}
   \caption{An example of ``contact common envelope channel'' when $\alpha\lambda=1.0$.}
   \label{fig: contact ce}
\end{figure*}
\begin{figure*}
   \begin{center}
      \includegraphics[width=\linewidth]{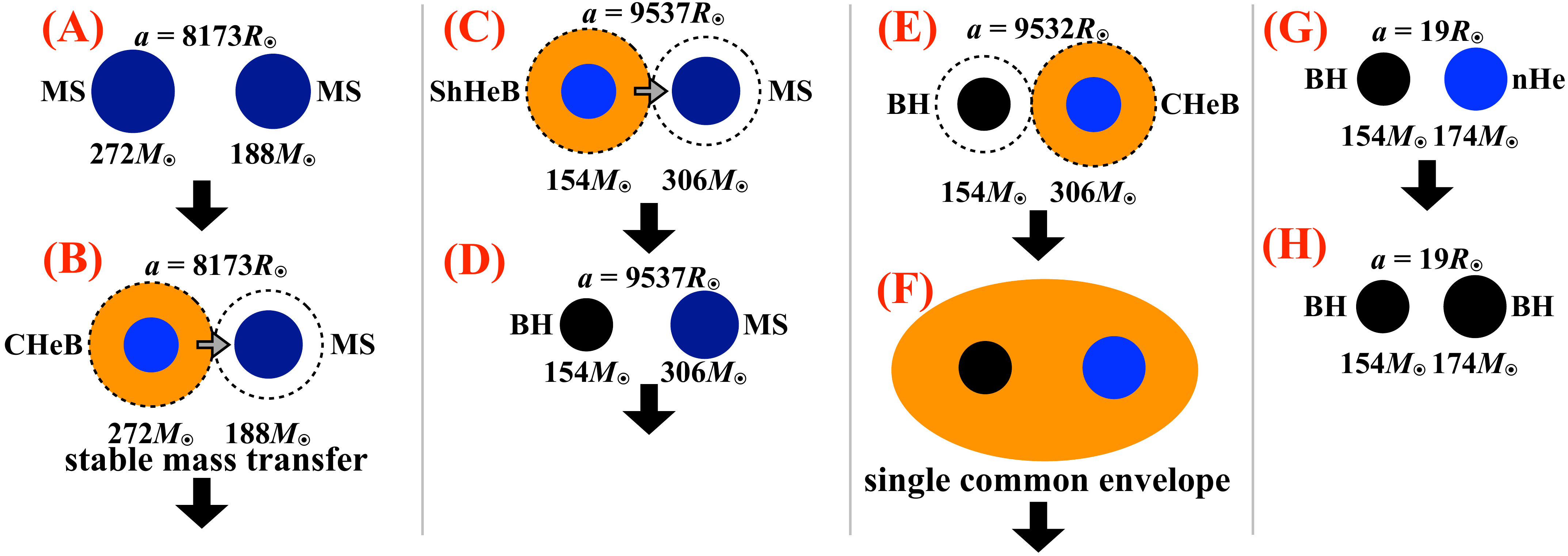}
   \end{center}
   \caption{An example of ``single common envelope channel (2)'' when $\alpha\lambda=0.01$.
   ``ShHeB'' stands for the shell helium burning phase.}
   \label{fig: sce}
\end{figure*}
First, we introduce the dominant formation channel of ``high mass $+$ high mass''
for each common envelope parameter $\alpha\lambda$.
In our calculation, when $\alpha\lambda=$ 0.5, 1.0, 2,0 and 5.0, ``high mass $+$ high mass'' BBHs are 
formed through ``double common envelope channel'' and ``contact common envelope channel''.
The definitions of these evolutionary channels are that
a binary system pass through one ``double common envelope'' and ``contact common envelope'', respectively 
(see Section \ref{subsec: single and binary evolution}).
An example of ``double common envelope channel'' and ``contact common envelope channel''
are shown in Figure \ref{fig: double ce} and \ref{fig: contact ce} (for $\alpha\lambda=1.0$).
When $\alpha\lambda=$ 0.1, ``single common envelope channel (2)'' emerges and becomes dominant
in addition to ``double common envelope channel'' and ``contact common envelope channel''.
``Single common envelope channel (2)'' is the evolutionary channel that passes through one ``single common envelope''
whose donor star is the secondary star at the ZAMS time, i.e., the initially lighter star.
When $\alpha\lambda=0.01$, only ``single common envelope channel (2)'' can form a merging BBH.
An example channel is shown in Figure \ref{fig: sce} (for $\alpha\lambda=0.01$).
Almost all BBH formed through ``single common envelope channel (2)'' are mass inverted.

In general, it is easier to shrink through ``double common envelope channel'' and 
``contact common envelope channel'' than through ``single common envelope channel (2)''
because more orbital energy are used to strip the stellar envelope through the 
former common envelope than the latter common envelope.
Therefore, if $\alpha\lambda$ is too small such as 0.01, 
the binary system entering ``double common envelope channel''
and ``contact common envelope channel'' coalesces after the common envelope phase,
and thus they cannot form a merging BBH.
On the other hand, when $\alpha\lambda=$ 0.5, 1.0, 2.0,
it is impossible to shrink the orbital separation sufficiently 
to merge within a Hubble time through ``single common envelope channel (2)''.
When $\alpha\lambda=0.1$, both formation channels can occur.

\subsubsection{primary BH mass distribution} \label{subsubsec: primary mass hh}
\begin{figure*}
   \begin{minipage}{\hsize}
      \begin{center}
         \includegraphics[width=\linewidth]{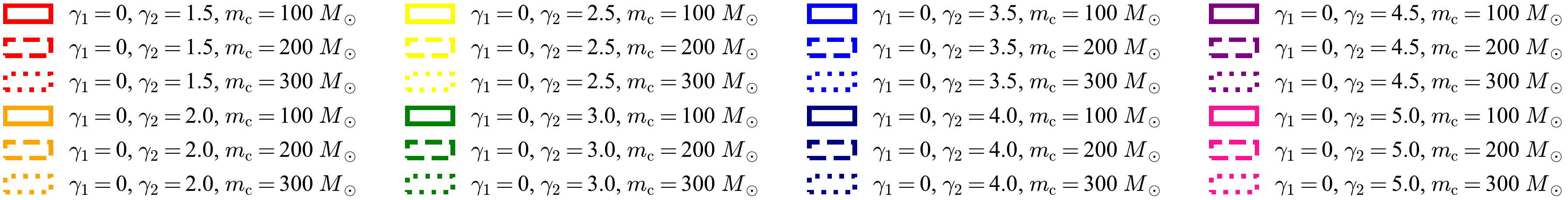}
      \end{center}
   \end{minipage}
   \begin{minipage}{\hsize}
      \begin{center}
         \includegraphics[width=\linewidth]{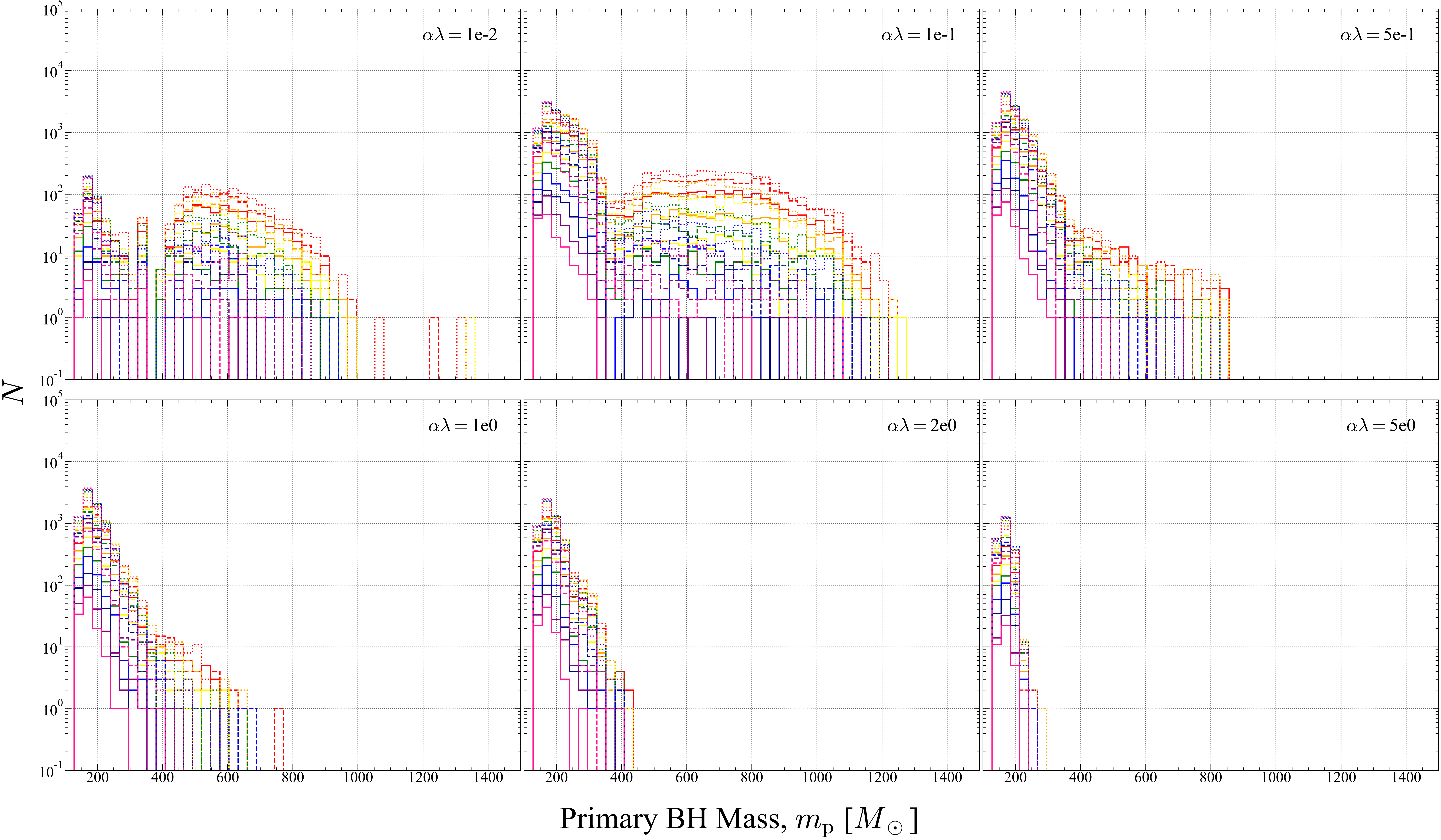}
      \end{center}
   \end{minipage}
   \caption{The primary BH mass distributions of ``high mass $+$ high mass''
      BBH mergers which merge within a Hubble
      time. 
      From the upper left to upper right panel, $\alpha\lambda=$ 0.01, 0.1 and 0.5, and
      from the lower left to lower right panel, $\alpha\lambda=$ 1.0, 2.0 and 5.0.
      Each line corresponds to each IMF model.
      The normal lines and relatively transparent lines correspond to $\gamma_1=0.0$
      and 1.0, respectively.
      The red, orange, yellow, green, blue, navy, purple and pink lines correspond to
      $\gamma_2=1.5$, 2.0, 2.5, 3.0, 3.5, 4.0, 4.5 and 5.0, respectively.
      The solid, dashed and dotted lines correspond to $m_\mathrm{crit}=100\msun$,
      $200\msun$ and $300\msun$, respectively.}
   \label{fig: primary bh mass hh}
\end{figure*}
The primary BH mass distributions of ``high mass $+$ high mass''
BBH mergers which merge within a Hubble
time are shown in Figure \ref{fig: primary bh mass hh}.
For simplicity, we only show models with $\gamma_1=0.0$, but 
models with $\gamma_1=1.0$ also have a similar distribution.
The most distinctive feature is that the primary BH mass distributions have an
upper limit. 
This upper limit decreases as the common envelope parameter
$\alpha\lambda$ increases. The reason for this is as follows.
If the ZAMS mass of Pop. III stars is larger than $\sim600\msun$, they 
expand extremely, become a red super giant, and have a convective 
envelope when they still in their MS phase.
Therefore, if such a massive Pop. III MS star fills its Roche lobe, the 
mass transfer may be unstable, and the binary system enters a common 
envelope phase.
After that, the binary system always coalesces because the core--envelope
structure of MS star is not clear, and it will no longer evolve to a BBH.
In order to avoid this common envelope episode, the initial orbital 
separation needs to be large enough.

Hereafter, in order to understand why there is an upper limit on the
primary ZAMS mass of merging ``high mass $+$ high mass'' BBHs, 
we consider the dependence
of the merging timescale of the binary system on the primary ZAMS
mass.
The merging timescale $t$ is proportional to $a^4m_\mathrm{BH,p}^{-1}m_\mathrm{BH,s}^{-1}m_\mathrm{tot}^{-1}$, where $a$ is the orbital separation when BBH system is formed, $m_\mathrm{BH,p}$ is the primary BH mass, $m_\mathrm{BH,s}$ is the secondary BH mass and $m_\mathrm{tot}$ is the total mass of the BBH.
Here, if we fix the secondary BH mass, $m_\mathrm{BH,s}$, then the merging timescale $t$ is roughly proportional to $a^4m_\mathrm{BH,p}^{-2}$.
The orbital separation, $a$, is proportional to the radius $r$ of a Pop. III giant star heavier than
$\simeq 600\msun$, and the primary BH mass, $m_\mathrm{BH,p}$, is proportional to the ZAMS mass, $m$.
Since there is the relation $r\propto m^{0.6}$, $t\propto m^{0.4}$.
Therefore,  as the primary ZAMS mass increases, the merging timescale also increases, and
exceeds a Hubble time at a critical ZAMS mass. 
That is why there is an upper limit on the primary ZAMS mass of merging BBHs.

If the value of $\alpha\lambda$ is smaller, the efficiency of envelope stripping
through common envelope gets lower, and thus 
the binary system can shrink more during the common envelope phase.
Therefore, the more massive Pop. III stars can merge within a Hubble time
with the smaller $\alpha\lambda$, and the upper limit of primary BH mass
increases when the value of $\alpha\lambda$ decreases.

\subsubsection{mass ratio distribution} \label{subsubsec: mass ratio hh}
\begin{figure*}
   \begin{center}
      \includegraphics[width=\linewidth]{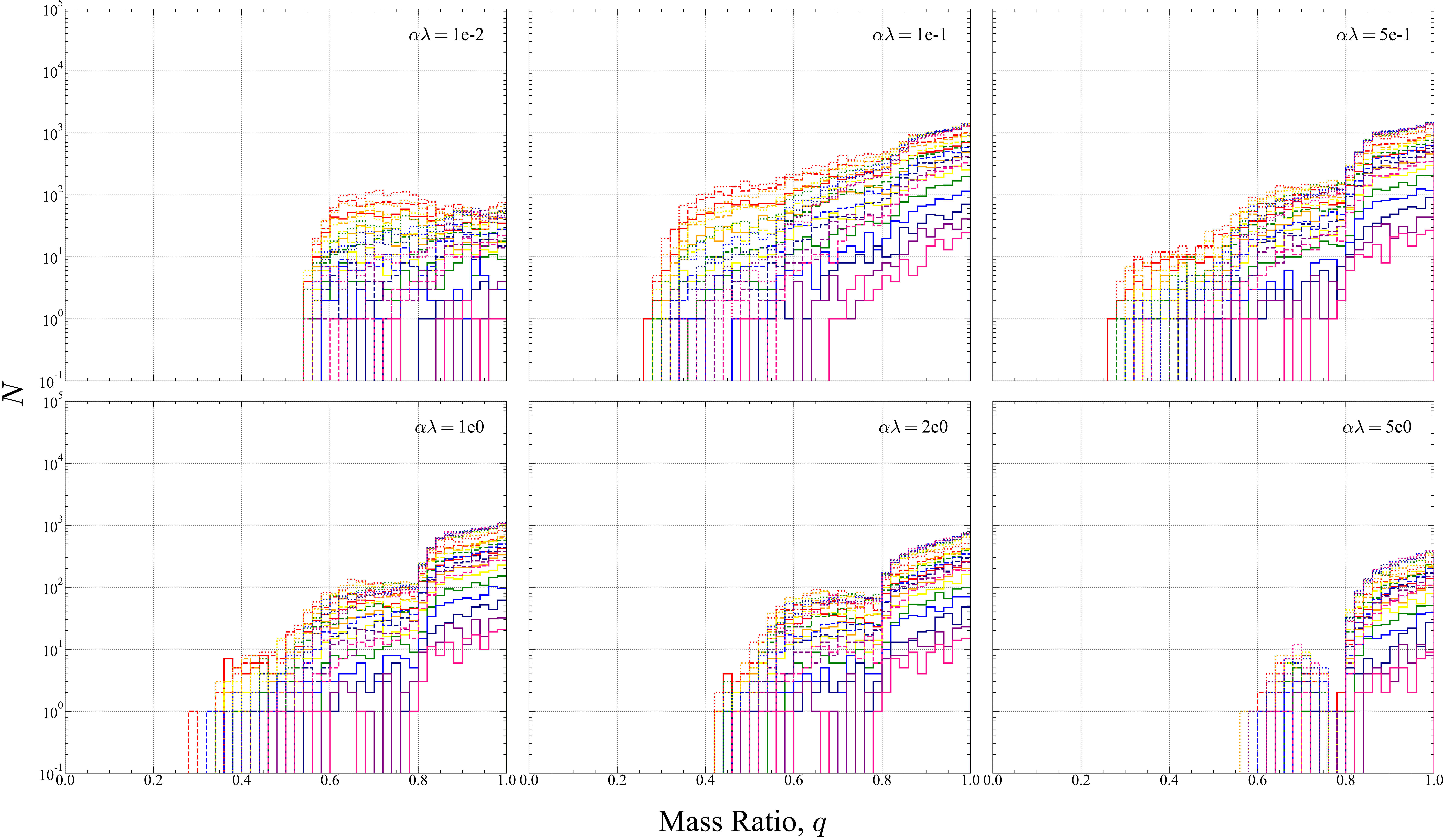}
   \end{center}
   \caption{The mass ratio distributions of ``high mass $+$ high mass''
   BBH mergers which merge within a Hubble time.
   From the upper left to upper right panel, $\alpha\lambda=$ 0.01, 0.1 and 0.5, and
   from the lower left to lower right panel, $\alpha\lambda=$1.0, 2.0 and 5.0.
   The legends are same as that of Figure \ref{fig: primary bh mass hh}.}
   \label{fig: mass ratio hh}
\end{figure*}
The mass ratio distributions of ``high mass $+$ high mass'' BBH mergers which merge within
a Hubble time are shown in Figure \ref{fig: mass ratio hh}.
When $\alpha\lambda\gtrsim$ 0.5, ``high mass $+$ high mass'' BBH mergers with the mass ratio $>0.8$ and $<0.8$
are formed through
``contact common envelope channel'' and ``double common envelope channel'', respectively.
If the binary is initially symmetric (initial mass ratio $\gtrsim0.8$), 
the evolutionary timescale is similar to each other and initially lighter star fills its Roche lobe
while the initially heavier star fills its Roche lobe and stable mass transfer proceeds.
After that, a ``contact common envelope'' occurs, only helium cores remain
and the binary system becomes a binary naked helium star.
Since the helium core mass ratio is roughly equal to initial mass ratio,
the mass ratio of BBH formed through ``contact common envelope channel'' is $\gtrsim0.8$.

If the initial mass ratio is $\lesssim0.8$, the evolutionary timescale of both stars are somewhat different.
Therefore, while the initially heavier star fills its Roche lobe and stable mass transfer
proceeds, the initially lighter star does not yet expand significantly.
After the initially lighter star becomes the post MS star but before it fills its Roche lobe,
the initially heavier star evolves and its envelope becomes a convective, and then the binary system
enters a common envelope phase.
Like ``contact common envelope channel'', the binary system becomes a binary naked helium star
after the common envelope phase, and thus the mass ratio of BBHs formed through
``double common envelope channel'' is roughly same as initial mass ratio.
If the initial mass ratio gets smaller, the wider orbital separation is needed for 
``double common envelope channel''. The reason is as follows.
In order to cause a ``double common envelope'', the initially heavier star fills its Roche lobe 
after the initially lighter star becomes a giant.
If the initial mass ratio gets smaller, the difference between each evolutionary timescale gets larger,
and thus the initially lighter star takes a relatively longer period to become a giant star.
In order for the initially heavier star not to fill its Roche lobe during this relatively longer period,
the wider initial orbital separation is needed.
If the initial orbital separation is too wide, the delay time exceeds a Hubble time.
In this way the minimum mass ratio of BBHs formed through ``double common envelope channel''
is determined.
If the $\alpha\lambda$ gets larger, the orbital separation just after the common envelope gets wider
(see Equation \ref{eq: dce}),
and thus the minimum mass ratio gets larger.
When $\alpha\lambda=0.1$, unlike the $\alpha\lambda\gtrsim0.5$ case, 
there are ``high mass $+$ high mass'' BBHs formed through
``single common envelope channel (2)'', but the mechanism
determining the minimum value of mass ratio $q$ is the same.
When $\alpha\lambda=0.01$, all ``high mass $+$ high mass'' BBHs which merge within a Hubble time
are formed through ``single common envelope channel (2)'' (see Figure \ref{fig: sce}).
Almost all ``high mass $+$ high mass'' BBHs are mass inverted for $\alpha\lambda=0.01$,
and have the mass ratio $q$
roughly equal to the ratio of the helium core mass of initially heavier star to
the helium core mass of initially lighter star which gains significant mass from the
initially heavier star (phase B in Figure \ref{fig: sce}).
Therefore, as the initial mass ratio increases, the mass ratio $q$ decreases.
Since the initial mass ratio has an upper limit ($=1$), the mass ratio $q$ also has
lower limit.

\subsubsection{effective spin distribution} \label{subsubsec: spin distribution}
\begin{figure*}
   \begin{center}
      \includegraphics[width=\linewidth]{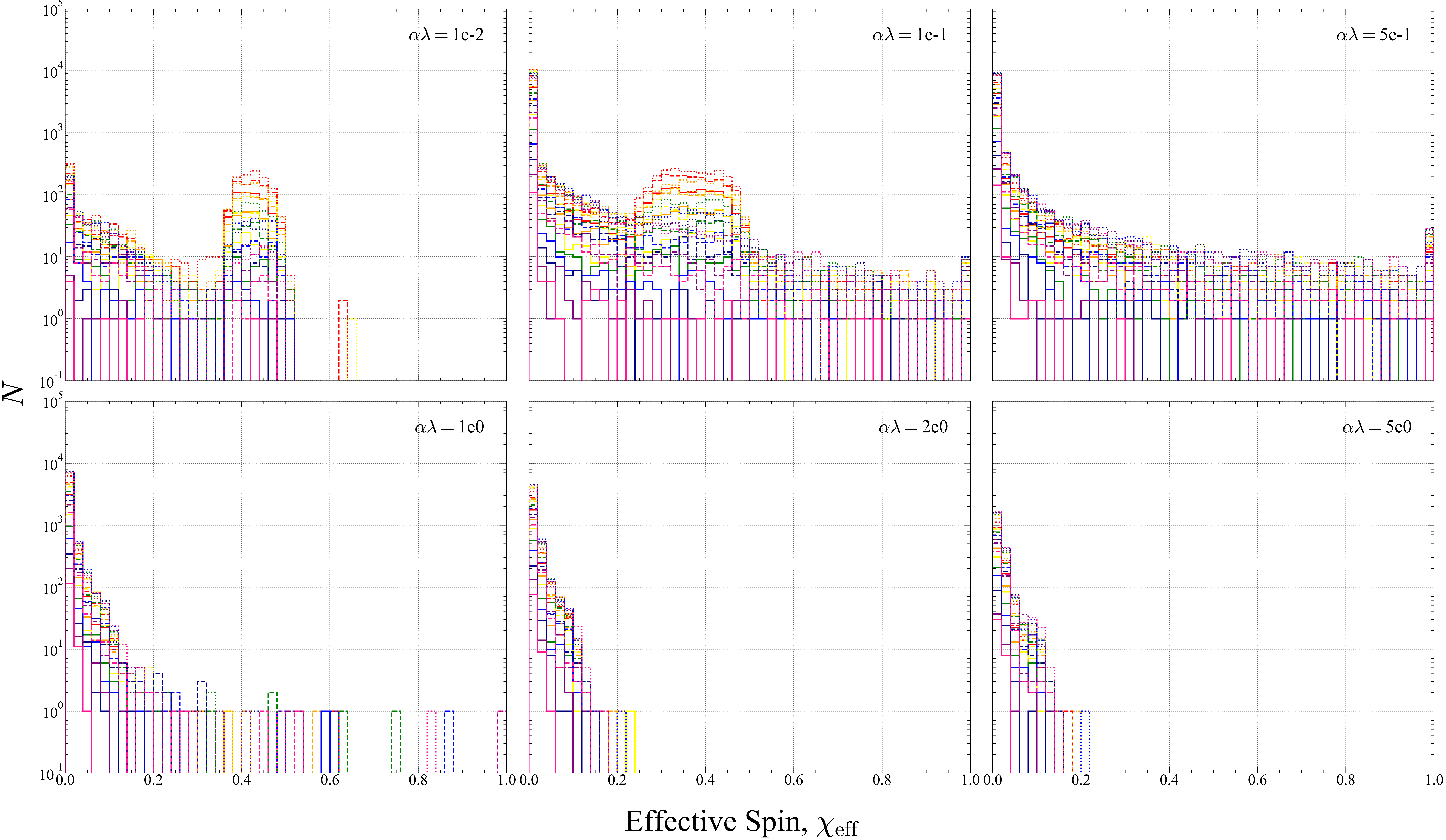}
   \end{center}
   \caption{The effective spin distribution of ``high mass $+$ high mass'' 
   BBH mergers which merge within a Hubble time.
   From the upper left to upper right panel, $\alpha\lambda=$ 0.01, 0.1 and 0.5, and
   from the lower left to lower right panel, $\alpha\lambda=$ 1.0, 2.0 and 5.0.
   The legends are same as that of Figure \ref{fig: primary bh mass hh}.}
   \label{fig: effective spin hh}
\end{figure*}
The effective spin distributions of ``high mass $+$ high mass'' BBH mergers which merge within
a Hubble time are shown in Figure \ref{fig: effective spin hh}.
The effective spin,
\begin{align}
   \chi_\mathrm{eff}=
   \frac{m_\mathrm{BH,p}\vec{\chi}_\mathrm{BH,p}+m_\mathrm{BH,s}\vec{\chi}_\mathrm{BH,s}}{m_\mathrm{BH,p}+m_\mathrm{BH,s}}
   \cdot\frac{\vec{L}_\mathrm{orb}}{|\vec{L}_\mathrm{orb}|},
\end{align}
is the mass weighted average of the dimensionless spin components perpendicular to the orbital plane.
Here, $m_\mathrm{BH,p}$ and $m_\mathrm{BH,s}$ are the primary and secondary BH mass,
$\vec{\chi}_\mathrm{BH,p}$ and $\vec{\chi}_\mathrm{BH,s}$ are the primary and secondary BH spin vector
and $\vec{L}_\mathrm{orb}$ is the orbital angular momentum vector.
Since we do not assume that the kick is imparted during the BH formation in this study, 
BHs always aligned with the orbital angular momentum vector.
Therefore, the effective spin can be written as
\begin{align}
   \chi_\mathrm{eff}&=
   \frac{m_\mathrm{BH,p}\chi_\mathrm{BH,p}+m_\mathrm{BH,s}\chi_\mathrm{BH,s}}{m_\mathrm{BH,p}+m_\mathrm{BH,s}}\notag\\
   &=\frac{\chi_\mathrm{BH,p}+q\chi_\mathrm{BH,s}}{1+q}.
\end{align}

When $\alpha\lambda\gtrsim$ 0.5, the ``high mass $+$ high mass'' BBH mergers are formed through
``contact common envelope channel'' or ``double common envelope channel''.
Through these channels, both stars lose their hydrogen envelope during the common envelope
phase and lose spin angular momenta at the same time.
Therefore, both BHs have very low spin ($\sim0$) and the effective spin distribution sharply peaks at 0.
If the orbital separation after the common envelope phase is too short,
the naked helium star can be spun up by tidal interaction.
In such a case, BHs have a non-zero spin and the effective spin of BBH mergers have a non-zero value.
If $\alpha\lambda$ gets larger, the orbital separation after the common envelope gets relatively wider,
and the degree of spin-up by tidal interaction gets smaller.
Therefore, the maximum effective spin gets smaller when $\alpha\lambda$ gets larger.
When $\alpha\lambda<$ 0.1, BBH mergers formed through ``single common envelope (2)''
emerges.
Through this channel, the progenitor of one of the BH lose its hydrogen envelope by the common
envelope, and thus this BH have a very low spin ($\sim0$).
On the other hand, the progenitor of the other BH did not undergo a significant mass loss process,
and thus the spin of this BH is high ($\sim1$).
Therefore, the effective spin is written as
\begin{align}
   \chi_\mathrm{eff}\sim\frac{q}{1+q}\in[0,0.5].
\end{align}
Therefore, the effective spin distribution of ``high mass $+$ high mass'' BBH mergers assemble
around non-zero value, 0.35--0.5 for $\alpha\lambda=$0.01 and 0.25--0.5 for $\alpha\lambda=$0.1.

\subsubsection{delay time distribution} \label{sebsubsec: delay time hh}
\begin{figure*}
   \begin{center}
      \includegraphics[width=\linewidth]{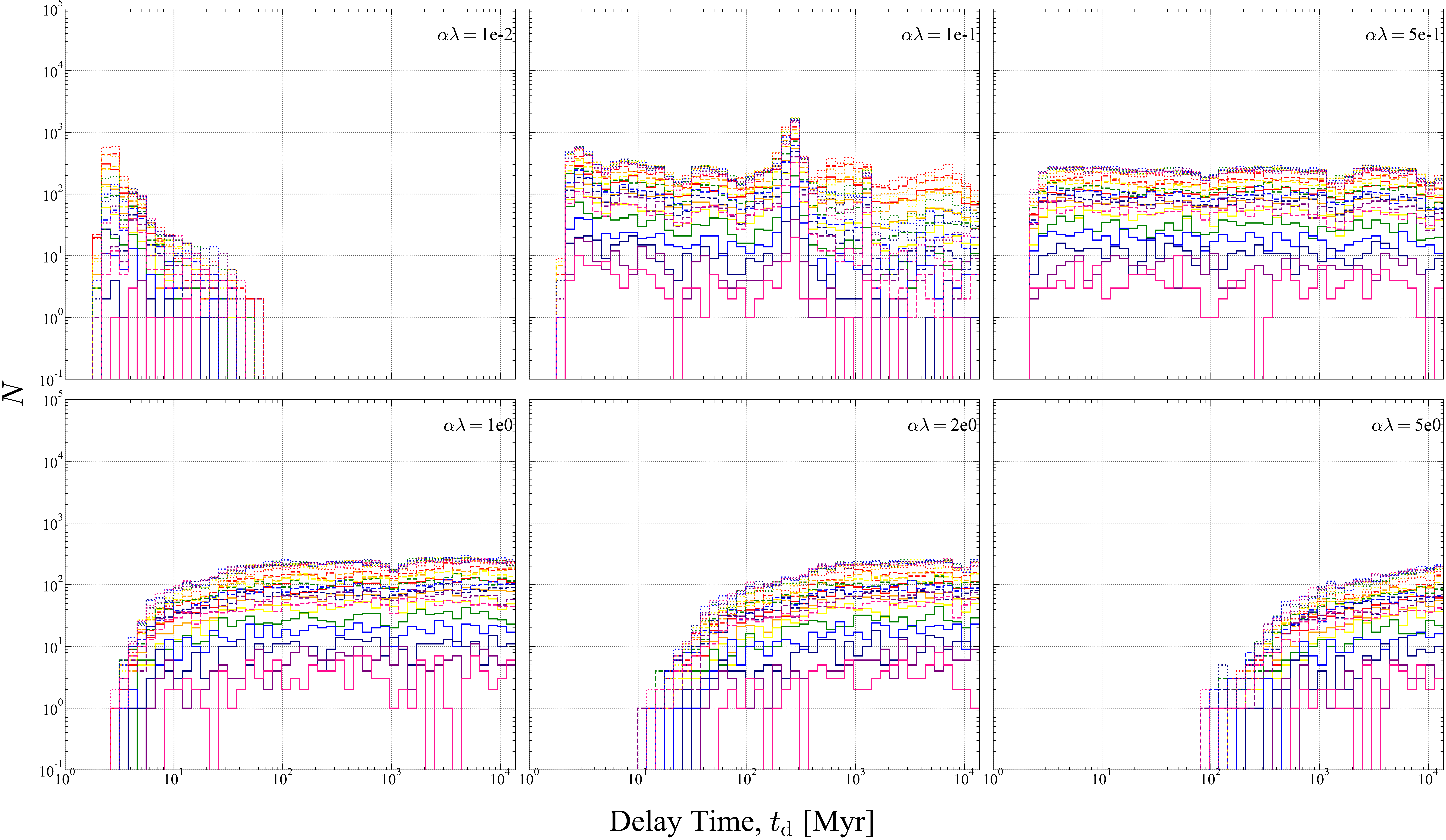}
   \end{center}
   \caption{The delay time distributions of ``high mass $+$ high mass'' BBH mergers 
   which merge within a Hubble time.
   From the upper left to upper right panel, $\alpha\lambda=$ 0.01, 0.1 and 0.5, and
   from the lower left to lower right panel, $\alpha\lambda=$ 1.0, 2.0 and 5.0.
   The legends are same as that of Figure \ref{fig: primary bh mass hh}.}
   \label{fig: delay time hh}
\end{figure*}
The delay time distributions of ``high mass $+$ high mass'' BBH mergers which merge within
a Hubble time are shown in Figure \ref{fig: delay time hh}.
The delay time $t_\mathrm{d}$ is roughly proportional to
$a^4m_\mathrm{BH,p}^{-1}m_\mathrm{BH,s}^{-1}m_\mathrm{tot}^{-1}$.
If the $\alpha\lambda$ gets larger, the orbital separation after the common envelope
gets wider, and thus the delay time gets longer.
Therefore, the minimum delay time increases when $\alpha\lambda$ gets larger,
and exceeds 10 and 100 Myr when $\alpha\lambda=$2.0 and 5.0.
When the value of $\alpha\lambda$ is too small such as 0.01, 
the orbital separation after the common envelope is too short, and thus
the maximum delay time is lower than 100 Myr.

\subsubsection{merger rate density} \label{subsubsec: merger rate hh}
\begin{figure*}
   \begin{center}
      \includegraphics[width=\linewidth]{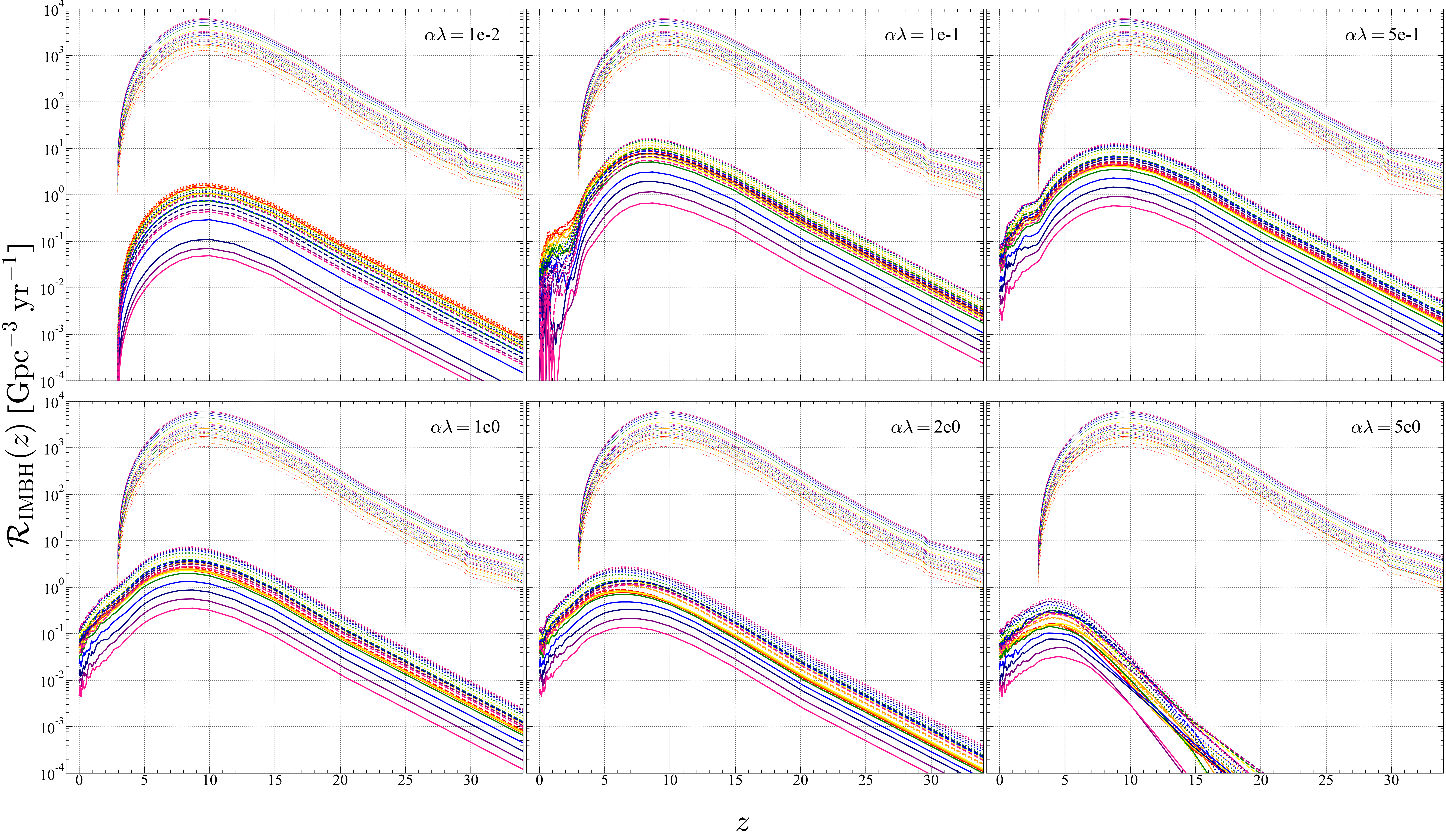}
   \end{center}
   \caption{The redshift evolution of merger rate density of ``high mass $+$ high mass''
   BBH mergers.
   From the upper left to upper right panel, $\alpha\lambda=$ 0.01, 0.1 and 0.5, and
   from the lower left to lower right panel, $\alpha\lambda=$ 1.0, 2.0 and 5.0.
   The legends are same as that of Figure \ref{fig: primary bh mass hh}.
   The lines at the bottom of each panel are the redshift evolution of BBH merger rate density.
   The relatively thin lines at the top of each panel indicate the binary 
   formation rate [$\mathrm{Gpc}^{-3}$ $\mathrm{yr}^{-1}$].
   The binary formation rate is the number density of Pop. III binaries born at
   each redshift and is calculated from star formation rate.}
   \label{fig: merger rate hh}
\end{figure*}
The redshift evolution of ``high mass $+$ high mass'' BBH merger rate density is shown
in Figure \ref{fig: merger rate hh}.
The merger rate densities roughly trace the star formation rate history \citep{desouza11}, and 
the peaks of merger rate densities are located near the redshift $z\sim10$.
However, when $\alpha\lambda=$ 2.0, the minimum delay time exceeds 10 Myr, hence the peak of 
merger rate density is shifted toward low redshift compared to that of $\alpha\lambda\lesssim$ 1.0.
When $\alpha\lambda=$ 5.0, the minimum delay time exceeds 100 Myr, hence the peak of 
merger rate density is significantly shifted toward low redshift compared to other cases.
Furthermore, the merger rate density becomes significantly smaller
at the redshift $z\gtrsim$ 15--20.
When $\alpha\lambda=0.01$, the maximum delay time is so short that the merger rate density rapidly decreases
after the end of Pop. III star formation.
Therefore, the merger rate densities are strictly 0 when $z\lesssim3$.

\subsection{low mass $+$ high mass} \label{subsec: lh}
\subsubsection{formation channel} \label{subsubsec: channel lh}
\begin{figure*}
   \begin{center}
      \includegraphics[width=\linewidth]{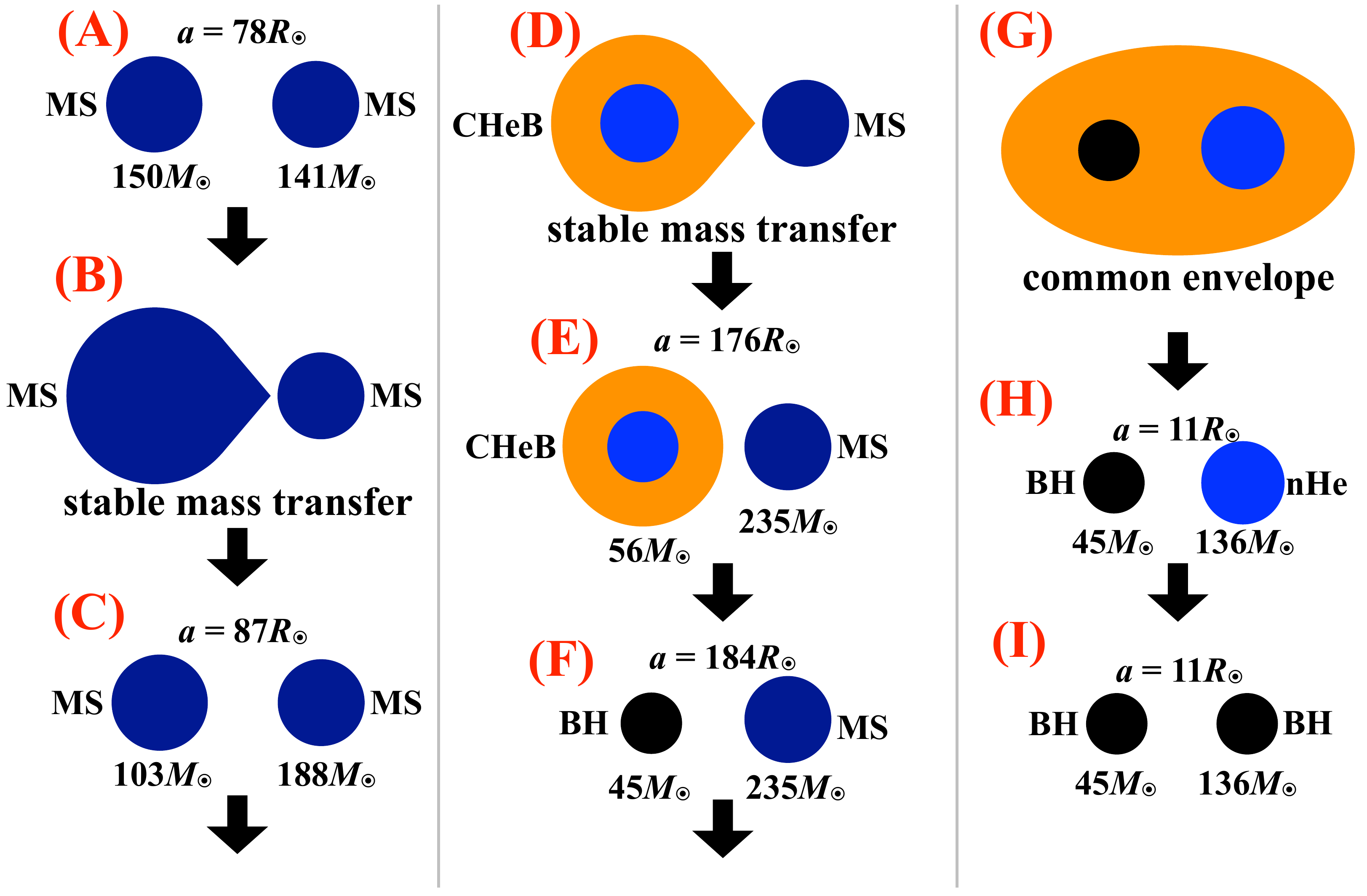}
   \end{center}
   \caption{The example channel of ``low mass $+$ high mass'' when $\alpha\lambda=1.0$.}
   \label{fig: channel lh}
\end{figure*}
The almost all ``low mass $+$ high mass'' BBH mergers are formed through
``single common envelope channel (2)'', and example formation channel is
shown in Figure \ref{fig: channel lh}.
Both stars would cause the pair-instability supernova if they were not in the binary
system, but thanks to the binary interaction such as stable mass transfer one of the 
star gains the mass and becomes a ``high mass BH''.
On the other hand, the other star loses the mass and becomes a ``low mass BH''.
In order to significantly transfer the stellar mass to avoid the pair-instability supernova,
the mass transfer rate needs to be high and the initial orbital separation needs to be short.
If the initial orbital separation is short and $\alpha\lambda$ is smaller, the binary system coalesces after the common
envelope phase, and thus ``low mass $+$ high mass'' BBH mergers cannot be formed when $\alpha\lambda\lesssim$ 0.5.

\subsubsection{primary BH mass distribution} \label{subsubsec: primary mass lh}
\begin{figure*}
   \begin{center}
      \includegraphics[width=\linewidth]{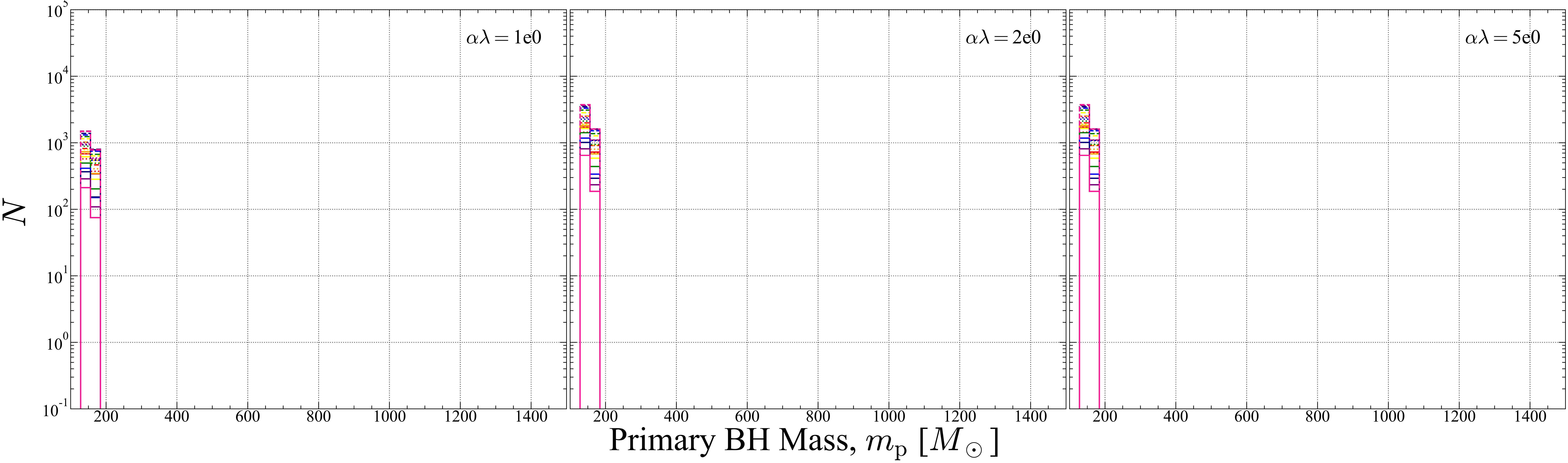}
   \end{center}
   \caption{The primary BH mass distribution of ``low mass $+$ high mass''
   BBH mergers which merge within a Hubble time.
   From the left to right panel, $\alpha\lambda=$ 1.0, 2.0 and 5.0.
   The legends are same as that of Figure \ref{fig: primary bh mass hh}.}
   \label{fig: primary bh mass lh}
\end{figure*}
The primary BH mass distributions of ``low mass $+$ high mass''
BBH mergers which merge within a Hubble
time are shown in Figure \ref{fig: primary bh mass lh}.
The primary BH mass distributions are almost same for the different $\alpha\lambda$ values.
Since both stars are initially 100--200 $\msun$ and the progenitor of primary BH
can only increase its mass up to $\sim300\msun$, the primary BH mass is from 135 $\msun$ to $\sim200\msun$.
135 $\msun$ is the upper end of pair-instability mass gap in this study.

\subsubsection{mass ratio distribution} \label{subsubsec: mass ratio lh}
\begin{figure*}
   \begin{center}
      \includegraphics[width=\linewidth]{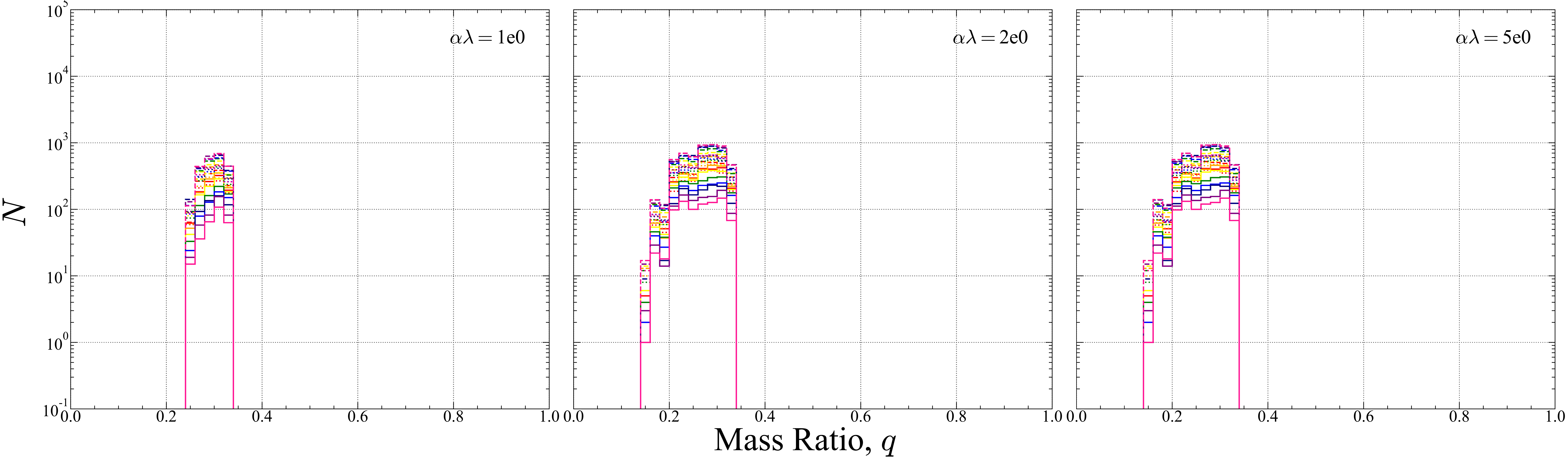}
   \end{center}
   \caption{The mass ratio distribution of ``low mass $+$ high mass''
   BBH mergers which merge within a Hubble time.
   From the left to right panel, $\alpha\lambda=$ 1.0, 2.0 and 5.0.
   The legends are same as that of Figure \ref{fig: primary bh mass hh}.}
   \label{fig: mass ratio lh}
\end{figure*}
The mass ratio distributions of ``low mass $+$ high mass'' BBH mergers which merge within
a Hubble time are shown in Figure \ref{fig: mass ratio lh}.
The mass ratio distributions are roughly same for the different $\alpha\lambda$ values.
The maximum mass ratio is equal to the ratio of lower end of pair-instability mass gap ($=45\msun$)
to the upper end ($=135\msun$), so that 0.33.
When the $\alpha\lambda=$ 2.0 and 5.0, the orbital separation does not shrink much through
common envelope phase compared to when $\alpha\lambda=1.0$.
Therefore, the initially shorter orbital separation are allowed for $\alpha\lambda=$ 2.0 and 5.0,
and in these cases the stellar mass is transferred more significantly
(phase B and D in Figure \ref{fig: channel lh}).
That is why minimum mass ratio gets lower when $\alpha\lambda=$ 2.0 and 5.0.

\subsubsection{effective spin distribution} \label{subsubsec: effective spin lh}
\begin{figure*}
   \begin{center}
      \includegraphics[width=\linewidth]{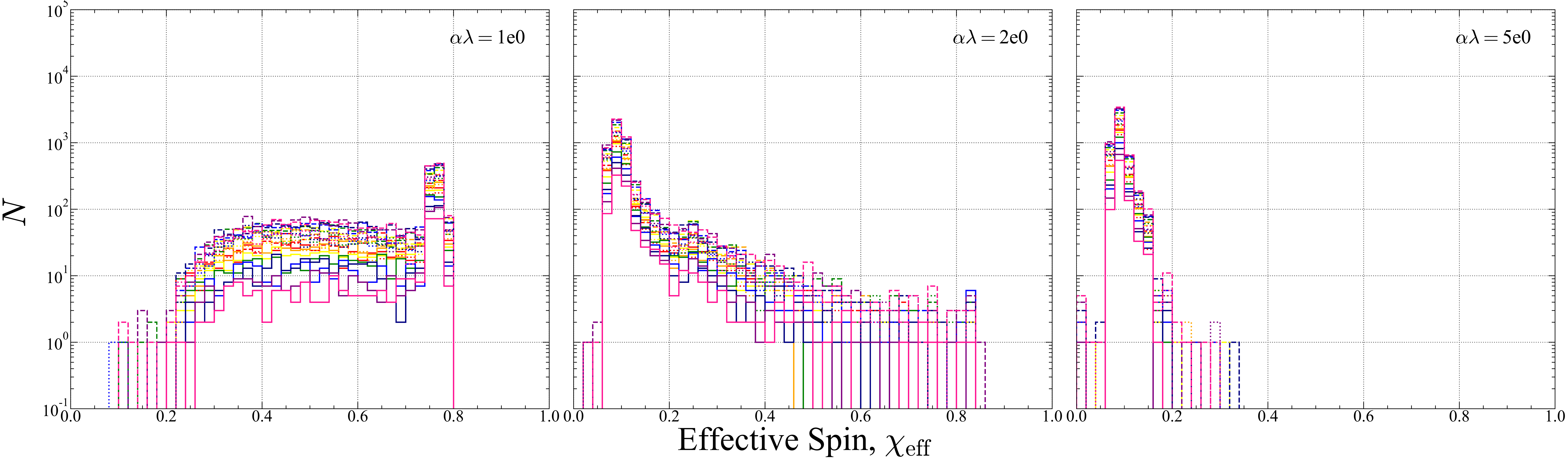}
   \end{center}
   \caption{The effective spin distribution of ``low mass $+$ high mass''
   BBH mergers which merge within a Hubble time.
   From the left to right panel, $\alpha\lambda=$ 1.0, 2.0 and 5.0.
   The legends are same as that of Figure \ref{fig: primary bh mass hh}.}
   \label{fig: effective spin lh}
\end{figure*}
The effective spin distributions of ``low mass $+$ high mass'' BBH mergers which merge within
a Hubble time are shown in Figure \ref{fig: effective spin lh}.
The progenitor of secondary BH loses its spin angular momentum through significant mass
transfer process (phase B and D in Figure \ref{fig: channel lh}), and thus
the BH also has very low spin ($\sim0$).
The progenitor of the primary BH also loses its spin angular momentum during the common
envelope phase (phase G in Figure \ref{fig: channel lh}).
However, the orbital separation after the common envelope is so short that
the progenitor of the primary BH is highly spun up by tidal interaction.
Thus, the spin of primary BH is high.
As the value of $\alpha\lambda$ increases, the degree of spin-up decreases,
and thus the effective spin distribution shifted toward lower value.

\subsubsection{delay time distribution} \label{subsubsec: delay time lh}
\begin{figure*}
   \begin{center}
      \includegraphics[width=\linewidth]{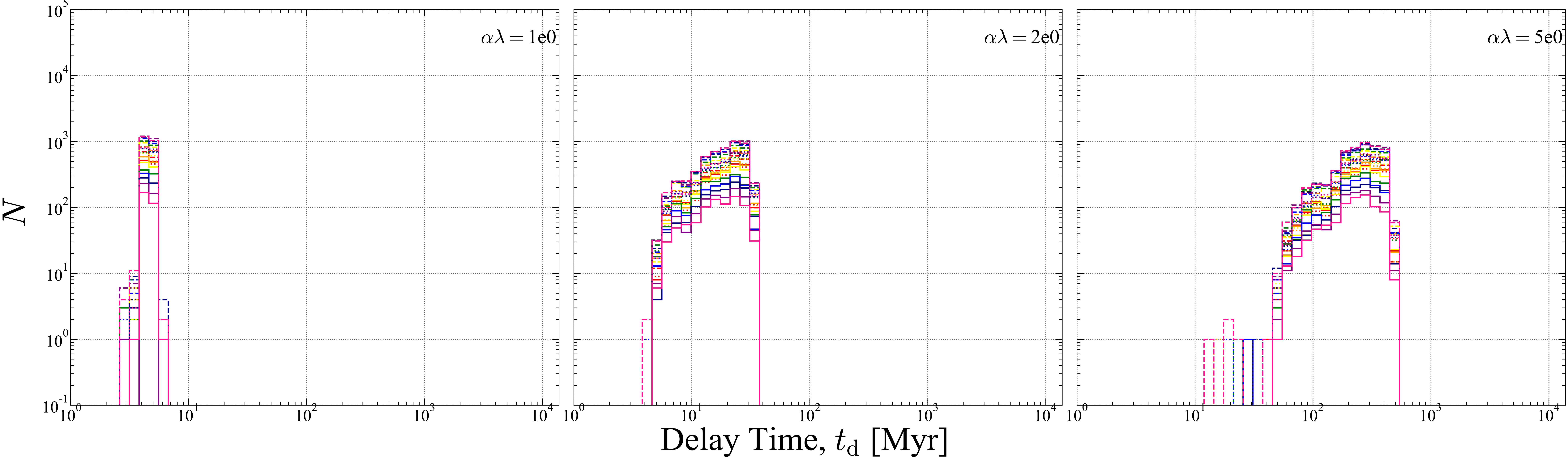}
   \end{center}
   \caption{The delay time distribution of ``low mass $+$ high mass''
   BBH mergers which merge within a Hubble time.
   From the left to right panel, $\alpha\lambda=$ 1.0, 2.0 and 5.0.
   The legends are same as that of Figure \ref{fig: primary bh mass hh}.}
   \label{fig: delay time lh}
\end{figure*}
The delay time distribution of ``low mass $+$ high mass'' BBH mergers which merge within
a Hubble time are shown in Figure \ref{fig: delay time lh}.
As the value of $\alpha\lambda$ increases, the orbital separation after the common envelope
widens, and thus the delay time distribution shifted toward the higher value.
Since the initial orbital separation is short in order to form ``low mass $+$ high mass''
BBHs, the delay time also becomes short, and
does not exceeds $10^3$ Myr for $\alpha\lambda=$ 1.0, 2.0 and 5.0.

\subsubsection{merger rate density} \label{subsubsec: merger rate lh}
\begin{figure*}
   \begin{center}
      \includegraphics[width=\linewidth]{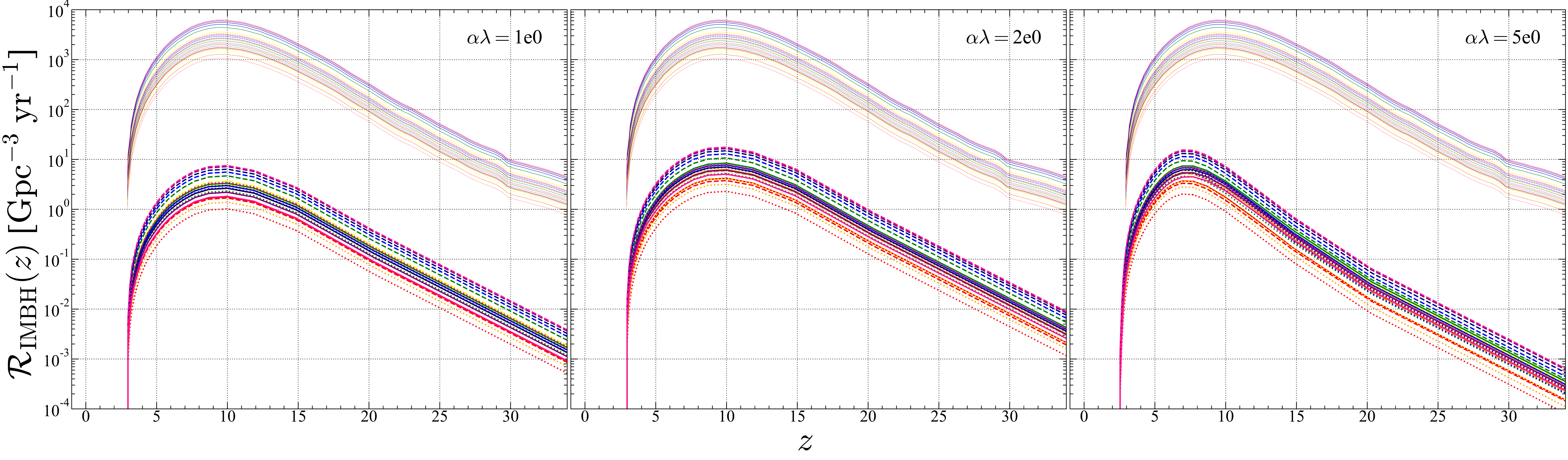}
   \end{center}
   \caption{The redshift evolution of merger rate density of ``low mass $+$ high mass'' BBH.
   From the left to right, for $\alpha\lambda=$ 1.0, 2.0 and 5.0.
   The legends are same as that of Figure \ref{fig: primary bh mass hh}.
   The lines at the bottom of each panel are the redshift evolution of BBH merger rate density.
   The relatively thin lines at the top of each panel indicate the binary 
   formation rate [$\mathrm{Gpc}^{-3}$ $\mathrm{yr}^{-1}$].}
   \label{fig: merger rate lh}
\end{figure*}
The redshift evolution of ``low mass $+$ high mass'' BBH merger rate density is shown
in Figure \ref{fig: merger rate lh}.
The merger rate densities roughly trace the star formation rate history \citep{desouza11}, and 
the peaks of merger rate densities are located near the redshift $z\sim10$.
However, when $\alpha\lambda=$ 5.0, the minimum delay time exceeds 10 Myr, hence the peak of 
merger rate density is shifted toward low redshift compared to that of $\alpha\lambda\lesssim$ 2.0.
The maximum delay time is so short that the merger rate density rapidly decreases
after the end of Pop. III star formation.
Therefore, the merger rate densities are strictly 0 when $z\lesssim3$.

\section{Discussion} \label{sec: discussion}
So far, the BBH mergers with IMBHs have not been discovered by GW observation.
Thus, the upper limit on the merger rate density of the BBH mergers with IMBHs is obtained,
and the value is $\sim0.01$ $\mathrm{Gpc}^{-3}$ $\mathrm{yr}^{-1}$ \citep{lvkc21,mariaEzquiaga21}.
In Section \ref{subsec: comparison}, we compare the merger rate density of BBHs 
with IMBHs with this upper limit and investigate what parameter set meets the 
criterion obtained from the non-detection.
In Section \ref{subsec: future}, using these parameter sets, we infer the 
detection rate by future GW detectors.
Furthermore, we show the likelihood ratio for each model assuming a BBH with IMBHs
is observed and explain how we can limit the value of $\alpha\lambda$ and IMF models.

\subsection{comparison with the current GW observation} \label{subsec: comparison}
The merger rate density of BBHs with IMBHs at the redshift $z=0$ are 
shown in Table \ref{table: mergerrate_01}--\ref{table: mergerrate_50} for $\alpha\lambda=$ 0.1--5.0.
The value lower than the upper limit, 0.01 $\mathrm{Gpc}^{-3}$ $\mathrm{yr}^{-1}$,
is written in bold.
The BBHs with IMBHs are classified as ``high mass $+$ high mass'' BBHs and 
``low mass $+$ high mass'' BBHs.
The merger rate density of ``low mass $+$ high mass'' BBHs is 0
at the redshift $z=0$ (see Section \ref{subsubsec: merger rate lh}), hence
the merger rate density shown in the Table \ref{table: mergerrate_01}--\ref{table: mergerrate_50}
are that of ``high mass $+$ high mass'' BBH mergers.
When $\alpha\lambda=0.01$, the merger rate density of ``high mass $+$ high mass'' BBHs
is also 0 at the redshift $z=0$ (see Section \ref{subsubsec: merger rate hh}), hence
the merger rate density of BBHs with IMBHs is strictly 0.

Since the delay time is so short that the Pop. III BBHs with IMBHs cannot merge at the local Universe and the merger rate is 0 when $\alpha\lambda=0.01$, any IMF is allowed.
Therefore it is impossible to impose any restrictions on Pop. III IMF.
When $\alpha\lambda=0.1$, the contribution from the Pop. III star heavier than $\sim400\msun$
is significant.
Therefore, an IMF with the sufficiently large $\gamma_2$ is allowed for 
$\gamma_1=$ 0 and 1 and $m_\mathrm{crit}=$ 100--300 $\msun$ (see Table \ref{table: mergerrate_01}).
When $\alpha\lambda\gtrsim0.5$, the contribution from the Pop. III star $\sim200\msun$
is significant.
In this study, the Pop. III IMF is relatively top heavy below the critical mass $m_\mathrm{crit}$.
Therefore, when $m_\mathrm{crit}=$ 200 and 300 $\msun$, it is impossible to decrease the
merger rate density by steeping the IMF slope (i.e., increasing the value of $\gamma_2$) above the critical mass,
and thus the Pop. III IMF with $m_\mathrm{crit}=$ 200 and 300 $\msun$ 
are not allowed (see Table \ref{table: mergerrate_05}--\ref{table: mergerrate_50}).

\begin{table*}
   \caption{The merger rate density [$\mathrm{Gpc}^{-3}$ $\mathrm{yr}^{-1}$] of BBHs with IMBHs at the redshift $z=0$ for $\alpha\lambda=0.1$. The value lower than the upper limit, 0.01 $\mathrm{Gpc}^{-3}$ $\mathrm{yr}^{-1}$,
is written in bold.}
   \begin{center}
     \begin{tabular}[c]{cc|cccccccc}\hline\hline
      \multicolumn{2}{c|}{} & \multicolumn{8}{c}{$\gamma_2$}\\
      \cline{3-10}
        $\gamma_1$ & $m_\mathrm{crit}$ [$\msun$] & 1.5 & 2.0 & 2.5 & 3.0 & 3.5 & 4.0 & 4.5 & 5.0\\\hline
        0.0 & 100 & 2.44e-02 & 1.09e-02 & \textbf{9.83e-03} & 1.23e-02 & \textbf{3.12e-03} & \textbf{4.77e-03} & \textbf{2.40e-04} & \textbf{3.79e-04} \\
        0.0 & 200 & 1.82e-02 & 2.33e-02 & 1.42e-02 & 2.29e-02 & 1.34e-02 & \textbf{6.04e-03} & \textbf{8.70e-03} & \textbf{2.38e-03} \\
        0.0 & 300 & 2.29e-02 & 2.74e-02 & 2.91e-02 & 2.60e-02 & 2.30e-02 & 1.50e-02 & \textbf{9.06e-03} & \textbf{9.98e-03} \\
        1.0 & 100 & 2.69e-02 & 1.36e-02 & 1.43e-02 & \textbf{6.08e-03} & \textbf{8.34e-03} & \textbf{1.50e-04} & \textbf{8.01e-04} & \textbf{5.68e-05} \\
        1.0 & 200 & 2.42e-02 & 2.08e-02 & 1.86e-02 & \textbf{1.60e-03} & \textbf{6.40e-03} & \textbf{2.74e-03} & \textbf{2.41e-03} & \textbf{1.79e-03} \\
        1.0 & 300 & 2.18e-02 & 1.49e-02 & 1.91e-02 & 1.29e-02 & 1.08e-02 & \textbf{8.18e-03} & \textbf{9.28e-03} & \textbf{7.72e-03} \\\hline        
     \end{tabular}
  \end{center}
  \label{table: mergerrate_01}
\end{table*}
\begin{table*}
   \caption{Same as Table \ref{table: mergerrate_01} but for $\alpha\lambda=0.5$.}
   \begin{center}
     \begin{tabular}{cc|cccccccc}\hline\hline
      \multicolumn{2}{c|}{} & \multicolumn{8}{c}{$\gamma_2$}\\
      \cline{3-10}
        $\gamma_1$ & $m_\mathrm{crit}$ [$\msun$] & 1.5 & 2.0 & 2.5 & 3.0 & 3.5 & 4.0 & 4.5 & 5.0\\\hline
        0.0 & 100 & 3.97e-02 & 3.77e-02 & 3.79e-02 & 3.38e-02 & 1.38e-02 & \textbf{8.73e-03} & \textbf{6.69e-03} & \textbf{3.16e-03} \\
        0.0 & 200 & 4.37e-02 & 5.06e-02 & 5.33e-02 & 4.76e-02 & 3.99e-02 & 4.95e-02 & 3.81e-02 & 3.04e-02 \\
        0.0 & 300 & 5.43e-02 & 6.29e-02 & 7.86e-02 & 7.33e-02 & 8.46e-02 & 7.89e-02 & 7.03e-02 & 9.14e-02 \\
        1.0 & 100 & 3.80e-02 & 3.95e-02 & 2.67e-02 & 1.93e-02 & \textbf{9.67e-03} & \textbf{7.76e-03} & \textbf{4.48e-03} & \textbf{2.09e-03} \\
        1.0 & 200 & 4.28e-02 & 3.89e-02 & 3.57e-02 & 4.34e-02 & 3.19e-02 & 2.36e-02 & 2.10e-02 & 1.31e-02 \\
        1.0 & 300 & 3.95e-02 & 4.72e-02 & 4.85e-02 & 6.53e-02 & 6.30e-02 & 4.98e-02 & 5.99e-02 & 4.95e-02 \\
         \hline        
     \end{tabular}
  \end{center}
  \label{table: mergerrate_05}
\end{table*}
\begin{table*}
   \caption{Same as Table \ref{table: mergerrate_01} but for $\alpha\lambda=1.0$.}
   \begin{center}
     \begin{tabular}{cc|cccccccc}\hline\hline
      \multicolumn{2}{c|}{} & \multicolumn{8}{c}{$\gamma_2$}\\
      \cline{3-10}
        $\gamma_1$ & $m_\mathrm{crit}$ [$\msun$] & 1.5 & 2.0 & 2.5 & 3.0 & 3.5 & 4.0 & 4.5 & 5.0\\\hline
        0.0 & 100 & 5.32e-02 & 5.30e-02 & 4.16e-02 & 3.03e-02 & 2.51e-02 & 1.32e-02 & 1.03e-02 & \textbf{4.85e-03} \\
        0.0 & 200 & 5.68e-02 & 7.11e-02 & 6.59e-02 & 6.69e-02 & 6.61e-02 & 6.22e-02 & 5.57e-02 & 3.76e-02 \\
        0.0 & 300 & 5.92e-02 & 7.65e-02 & 9.86e-02 & 1.08e-01 & 1.16e-01 & 1.21e-01 & 1.17e-01 & 1.19e-01 \\
        1.0 & 100 & 4.19e-02 & 4.32e-02 & 3.90e-02 & 2.22e-02 & 1.59e-02 & 1.02e-02 & \textbf{6.22e-03} & \textbf{3.90e-03} \\
        1.0 & 200 & 5.04e-02 & 5.36e-02 & 3.86e-02 & 4.56e-02 & 4.56e-02 & 3.35e-02 & 3.69e-02 & 3.32e-02 \\
        1.0 & 300 & 4.85e-02 & 6.47e-02 & 7.73e-02 & 7.43e-02 & 7.41e-02 & 6.96e-02 & 7.63e-02 & 7.73e-02 \\
         \hline
     \end{tabular}
  \end{center}
  \label{table: mergerrate_10}
\end{table*}
\begin{table*}
   \caption{Same as Table \ref{table: mergerrate_01} but for $\alpha\lambda=2.0$.}
   \begin{center}
     \begin{tabular}{cc|cccccccc}\hline\hline
      \multicolumn{2}{c|}{} & \multicolumn{8}{c}{$\gamma_2$}\\
      \cline{3-10}
        $\gamma_1$ & $m_\mathrm{crit}$ [$\msun$] & 1.5 & 2.0 & 2.5 & 3.0 & 3.5 & 4.0 & 4.5 & 5.0\\\hline
        0.0 & 100 & 4.28e-02 & 4.95e-02 & 4.76e-02 & 4.19e-02 & 2.05e-02 & 1.50e-02 & 1.00e-02 & \textbf{5.87e-03} \\
        0.0 & 200 & 4.92e-02 & 6.40e-02 & 7.69e-02 & 6.86e-02 & 6.32e-02 & 6.37e-02 & 5.83e-02 & 4.89e-02 \\
        0.0 & 300 & 5.19e-02 & 7.84e-02 & 8.36e-02 & 1.06e-01 & 1.13e-01 & 1.20e-01 & 1.13e-01 & 1.33e-01 \\
        1.0 & 100 & 4.27e-02 & 4.21e-02 & 3.67e-02 & 2.33e-02 & 1.79e-02 & 1.10e-02 & \textbf{6.06e-03} & \textbf{3.30e-03} \\
        1.0 & 200 & 5.44e-02 & 6.12e-02 & 5.09e-02 & 5.03e-02 & 4.99e-02 & 4.07e-02 & 4.03e-02 & 2.92e-02 \\
        1.0 & 300 & 5.48e-02 & 6.41e-02 & 6.45e-02 & 7.90e-02 & 7.25e-02 & 7.56e-02 & 8.26e-02 & 8.28e-02 \\
         \hline        
     \end{tabular}
  \end{center}
  \label{table: mergerrate_20}
\end{table*}
\begin{table*}
   \caption{Same as Table \ref{table: mergerrate_01} but for $\alpha\lambda=5.0$.}
   \begin{center}
     \begin{tabular}{cc|cccccccc}\hline\hline
        \multicolumn{2}{c|}{} & \multicolumn{8}{c}{$\gamma_2$}\\
        \cline{3-10}
        $\gamma_1$ & $m_\mathrm{crit}$ [$\msun$] & 1.5 & 2.0 & 2.5 & 3.0 & 3.5 & 4.0 & 4.5 & 5.0\\\hline
        0.0 & 100 & 3.32e-02 & 4.47e-02 & 3.36e-02 & 3.16e-02 & 1.90e-02 & \textbf{9.64e-03} & \textbf{8.18e-03} & \textbf{6.29e-03} \\
        0.0 & 200 & 3.59e-02 & 4.99e-02 & 6.03e-02 & 5.30e-02 & 5.99e-02 & 5.56e-02 & 5.61e-02 & 4.78e-02 \\
        0.0 & 300 & 4.39e-02 & 5.44e-02 & 7.18e-02 & 8.86e-02 & 1.03e-01 & 1.07e-01 & 1.01e-01 & 1.08e-01 \\
        1.0 & 100 & 3.73e-02 & 3.90e-02 & 3.33e-02 & 2.28e-02 & 1.39e-02 & 1.06e-02 & \textbf{6.21e-03} & \textbf{2.76e-03} \\
        1.0 & 200 & 3.12e-02 & 3.67e-02 & 4.56e-02 & 4.11e-02 & 4.06e-02 & 3.71e-02 & 2.95e-02 & 2.88e-02 \\
        1.0 & 300 & 3.96e-02 & 4.88e-02 & 4.80e-02 & 6.68e-02 & 7.08e-02 & 8.16e-02 & 7.09e-02 & 9.28e-02 \\
         \hline        
     \end{tabular}
  \end{center}
  \label{table: mergerrate_50}
\end{table*}

\subsection{future observation} \label{subsec: future}
\subsubsection{detection rate} \label{subsubsec: detection rate}
\begin{figure}
   \begin{center}
      \includegraphics[width=\columnwidth]{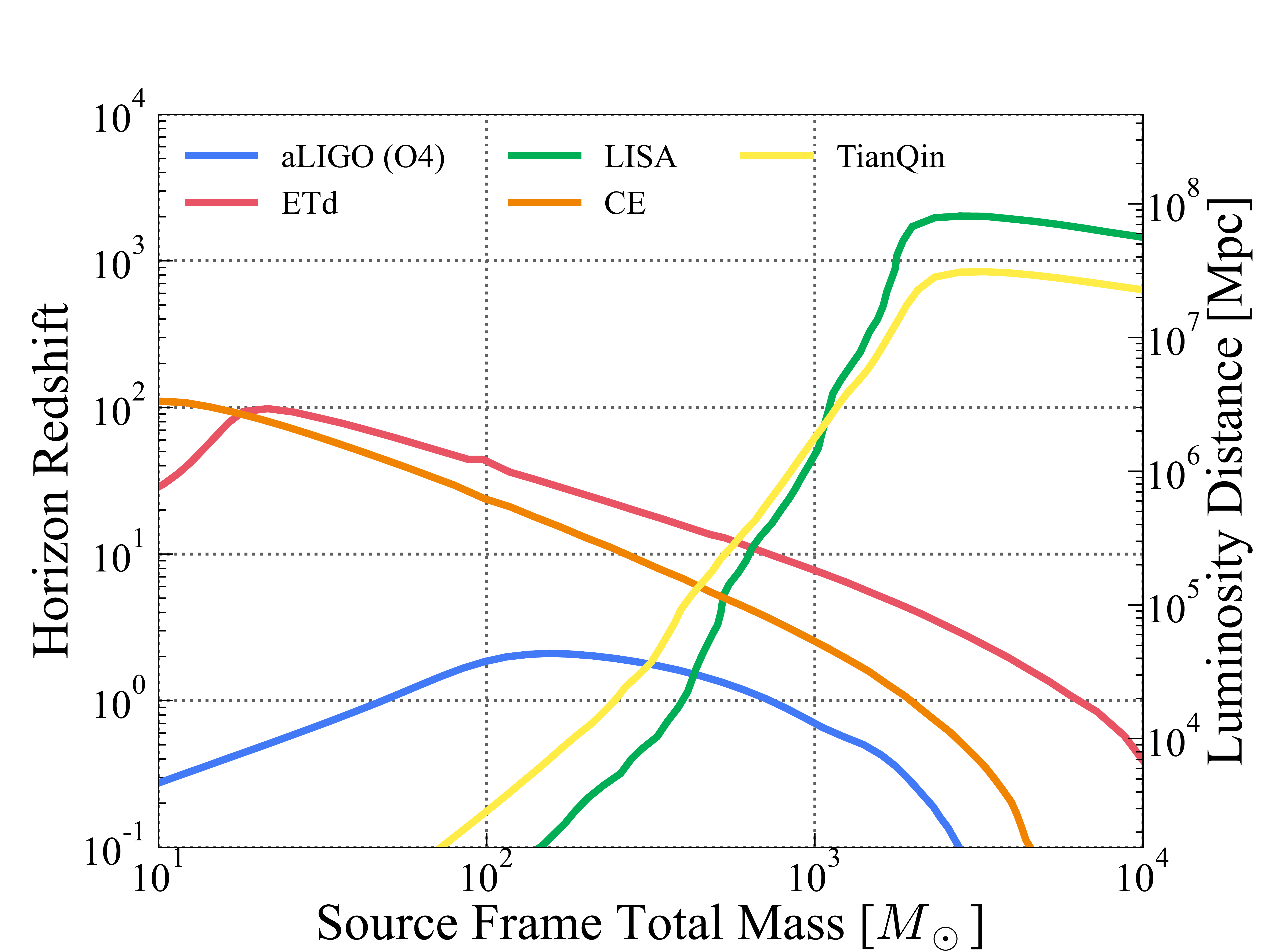}
   \end{center}
   \caption{The horizon redshift of future GW detectors. The blue, red, green, orange and yellow lines
   correspond to advanced LIGO (O4), Einstein telescope, LISA, Cosmic Explorer and TianQin respectively.}
   \label{fig: horizon}
\end{figure}
\begin{table*}
   \caption{The detection rate [$\mathrm{yr}^{-1}$] of BBHs with IMBHs for $\alpha\lambda=0.01$.
   The detection rate by advanced LIGO (O4), Einstein telescope and LISA are written in roman, \textit{italic} and \textbf{bold}
   font, respectively. If an IMF model does not meet the upper limit from the non-detection,
   only ``-'' is written in the cell.}
   \begin{center}
     \begin{tabular}[c]{cc|cccccccc}\hline\hline
      \multicolumn{2}{c|}{} & \multicolumn{8}{c}{$\gamma_2$}\\
      \cline{3-10}
        $\gamma_1$ & $m_\mathrm{crit}$ [$\msun$] & 1.5 & 2.0 & 2.5 & 3.0 & 3.5 & 4.0 & 4.5 & 5.0\\\hline
        0.0 & 100 & \begin{tabular}{c}0.0\\\textit{23.6}\\\textbf{213.8}\end{tabular} & \begin{tabular}{c}0.0\\\textit{24.0}\\\textbf{140.7}\end{tabular} & \begin{tabular}{c}0.0\\\textit{22.1}\\\textbf{79.3}\end{tabular} & \begin{tabular}{c}0.0\\\textit{22.1}\\\textbf{79.3}\end{tabular} & \begin{tabular}{c}0.0\\\textit{11.5}\\\textbf{21.8}\end{tabular} & \begin{tabular}{c}0.0\\\textit{6.0}\\\textbf{5.4}\end{tabular} & \begin{tabular}{c}0.0\\\textit{4.6}\\\textbf{1.4}\end{tabular} & \begin{tabular}{c}0.0\\\textit{3.4}\\\textbf{1.0}\end{tabular} \\\hline
        0.0 & 200 & \begin{tabular}{c}0.0\\\textit{25.3}\\\textbf{240.5}\end{tabular} & \begin{tabular}{c}0.0\\\textit{30.6}\\\textbf{192.4}\end{tabular} & \begin{tabular}{c}0.0\\\textit{34.0}\\\textbf{115.2}\end{tabular} & \begin{tabular}{c}0.0\\\textit{35.1}\\\textbf{87.0}\end{tabular} & \begin{tabular}{c}0.0\\\textit{36.7}\\\textbf{54.1}\end{tabular} & \begin{tabular}{c}0.0\\\textit{36.9}\\\textbf{27.5}\end{tabular} & \begin{tabular}{c}0.0\\\textit{30.6}\\\textbf{15.4}\end{tabular} & \begin{tabular}{c}0.0\\\textit{29.6}\\\textbf{7.8}\end{tabular} \\\hline
        0.0 & 300 & \begin{tabular}{c}0.0\\\textit{25.1}\\\textbf{263.3}\end{tabular} & \begin{tabular}{c}0.0\\\textit{30.2}\\\textbf{231.9}\end{tabular} & \begin{tabular}{c}0.0\\\textit{33.9}\\\textbf{170.3}\end{tabular} & \begin{tabular}{c}0.0\\\textit{42.9}\\\textbf{125.6}\end{tabular} & \begin{tabular}{c}0.0\\\textit{49.6}\\\textbf{93.5}\end{tabular} & \begin{tabular}{c}0.0\\\textit{58.3}\\\textbf{64.6}\end{tabular} & \begin{tabular}{c}0.0\\\textit{58.8}\\\textbf{38.5}\end{tabular} & \begin{tabular}{c}0.0\\\textit{57.4}\\\textbf{32.4}\end{tabular} \\\hline
        1.0 & 100 & \begin{tabular}{c}0.0\\\textit{18.9}\\\textbf{197.3}\end{tabular} & \begin{tabular}{c}0.0\\\textit{22.0}\\\textbf{138.5}\end{tabular} & \begin{tabular}{c}0.0\\\textit{16.8}\\\textbf{71.2}\end{tabular} & \begin{tabular}{c}0.0\\\textit{12.4}\\\textbf{32.6}\end{tabular} & \begin{tabular}{c}0.0\\\textit{11.2}\\\textbf{17.1}\end{tabular} & \begin{tabular}{c}0.0\\\textit{6.7}\\\textbf{3.3}\end{tabular} & \begin{tabular}{c}0.0\\\textit{3.4}\\\textbf{3.4}\end{tabular} & \begin{tabular}{c}0.0\\\textit{1.6}\\\textbf{0.0}\end{tabular} \\\hline
        1.0 & 200 & \begin{tabular}{c}0.0\\\textit{25.2}\\\textbf{228.0}\end{tabular} & \begin{tabular}{c}0.0\\\textit{27.0}\\\textbf{160.4}\end{tabular} & \begin{tabular}{c}0.0\\\textit{25.5}\\\textbf{104.8}\end{tabular} & \begin{tabular}{c}0.0\\\textit{28.9}\\\textbf{46.6}\end{tabular} & \begin{tabular}{c}0.0\\\textit{23.5}\\\textbf{37.6}\end{tabular} & \begin{tabular}{c}0.0\\\textit{29.5}\\\textbf{23.2}\end{tabular} & \begin{tabular}{c}0.0\\\textit{21.1}\\\textbf{9.1}\end{tabular} & \begin{tabular}{c}0.0\\\textit{19.1}\\\textbf{6.4}\end{tabular} \\\hline
        1.0 & 300 & \begin{tabular}{c}0.0\\\textit{20.8}\\\textbf{240.3}\end{tabular} & \begin{tabular}{c}0.0\\\textit{26.6}\\\textbf{174.9}\end{tabular} & \begin{tabular}{c}0.0\\\textit{27.7}\\\textbf{137.9}\end{tabular} & \begin{tabular}{c}0.0\\\textit{32.8}\\\textbf{85.1}\end{tabular} & \begin{tabular}{c}0.0\\\textit{38.1}\\\textbf{68.0}\end{tabular} & \begin{tabular}{c}0.0\\\textit{39.9}\\\textbf{39.5}\end{tabular} & \begin{tabular}{c}0.0\\\textit{42.4}\\\textbf{30.0}\end{tabular} & \begin{tabular}{c}0.0\\\textit{43.8}\\\textbf{15.9}\end{tabular} \\
        \hline
     \end{tabular}
  \end{center}
  \label{table: detectionrate_001}
\end{table*}

\begin{table*}
   \caption{Same as Table \ref{table: detectionrate_001} but for $\alpha\lambda=0.1$.}
   \begin{center}
     \begin{tabular}[c]{cc|cccccccc}\hline\hline
      \multicolumn{2}{c|}{} & \multicolumn{8}{c}{$\gamma_2$}\\
      \cline{3-10}
        $\gamma_1$ & $m_\mathrm{crit}$ [$\msun$] & 1.5 & 2.0 & 2.5 & 3.0 & 3.5 & 4.0 & 4.5 & 5.0\\\hline
        0.0 & 100 & - & - & \begin{tabular}{c}0.32\\\textit{604}\\\textbf{76}\end{tabular} & - & \begin{tabular}{c}0.10\\\textit{308}\\\textbf{15}\end{tabular} & \begin{tabular}{c}0.052\\\textit{195}\\\textbf{5.4}\end{tabular} & \begin{tabular}{c}0.033\\\textit{118}\\\textbf{1.5}\end{tabular} & \begin{tabular}{c}0.010\\\textit{71}\\\textbf{0.68}\end{tabular} \\\hline
        0.0 & 200 & - & - & - & - & - & \begin{tabular}{c}0.20\\\textit{785}\\\textbf{21}\end{tabular} & \begin{tabular}{c}0.14\\\textit{693}\\\textbf{13}\end{tabular} & \begin{tabular}{c}0.082\\\textit{587}\\\textbf{8.3}\end{tabular} \\\hline
        0.0 & 300 & - & - & - & - & - & - & \begin{tabular}{c}0.34\\\textit{1600}\\\textbf{31}\end{tabular} & \begin{tabular}{c}0.50\\\textit{1646}\\\textbf{21}\end{tabular} \\\hline
        1.0 & 100 & - & - & - & \begin{tabular}{c}0.13\\\textit{350}\\\textbf{25}\end{tabular} & \begin{tabular}{c}0.057\\\textit{222}\\\textbf{6.5}\end{tabular} & \begin{tabular}{c}0.017\\\textit{137}\\\textbf{0.77}\end{tabular} & \begin{tabular}{c}0.012\\\textit{84}\\\textbf{5.1}\end{tabular} & \begin{tabular}{c}0.005\\\textit{47}\\\textbf{0.014}\end{tabular} \\\hline
        1.0 & 200 & - & - & - & \begin{tabular}{c}0.30\\\textit{713}\\\textbf{69}\end{tabular} & \begin{tabular}{c}0.22\\\textit{623}\\\textbf{28}\end{tabular} & \begin{tabular}{c}0.20\\\textit{543}\\\textbf{22}\end{tabular} & \begin{tabular}{c}0.13\\\textit{456}\\\textbf{7.5}\end{tabular} & \begin{tabular}{c}0.069\\\textit{378}\\\textbf{6.9}\end{tabular} \\\hline
        1.0 & 300 & - & - & - & - & - & \begin{tabular}{c}0.32\\\textit{1077}\\\textbf{39}\end{tabular} & \begin{tabular}{c}0.26\\\textit{1083}\\\textbf{23}\end{tabular} & \begin{tabular}{c}0.28\\\textit{1085}\\\textbf{17}\end{tabular} \\
        \hline        
     \end{tabular}
  \end{center}
  \label{table: detectionrate_01}
\end{table*}

\begin{table*}
   \caption{Same as Table \ref{table: detectionrate_001} but for $\alpha\lambda=0.5$.}
   \begin{center}
     \begin{tabular}[c]{cc|cccccccc}\hline\hline
      \multicolumn{2}{c|}{} & \multicolumn{8}{c}{$\gamma_2$}\\
      \cline{3-10}
        $\gamma_1$ & $m_\mathrm{crit}$ [$\msun$] & 1.5 & 2.0 & 2.5 & 3.0 & 3.5 & 4.0 & 4.5 & 5.0\\\hline
        0.0 & 100 & - & - & - & - & - & \begin{tabular}{c}1.6\\\textit{212}\\\textbf{1.6}\end{tabular} & \begin{tabular}{c}0.8\\\textit{134}\\\textbf{0.7}\end{tabular} & \begin{tabular}{c}0.5\\\textit{83}\\\textbf{0.4}\end{tabular} \\\hline
        0.0 & 200 & - & - & - & - & - & - & - & - \\\hline
        0.0 & 300 & - & - & - & - & - & - & - & - \\\hline
        1.0 & 100 & - & - & - & - & \begin{tabular}{c}1.5\\\textit{240}\\\textbf{1.9}\end{tabular} & \begin{tabular}{c}1.0\\\textit{153}\\\textbf{0.9}\end{tabular} & \begin{tabular}{c}0.6\\\textit{90}\\\textbf{0.5}\end{tabular} & \begin{tabular}{c}0.4\\\textit{56}\\\textbf{0.3}\end{tabular} \\\hline
        1.0 & 200 & - & - & - & - & - & - & - & - \\\hline
        1.0 & 300 & - & - & - & - & - & - & - & - \\
        \hline
     \end{tabular}
  \end{center}
  \label{table: detectionrate_05}
\end{table*}

\begin{table*}
   \caption{Same as Table \ref{table: detectionrate_001} but for $\alpha\lambda=1.0$.}
   \begin{center}
     \begin{tabular}[c]{cc|cccccccc}\hline\hline
      \multicolumn{2}{c|}{} & \multicolumn{8}{c}{$\gamma_2$}\\
      \cline{3-10}
        $\gamma_1$ & $m_\mathrm{crit}$ [$\msun$] & 1.5 & 2.0 & 2.5 & 3.0 & 3.5 & 4.0 & 4.5 & 5.0\\\hline
        0.0 & 100 & - & - & - & - & - & - & - & \begin{tabular}{c}0.6\\\textit{143}\\\textbf{0.2}\end{tabular} \\\hline
        0.0 & 200 & - & - & - & - & - & - & - & - \\\hline
        0.0 & 300 & - & - & - & - & - & - & - & - \\\hline
        1.0 & 100 & - & - & - & - & - & - & \begin{tabular}{c}0.7\\\textit{136}\\\textbf{0.2}\end{tabular} & \begin{tabular}{c}0.5\\\textit{93}\\\textbf{0.1}\end{tabular} \\\hline
        1.0 & 200 & - & - & - & - & - & - & - & - \\\hline
        1.0 & 300 & - & - & - & - & - & - & - & - \\
        \hline        
     \end{tabular}
  \end{center}
  \label{table: detectionrate_10}
\end{table*}

\begin{table*}
   \caption{Same as Table \ref{table: detectionrate_001} but for $\alpha\lambda=2.0$.}
   \begin{center}
     \begin{tabular}[c]{cc|cccccccc}\hline\hline
      \multicolumn{2}{c|}{} & \multicolumn{8}{c}{$\gamma_2$}\\
      \cline{3-10}
        $\gamma_1$ & $m_\mathrm{crit}$ [$\msun$] & 1.5 & 2.0 & 2.5 & 3.0 & 3.5 & 4.0 & 4.5 & 5.0\\\hline
        0.0 & 100 & - & - & - & - & - & - & - & \begin{tabular}{c}0.65\\\textit{164}\\\textbf{0.12}\end{tabular} \\\hline
        0.0 & 200 & - & - & - & - & - & - & - & - \\\hline
        0.0 & 300 & - & - & - & - & - & - & - & - \\\hline
        1.0 & 100 & - & - & - & - & - & - & \begin{tabular}{c}0.71\\\textit{147}\\\textbf{0.12}\end{tabular} & \begin{tabular}{c}0.41\\\textit{104}\\\textbf{0.02}\end{tabular} \\\hline
        1.0 & 200 & - & - & - & - & - & - & - & - \\\hline
        1.0 & 300 & - & - & - & - & - & - & - & - \\
        \hline        
     \end{tabular}
  \end{center}
  \label{table: detectionrate_20}
\end{table*}

\begin{table*}
   \caption{Same as Table \ref{table: detectionrate_001} but for $\alpha\lambda=5.0$.}
   \begin{center}
     \begin{tabular}[c]{cc|cccccccc}\hline\hline
      \multicolumn{2}{c|}{} & \multicolumn{8}{c}{$\gamma_2$}\\
      \cline{3-10}
        $\gamma_1$ & $m_\mathrm{crit}$ [$\msun$] & 1.5 & 2.0 & 2.5 & 3.0 & 3.5 & 4.0 & 4.5 & 5.0\\\hline
        0.0 & 100 & - & - & - & - & - & \begin{tabular}{c}1.28\\\textit{216}\\\textbf{0.08}\end{tabular} & \begin{tabular}{c}0.84\\\textit{145}\\\textbf{0.05}\end{tabular} & \begin{tabular}{c}0.53\\\textit{133}\\\textbf{0.03}\end{tabular} \\\hline
        0.0 & 200 & - & - & - & - & - & - & - & - \\\hline
        0.0 & 300 & - & - & - & - & - & - & - & - \\\hline
        1.0 & 100 & - & - & - & - & - & - & \begin{tabular}{c}0.56\\\textit{114}\\\textbf{0.03}\end{tabular} & \begin{tabular}{c}0.37\\\textit{85.3}\\\textbf{0.02}\end{tabular} \\\hline
        1.0 & 200 & - & - & - & - & - & - & - & - \\\hline
        1.0 & 300 & - & - & - & - & - & - & - & - \\
        \hline        
     \end{tabular}
  \end{center}
  \label{table: detectionrate_50}
\end{table*}

In this section, we compute the detection rate by the second generation GW detector, 
advanced LIGO (O4), third generation detector, Einstein telescope
and the space-borne detector, LISA
for the models which meet the upper limit from the non-detection.
Figure \ref{fig: horizon} shows the horizon of each detector.
The detection rate of BBHs with IMBHs by advanced LIGO (O4), Einstein telescope
and LISA are shown in Table \ref{table: detectionrate_001}--\ref{table: detectionrate_50} 
for $\alpha\lambda=$ 0.01--5.0.
The detection rate by advanced LIGO (O4), Einstein telescope and LISA are written in roman, italic and bold
font, respectively.

Next, we choose the IMF model with the highest detection rate for each detector 
and for each $\alpha\lambda$, and show the detection rate as a function of mass.
If we choose different IMF model, the mass distribution does not change significantly.
Therefore, it is sufficient to show the mass distribution of one IMF model
for each detector and for each $\alpha\lambda$.

\begin{figure*}
      \begin{minipage}[t]{0.32\linewidth}
         \centering
         \includegraphics[width=\linewidth]{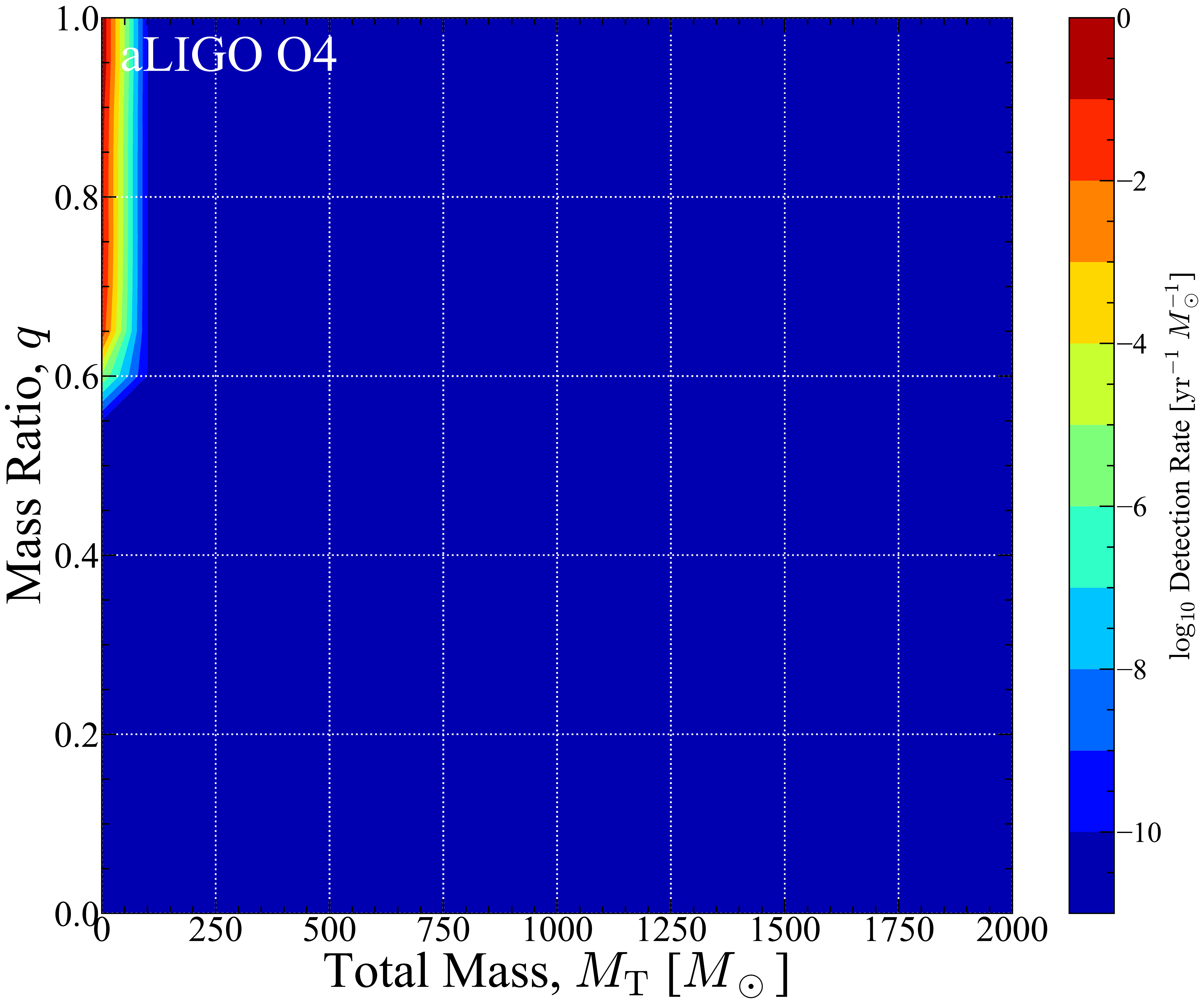}
         \subcaption{$\alpha\lambda=0.01$}
      \end{minipage}
      \begin{minipage}[t]{0.32\linewidth}
         \centering
         \includegraphics[width=\linewidth]{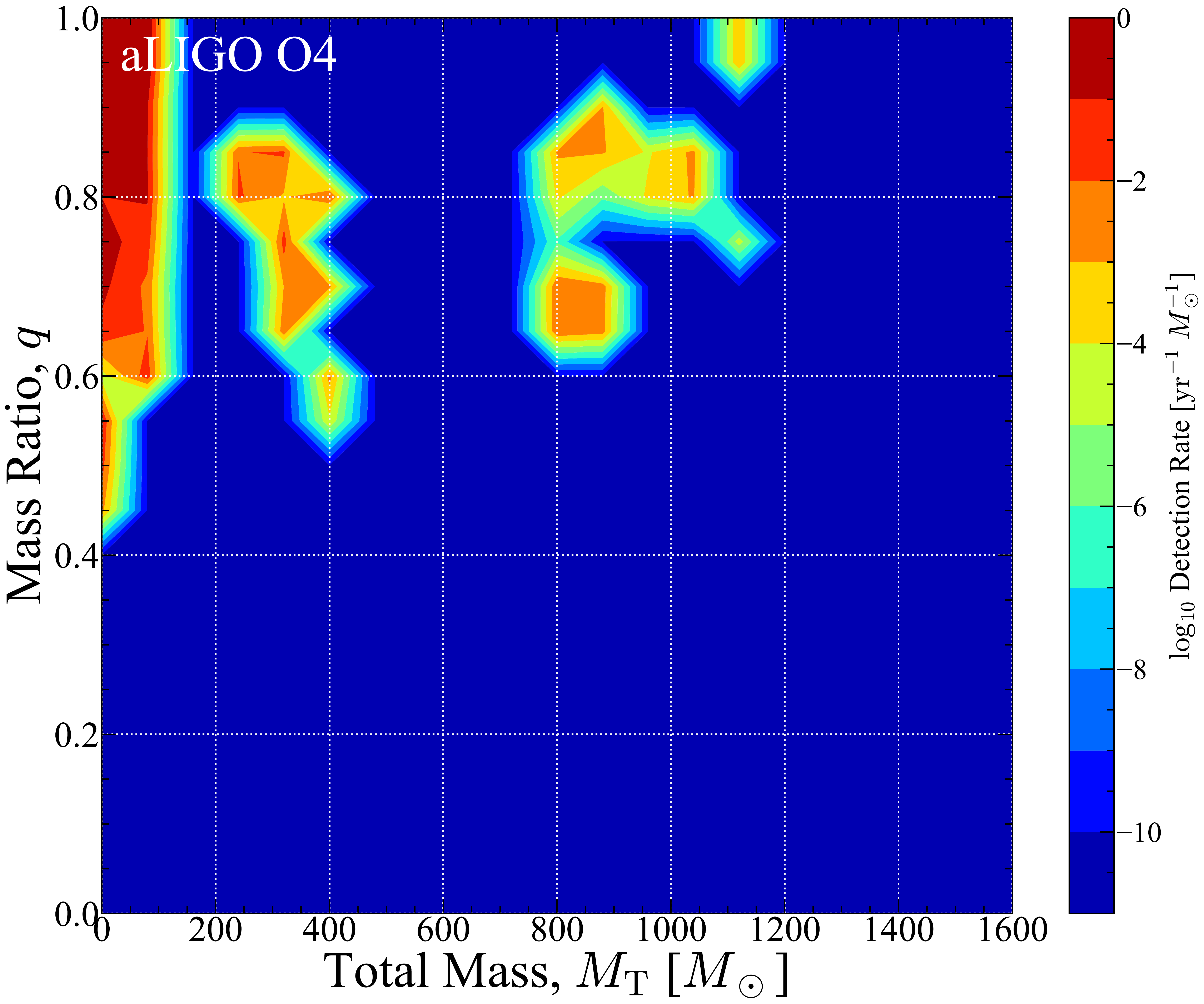}
         \subcaption{$\alpha\lambda=0.1$}
      \end{minipage}
      \begin{minipage}[t]{0.32\linewidth}
         \centering
         \includegraphics[width=\linewidth]{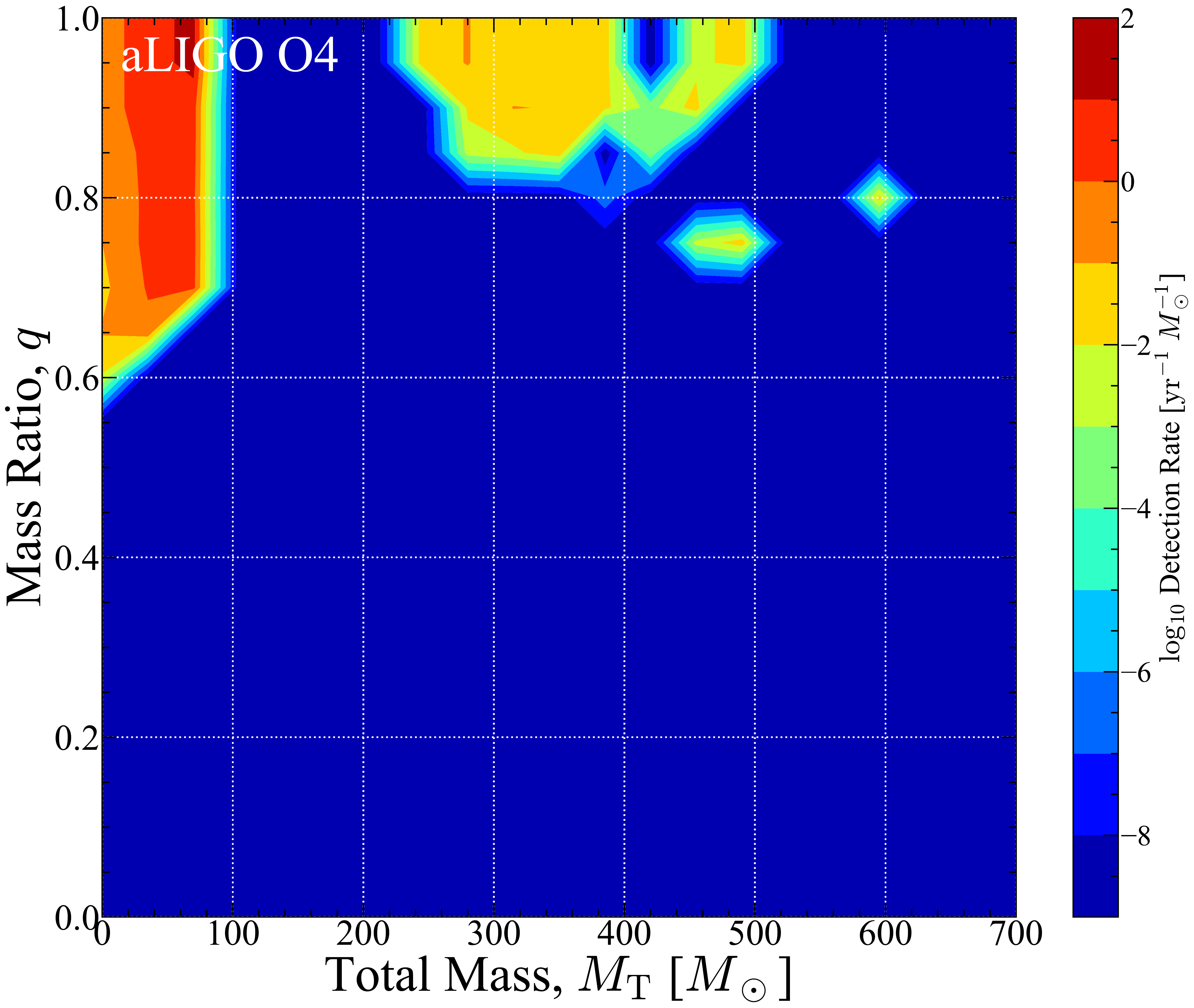}
         \subcaption{$\alpha\lambda=0.5$}
      \end{minipage} \\

      \begin{minipage}[t]{0.32\linewidth}
         \centering
         \includegraphics[width=\linewidth]{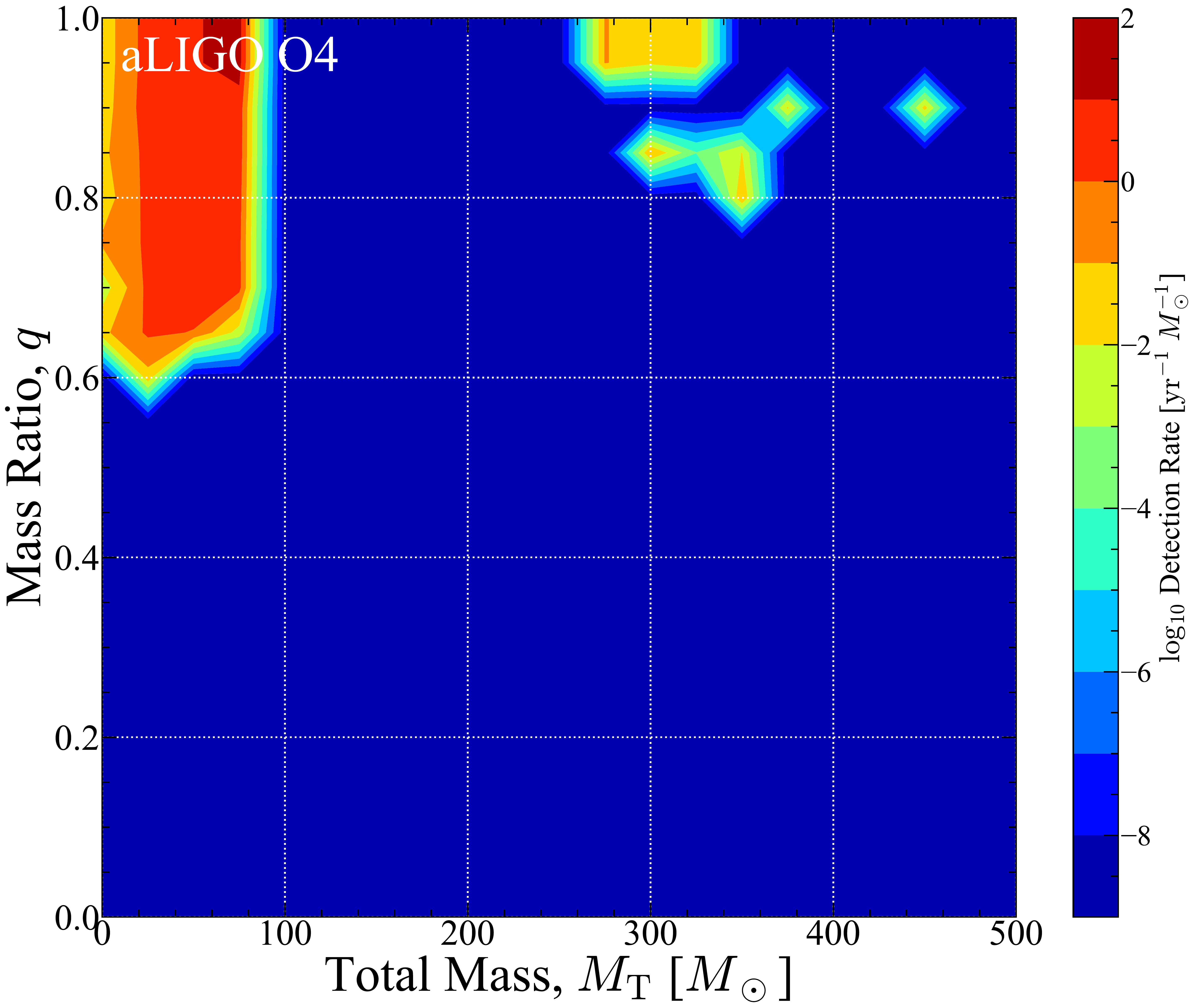}
         \subcaption{$\alpha\lambda=1.0$}
      \end{minipage}
      \begin{minipage}[t]{0.32\linewidth}
         \centering
         \includegraphics[width=\linewidth]{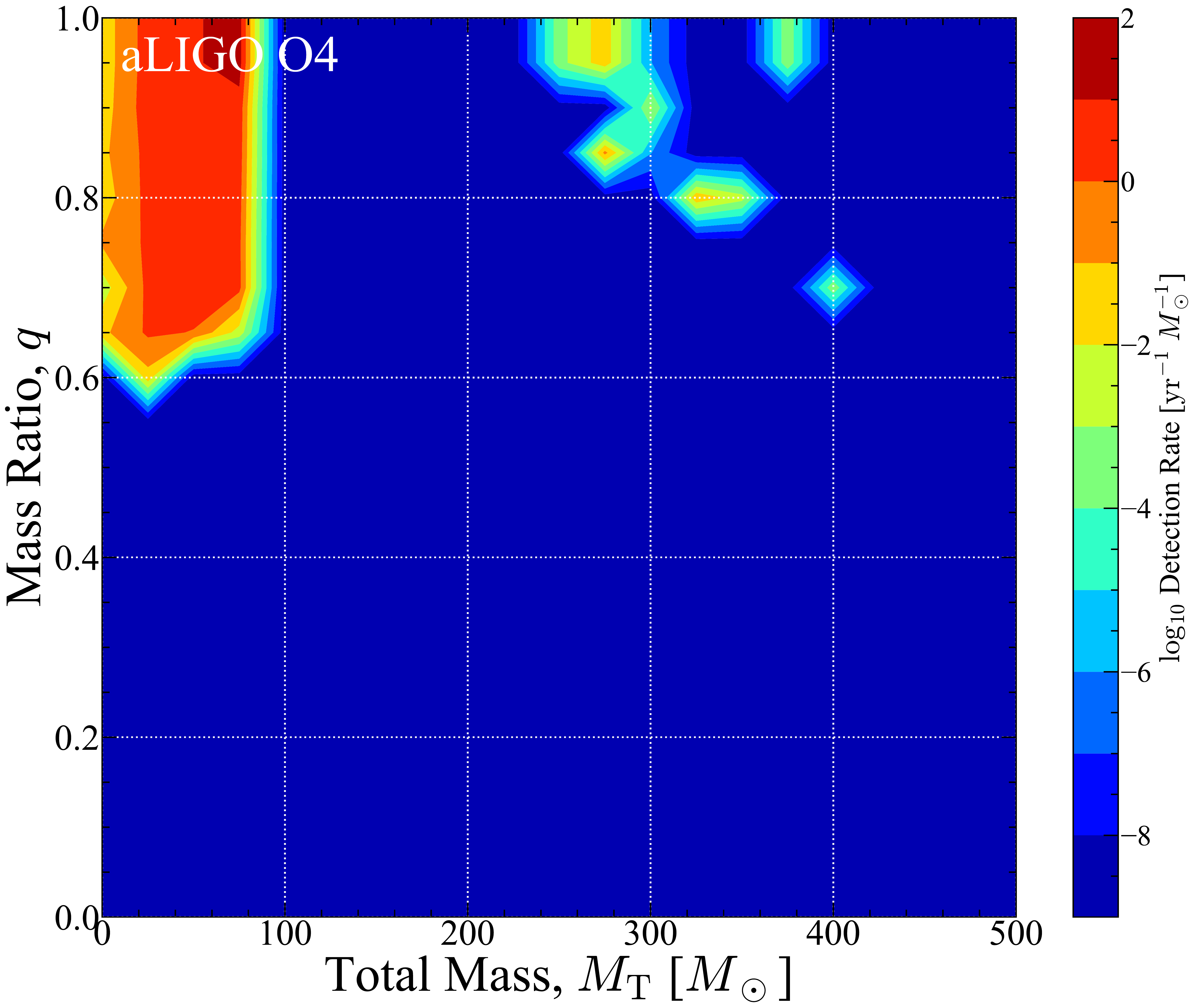}
         \subcaption{$\alpha\lambda=2.0$}
      \end{minipage}
      \begin{minipage}[t]{0.32\linewidth}
         \centering
         \includegraphics[width=\linewidth]{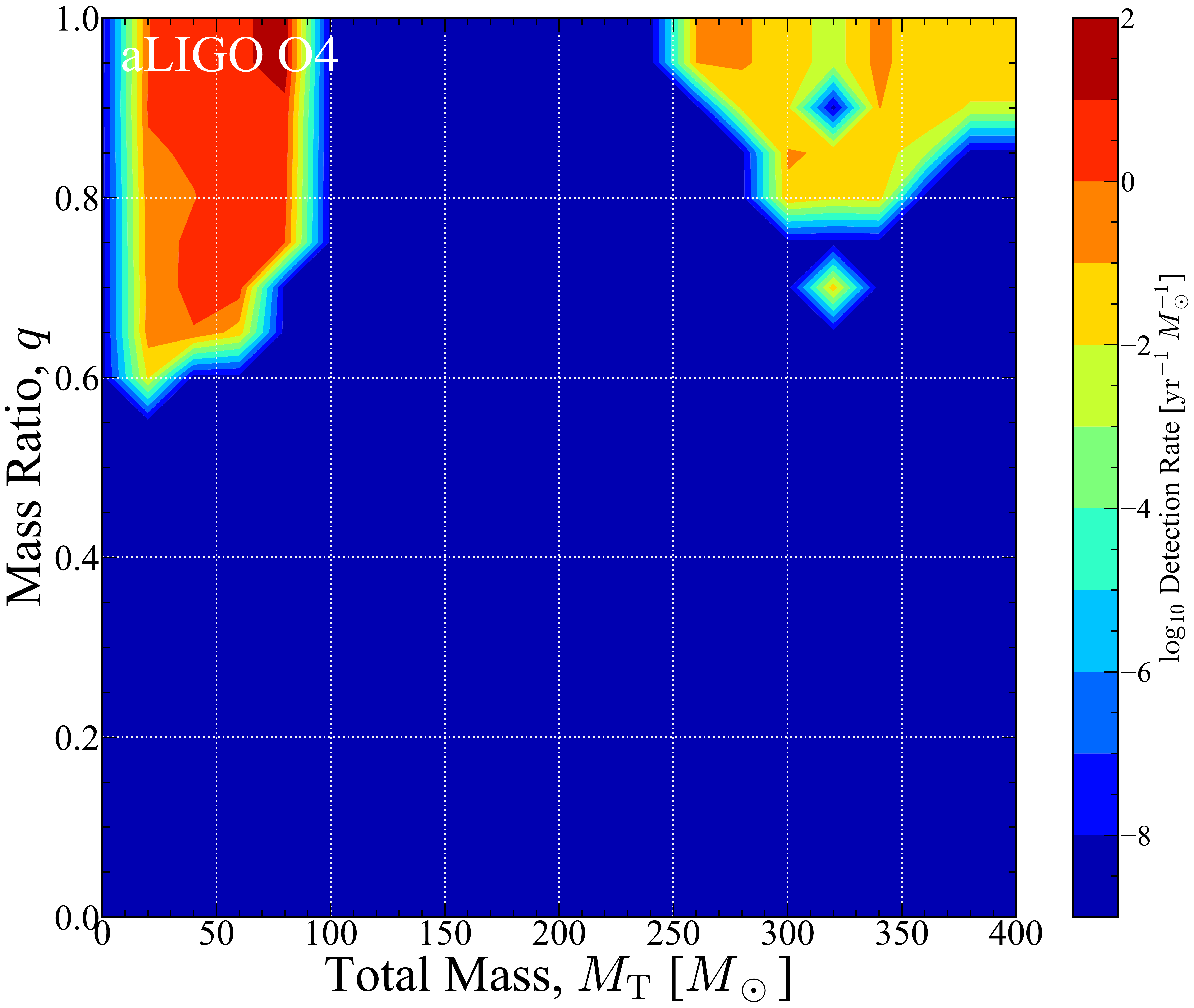}
         \subcaption{$\alpha\lambda=5.0$}
      \end{minipage} 
   \caption{The detection rate by advanced LIGO for each $\alpha\lambda$.
   Note that not all panels have the same color scale.}
   \label{fig: detection aligo}
\end{figure*}

\begin{figure*}
      \begin{minipage}[t]{0.32\linewidth}
         \centering
         \includegraphics[width=\linewidth]{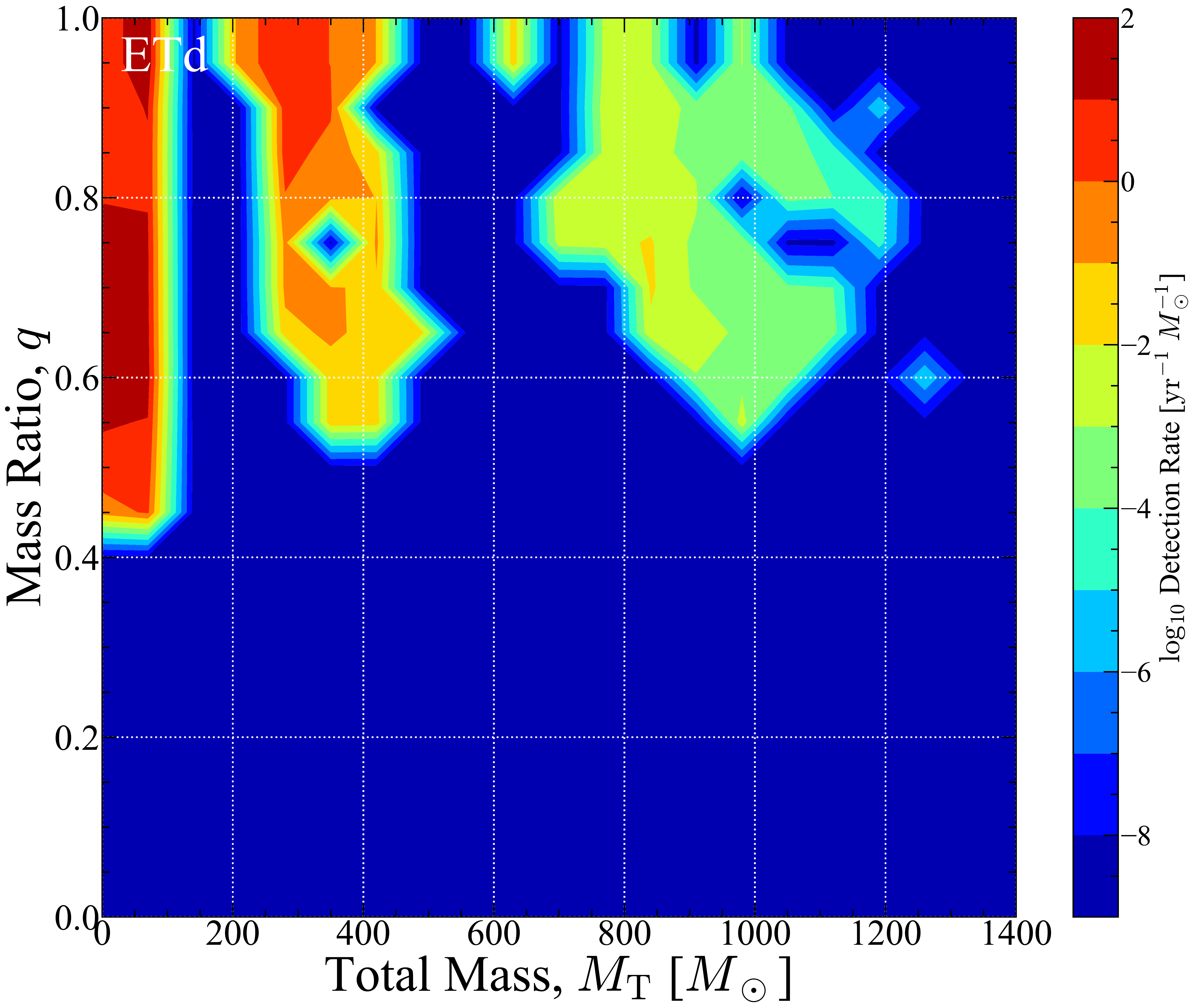}
         \subcaption{$\alpha\lambda=0.01$}
      \end{minipage}
      \begin{minipage}[t]{0.32\linewidth}
         \centering
         \includegraphics[width=\linewidth]{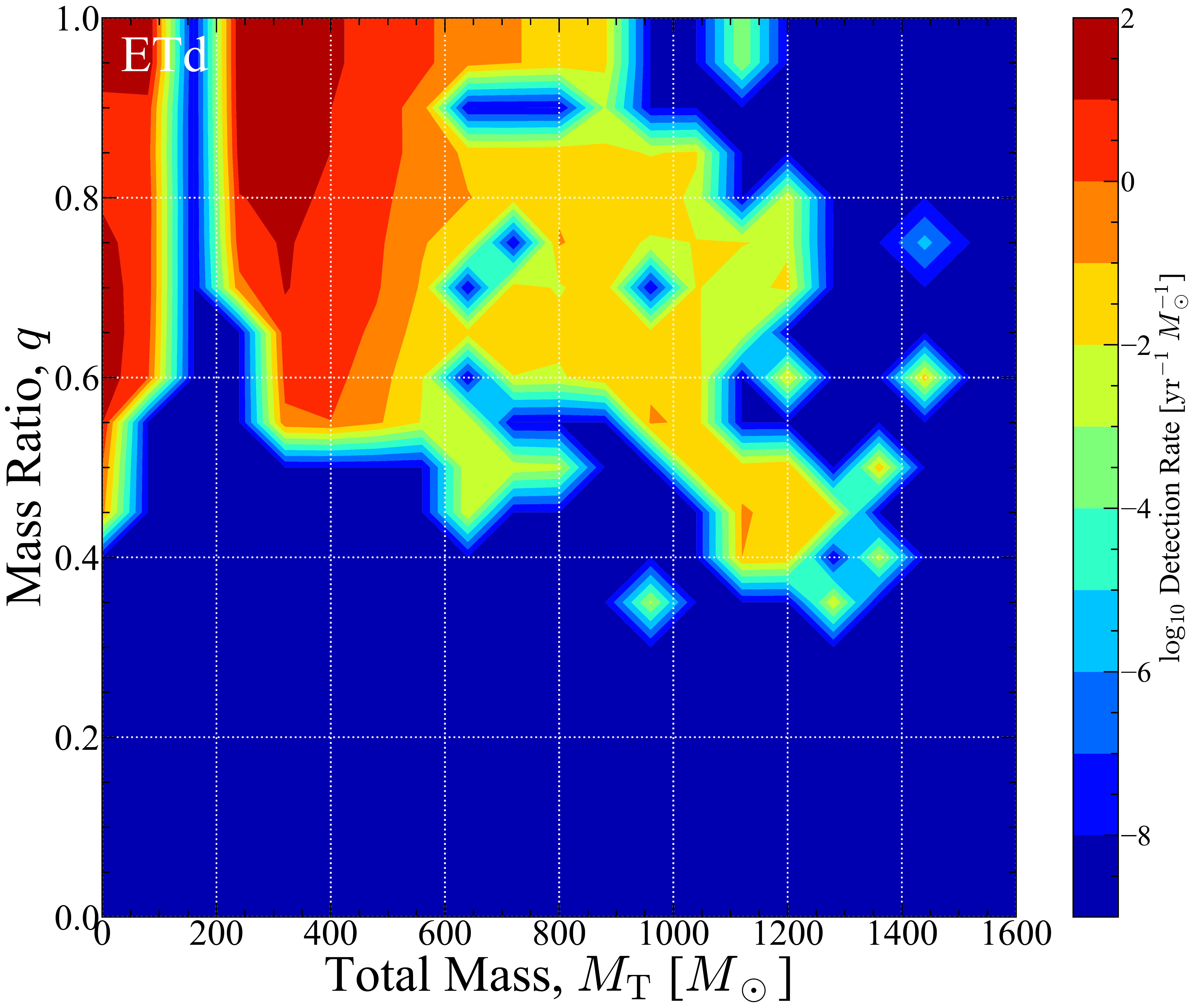}
         \subcaption{$\alpha\lambda=0.1$}
      \end{minipage}
      \begin{minipage}[t]{0.32\linewidth}
         \centering
         \includegraphics[width=\linewidth]{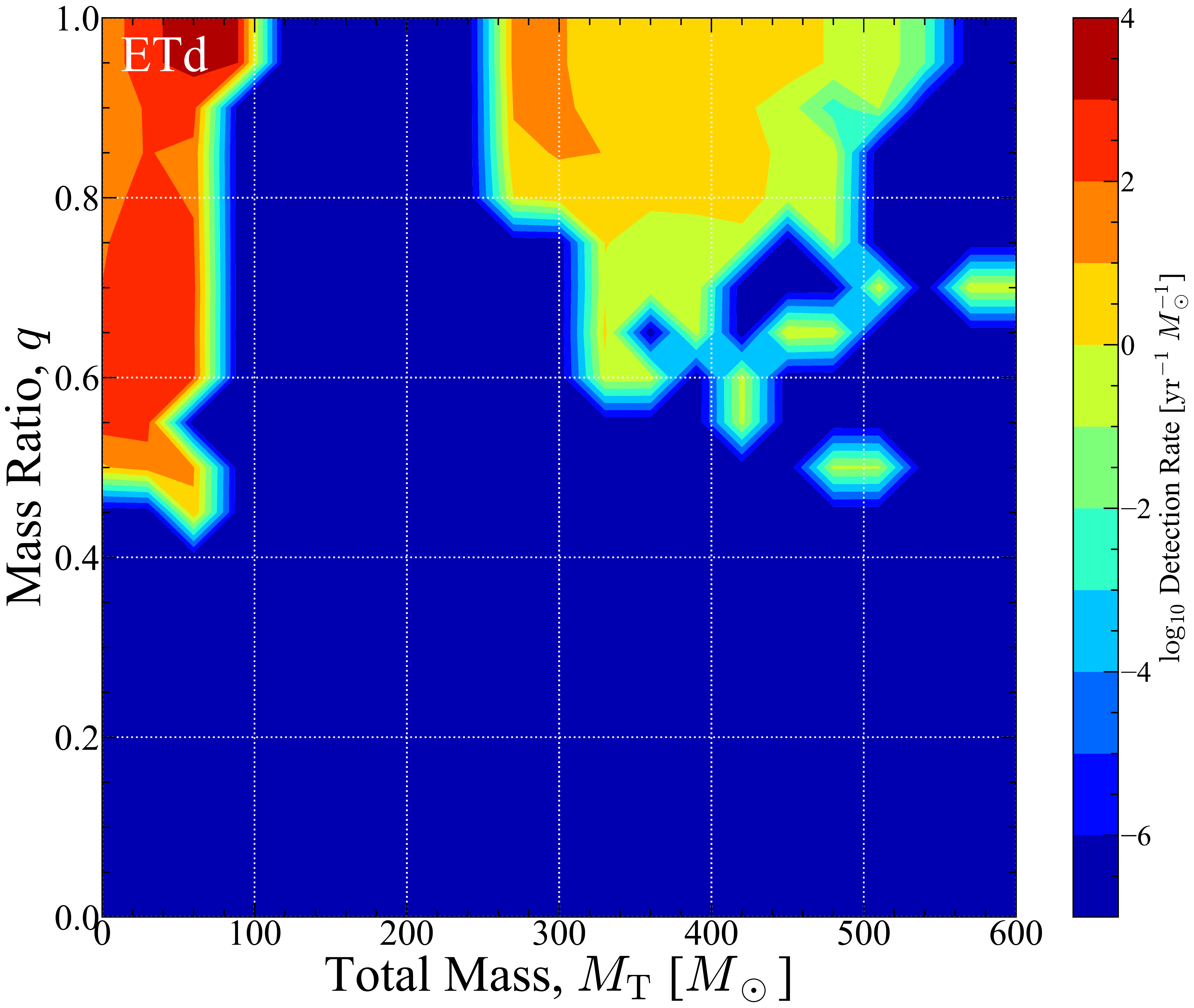}
         \subcaption{$\alpha\lambda=0.5$}
      \end{minipage} \\

      \begin{minipage}[t]{0.32\linewidth}
         \centering
         \includegraphics[width=\linewidth]{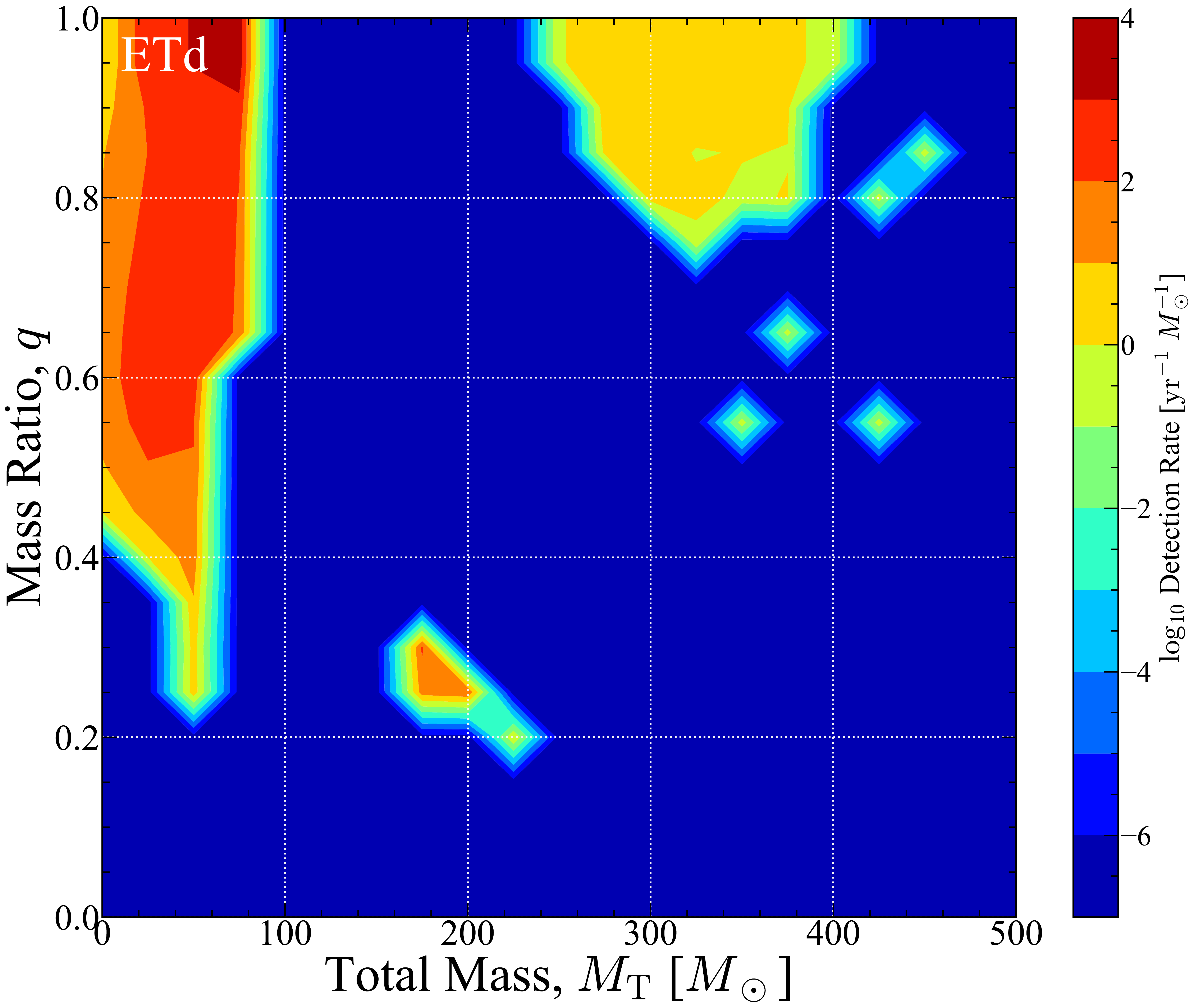}
         \subcaption{$\alpha\lambda=1.0$}
      \end{minipage}
      \begin{minipage}[t]{0.32\linewidth}
         \centering
         \includegraphics[width=\linewidth]{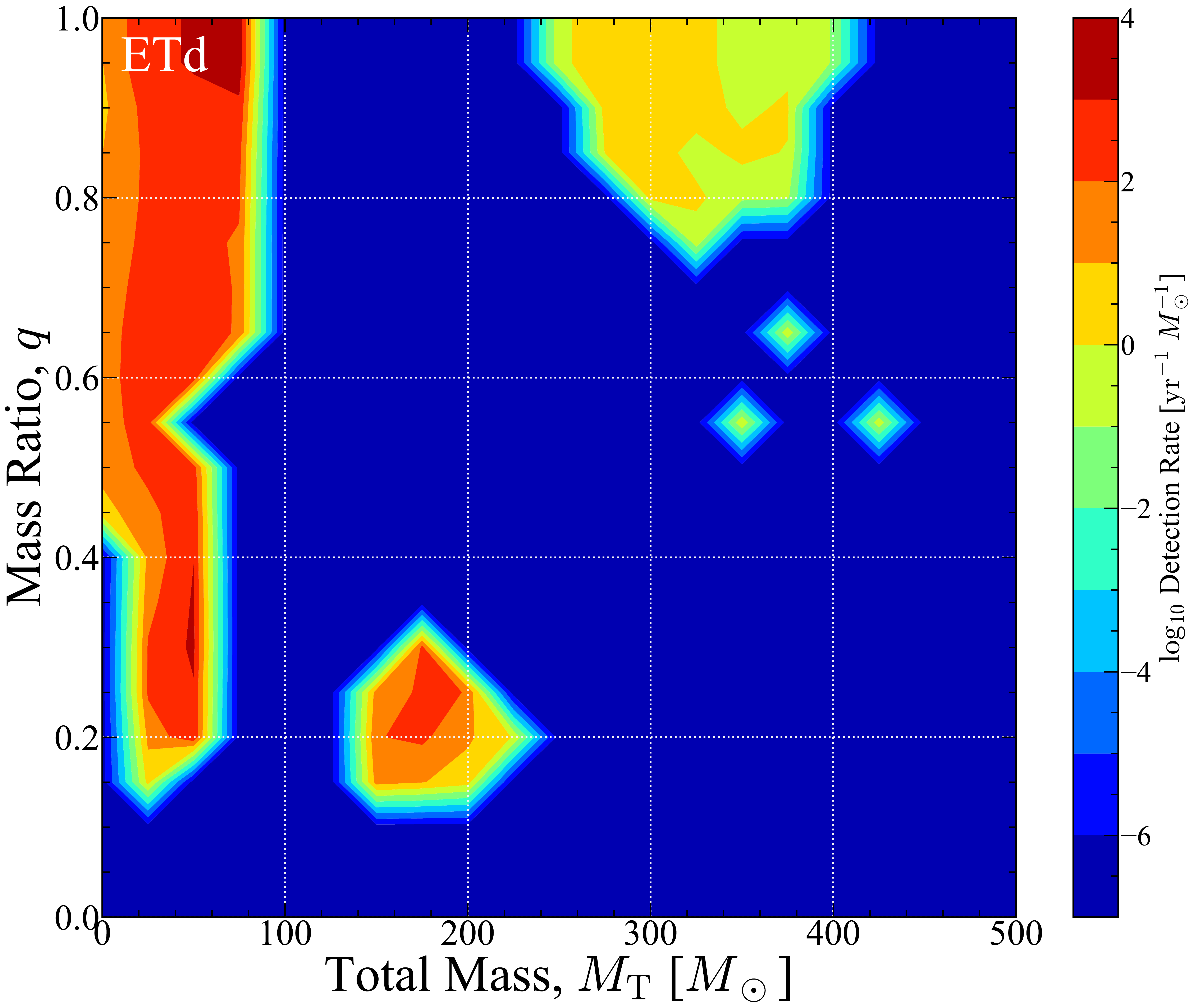}
         \subcaption{$\alpha\lambda=2.0$}
      \end{minipage}
      \begin{minipage}[t]{0.32\linewidth}
         \centering
         \includegraphics[width=\linewidth]{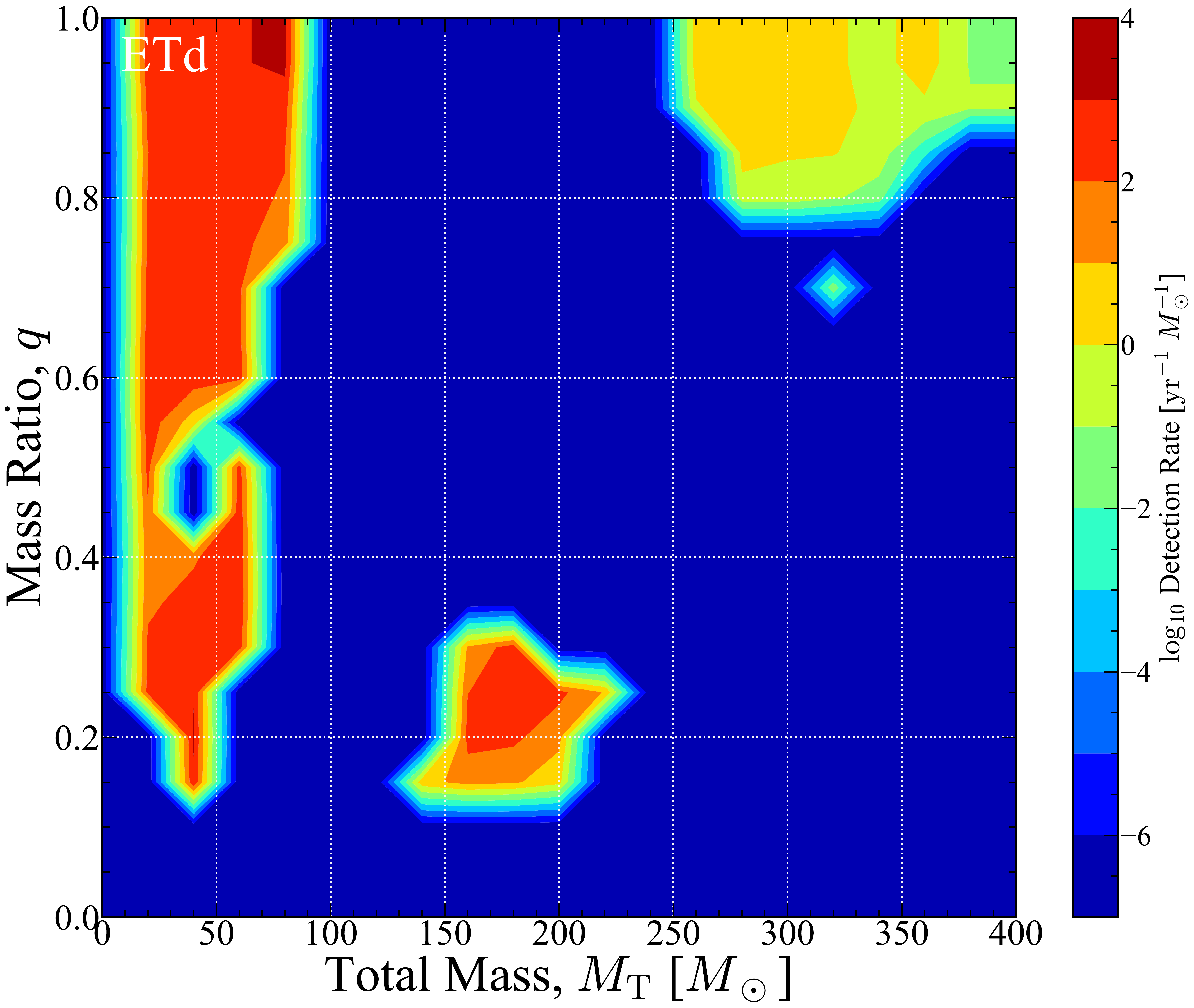}
         \subcaption{$\alpha\lambda=5.0$}
      \end{minipage} 
   \caption{The detection rate by Einstein telescope for each $\alpha\lambda$.
   Note that not all panels have the same color scale.}
   \label{fig: detection etd}
\end{figure*}

\begin{figure*}
      \begin{minipage}[t]{0.32\linewidth}
         \centering
         \includegraphics[width=\linewidth]{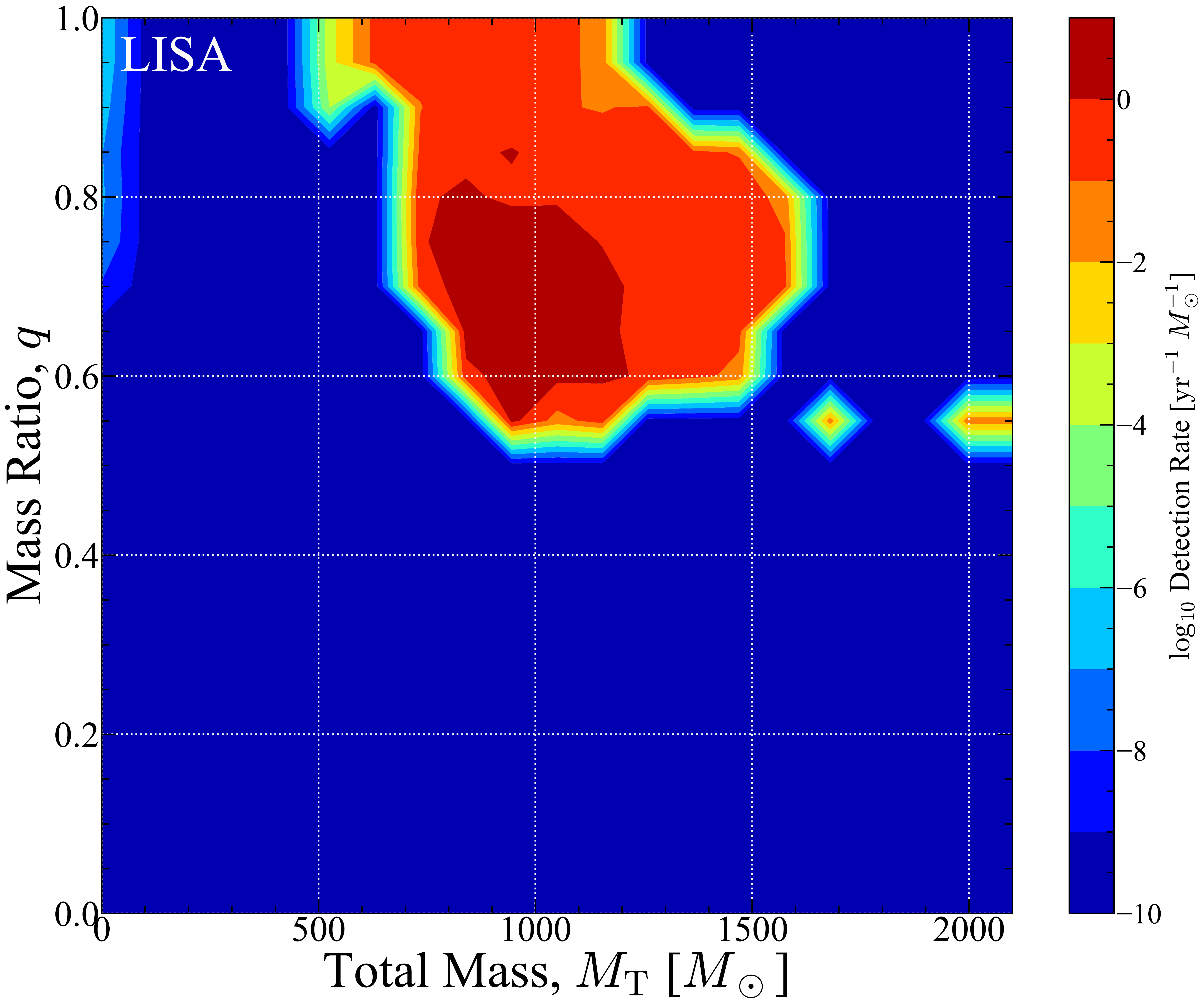}
         \subcaption{$\alpha\lambda=0.01$}
      \end{minipage}
      \begin{minipage}[t]{0.32\linewidth}
         \centering
         \includegraphics[width=\linewidth]{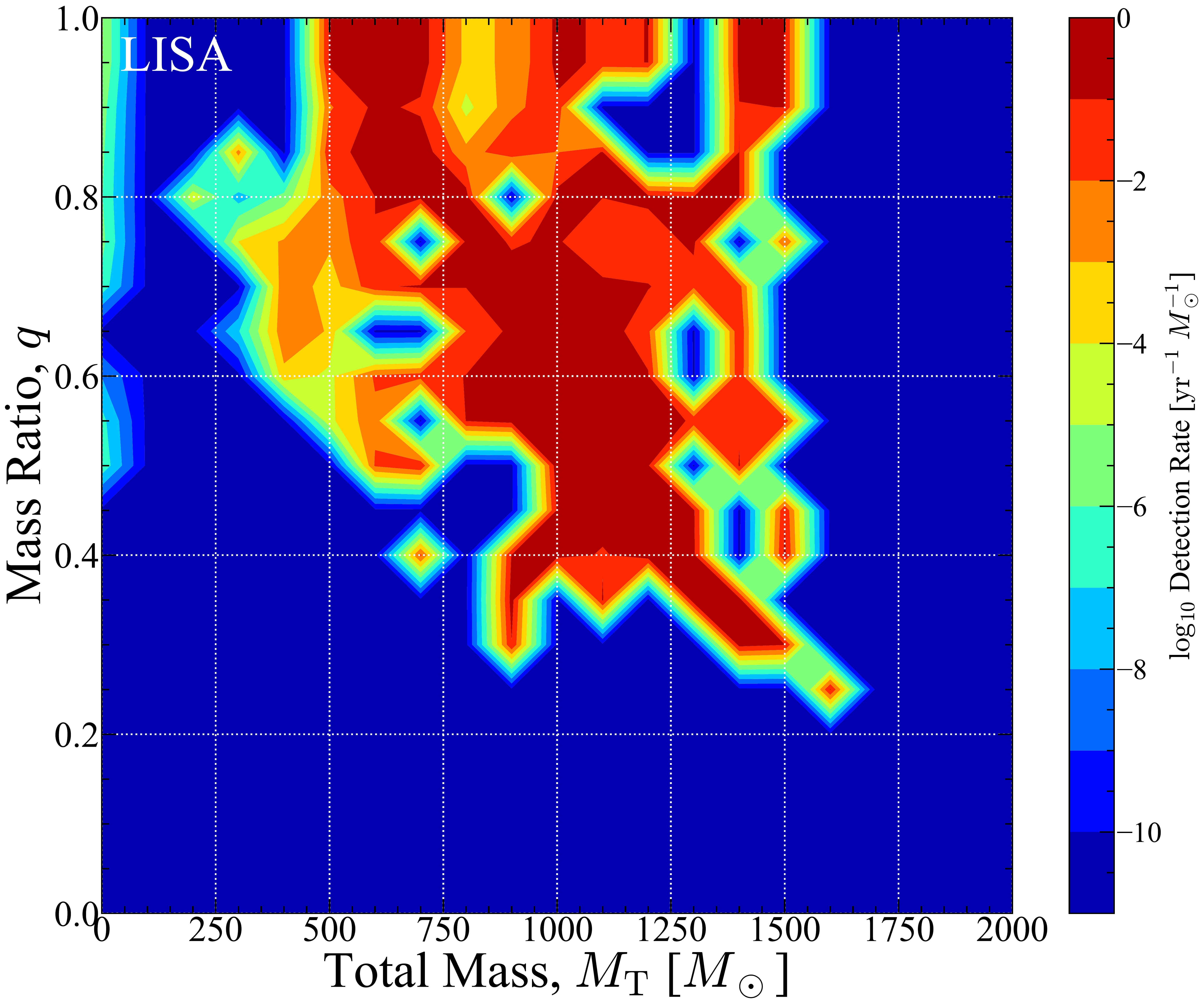}
         \subcaption{$\alpha\lambda=0.1$}
      \end{minipage}
      \begin{minipage}[t]{0.32\linewidth}
         \centering
         \includegraphics[width=\linewidth]{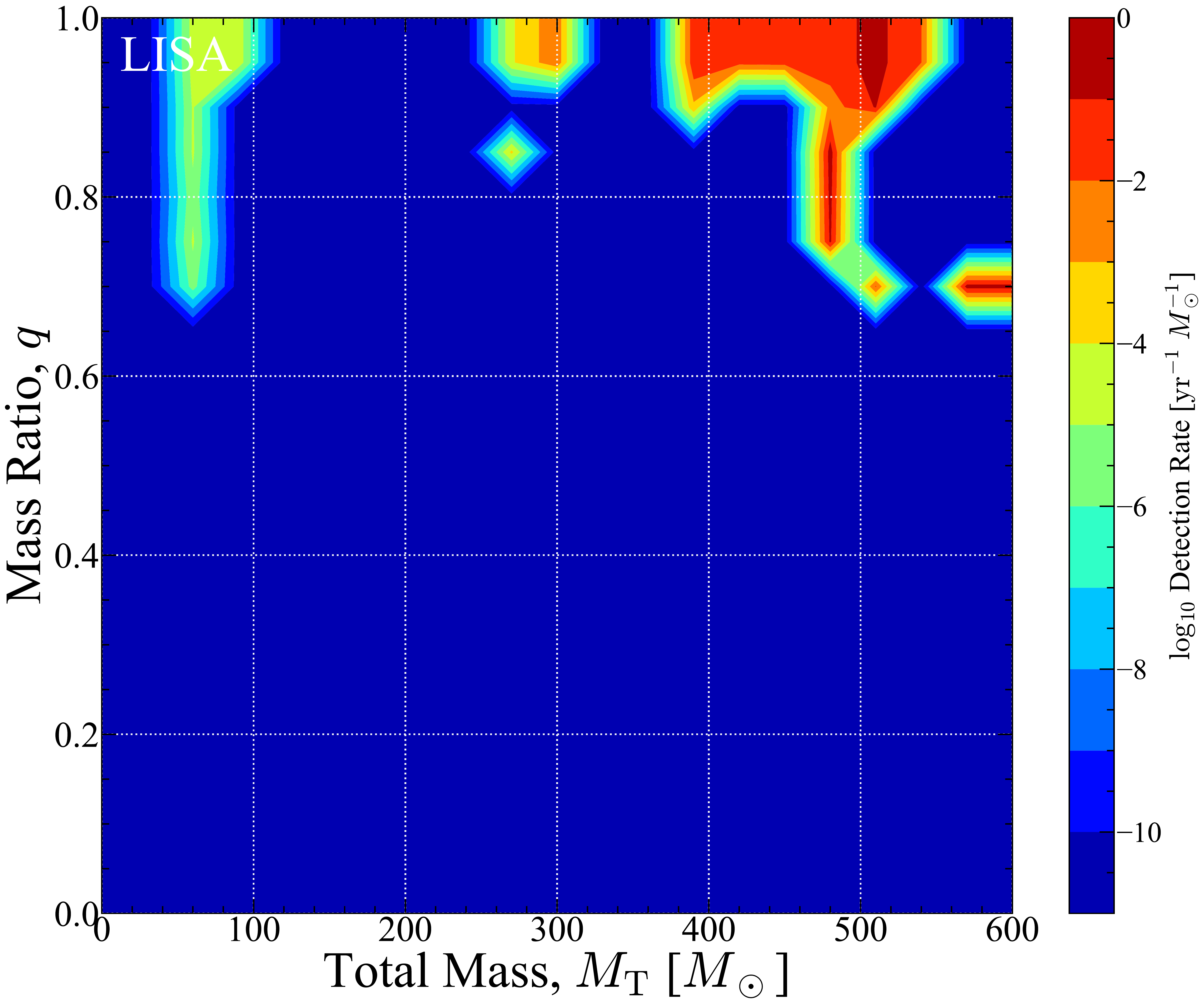}
         \subcaption{$\alpha\lambda=0.5$}
      \end{minipage} \\

      \begin{minipage}[t]{0.32\linewidth}
         \centering
         \includegraphics[width=\linewidth]{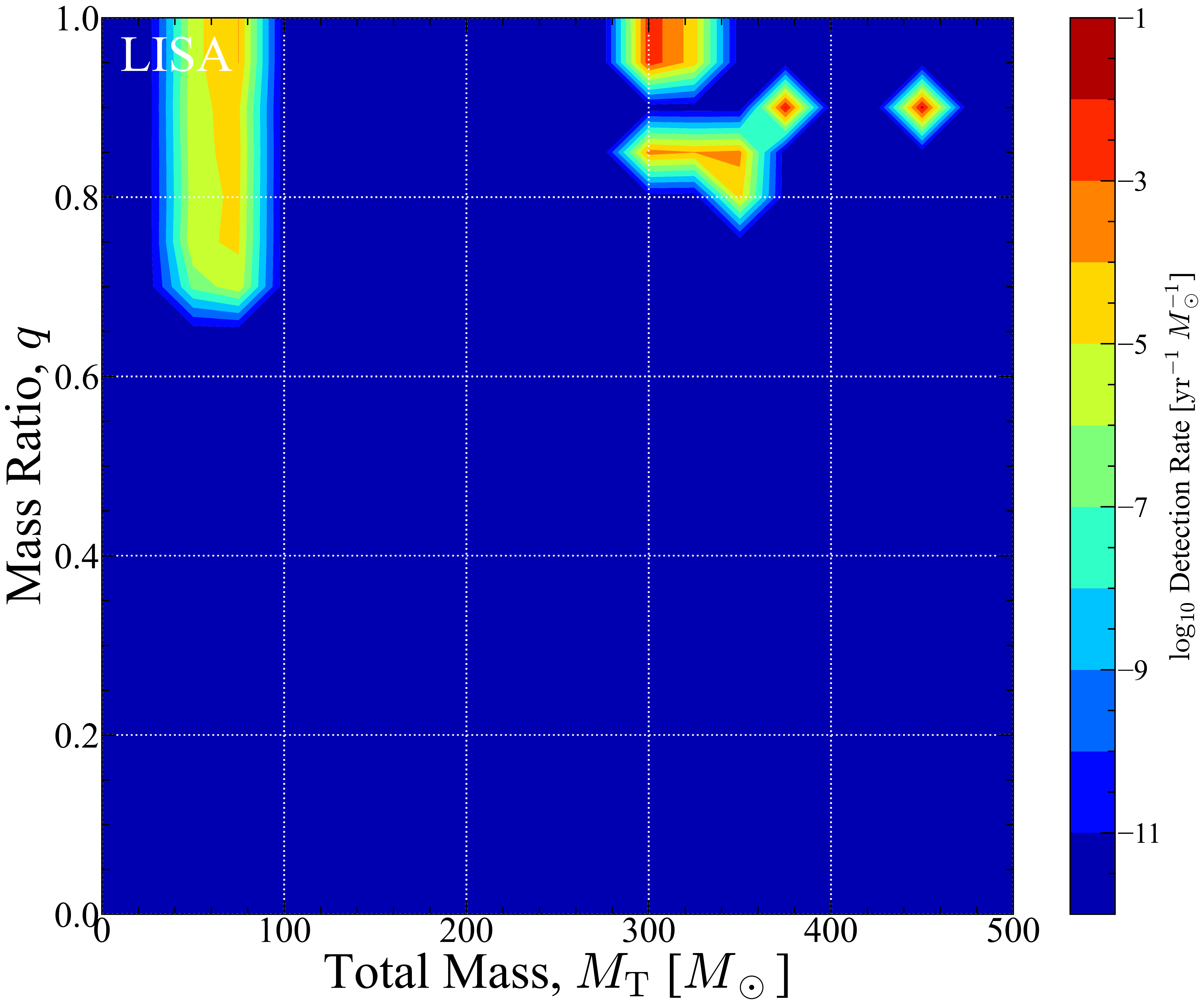}
         \subcaption{$\alpha\lambda=1.0$}
      \end{minipage}
      \begin{minipage}[t]{0.32\linewidth}
         \centering
         \includegraphics[width=\linewidth]{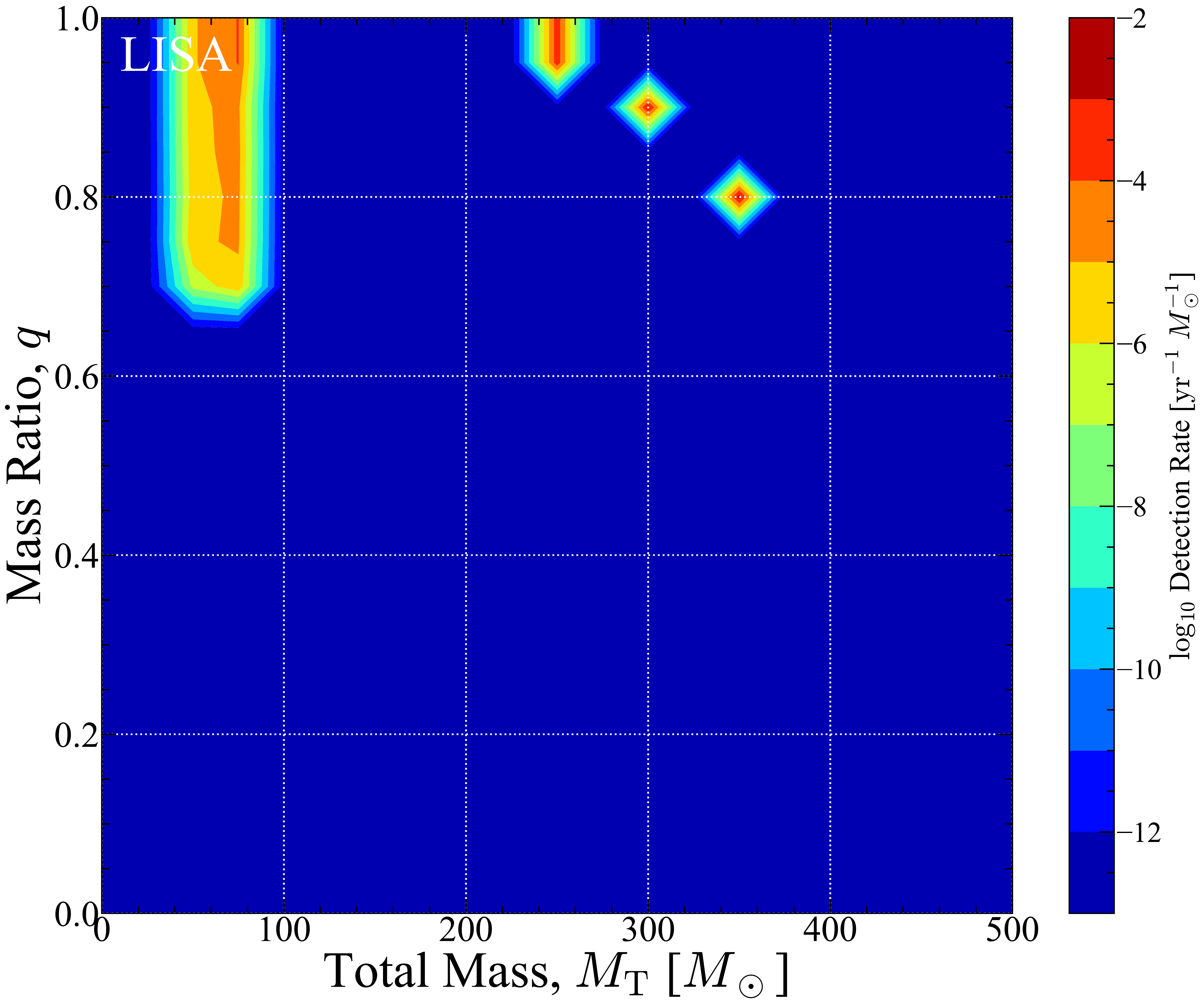}
         \subcaption{$\alpha\lambda=2.0$}
      \end{minipage}
      \begin{minipage}[t]{0.32\linewidth}
         \centering
         \includegraphics[width=\linewidth]{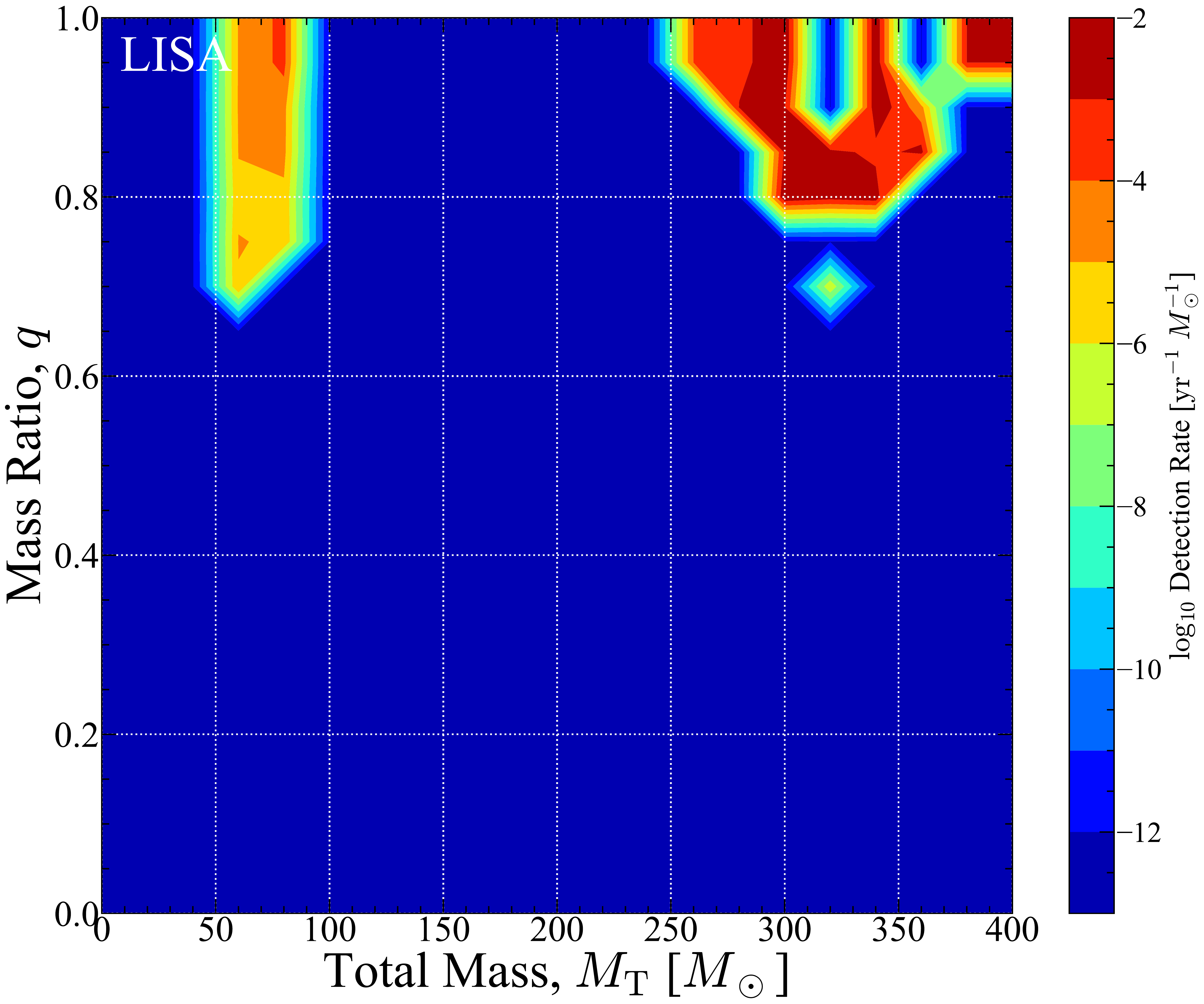}
         \subcaption{$\alpha\lambda=5.0$}
      \end{minipage} 
   \caption{The detection rate by LISA for each $\alpha\lambda$.
   Note that not all panels have the same color scale.}
   \label{fig: detection lisa}
\end{figure*}

The detection rate by advanced LIGO (O4) are shown in Figure \ref{fig: detection aligo}.
The IMF models with the highest detection rate by advanced LIGO (O4) are
$(\gamma_1=0.0, \gamma_2=1.5, m_\mathrm{crit}=100\msun)$, 
$(\gamma_1=0.0, \gamma_2=5.0, m_\mathrm{crit}=300\msun)$,
$(\gamma_1=0.0, \gamma_2=4.0, m_\mathrm{crit}=100\msun)$,
$(\gamma_1=1.0, \gamma_2=4.5, m_\mathrm{crit}=100\msun)$,
$(\gamma_1=1.0, \gamma_2=4.5, m_\mathrm{crit}=100\msun)$ and
$(\gamma_1=0.0, \gamma_2=4.0, m_\mathrm{crit}=100\msun)$ for
$\alpha\lambda=$0.01, 0.1, 0.5, 1.0, 2.0 and 5.0.
When the value of $\alpha\lambda$ is small, the delay time is also short.
Therefore, BBHs with IMBHs are hard to be detected by the advanced LIGO with the horizon $z\sim1$.
On the other hand, when $\alpha\lambda$ is larger than 0.5, BBH with IMBHs merges within $z\sim1$,
and the advanced LIGO have a potential to detect the BBHs with total mass $\sim300\msun$ and mass ratio $\sim 1$.

The detection rate by Einstein telescope are shown in Figure \ref{fig: detection etd}.
The IMF models with the highest detection rate by Einstein telescope are
$(\gamma_1=0.0, \gamma_2=4.5, m_\mathrm{crit}=300\msun)$, 
$(\gamma_1=0.0, \gamma_2=5.0, m_\mathrm{crit}=300\msun)$,
$(\gamma_1=1.0, \gamma_2=3.5, m_\mathrm{crit}=100\msun)$,
$(\gamma_1=0.0, \gamma_2=5.0, m_\mathrm{crit}=100\msun)$,
$(\gamma_1=0.0, \gamma_2=5.0, m_\mathrm{crit}=100\msun)$ and
$(\gamma_1=0.0, \gamma_2=4.0, m_\mathrm{crit}=100\msun)$ for
$\alpha\lambda=$0.01, 0.1, 0.5, 1.0, 2.0 and 5.0.
Einstein telescope can detect a BBH merger up to $z\sim$ 10--100 (see Figure \ref{fig: horizon}).
The total mass and mass ratio of observable ``high mass $+$ high mass'' BBHs are
$\sim300\msun$ and $\sim1$, respectively.
When $\alpha\lambda$ is larger than 1, 
``low mass $+$ high mass'' BBH mergers can occur (see Section ref{subsec: lh}), and can be 
detected by Einstein telescope. 
Their total mass and mass ratio are $\sim200\msun$ and $\sim0.2$, respectively.
The detection rate by Cosmic Explorer is roughly same as that by Einstein telescope.

The detection rate by LISA are shown in Figure \ref{fig: detection lisa}.
The IMF models with the highest detection rate by LISA are
$(\gamma_1=0.0, \gamma_2=1.5, m_\mathrm{crit}=300\msun)$, 
$(\gamma_1=0.0, \gamma_2=2.5, m_\mathrm{crit}=100\msun)$,
$(\gamma_1=1.0, \gamma_2=3.5, m_\mathrm{crit}=100\msun)$,
$(\gamma_1=1.0, \gamma_2=4.5, m_\mathrm{crit}=100\msun)$,
$(\gamma_1=1.0, \gamma_2=4.5, m_\mathrm{crit}=100\msun)$ and
$(\gamma_1=0.0, \gamma_2=4.0, m_\mathrm{crit}=100\msun)$ for
$\alpha\lambda=$0.01, 0.1, 0.5, 1.0, 2.0 and 5.0.
LISA can detect BBHs with total mass $\sim1000\msun$ up to $z\sim1000$,
but is hard to detect BBHs with total mass $\sim100\msun$ (see Figure \ref{fig: horizon}).
Since the maximum mass decreases when $\alpha\lambda$ increases (see Section \ref{subsubsec: primary mass hh}),
BBHs with IMBHs are hard to be detected by LISA when $\alpha\lambda$ is larger than 0.5.
On the other hand, when $\alpha\lambda$ is smaller than 0.1, BBHs with total mass $\sim1000\msun$
can form and merge. LISA can detect such very massive BBH mergers.
The detection rate by Tianqin is roughly same as that by LISA.

\subsubsection{constraint on the models} \label{subsubsec: constraint}
\begin{figure*}
      \begin{minipage}[t]{0.49\linewidth}
         \centering
         \includegraphics[width=\linewidth]{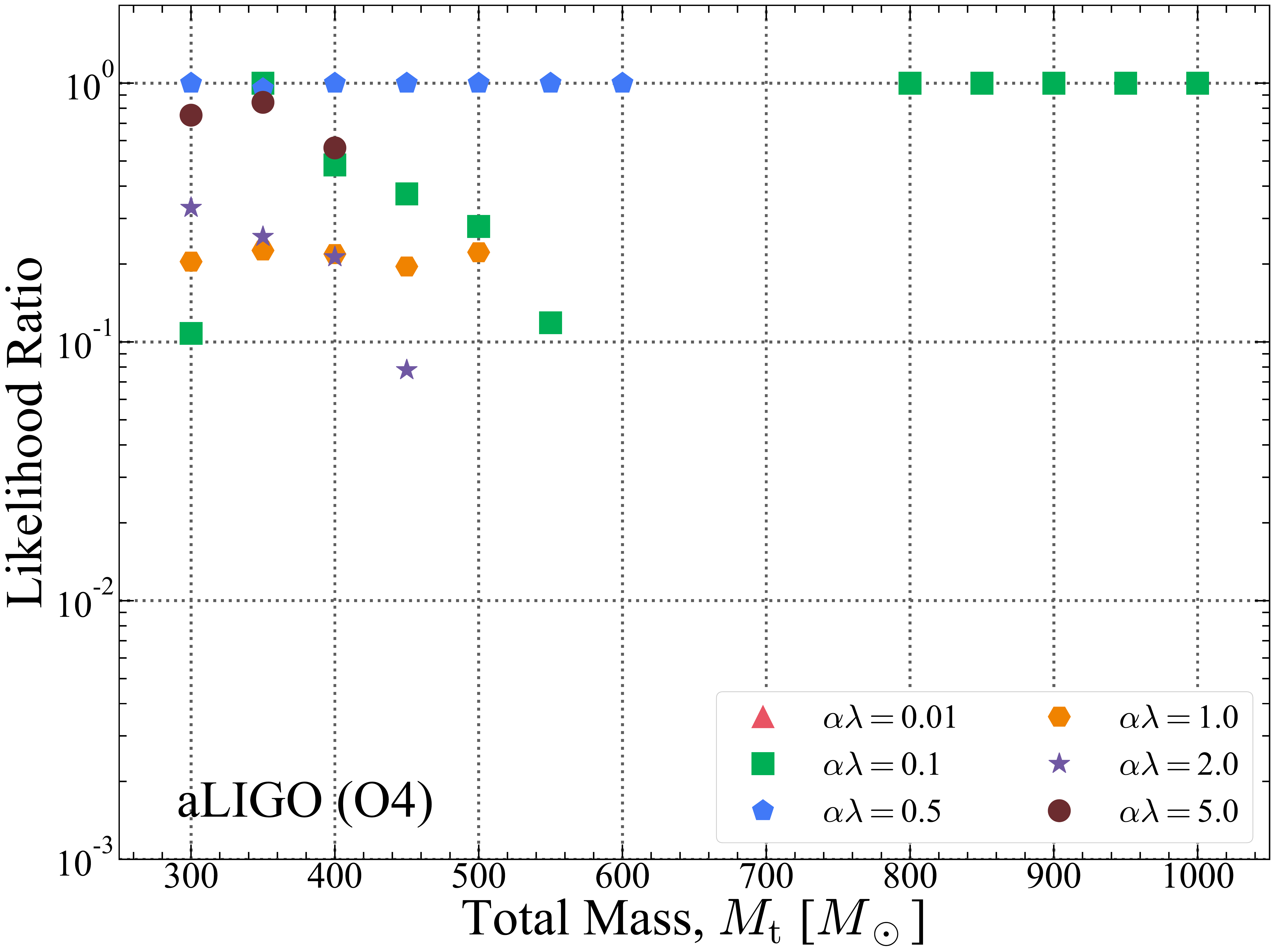}
         \subcaption{advanced LIGO (O4)}
      \end{minipage}
      \begin{minipage}[t]{0.49\linewidth}
         \centering
         \includegraphics[width=\linewidth]{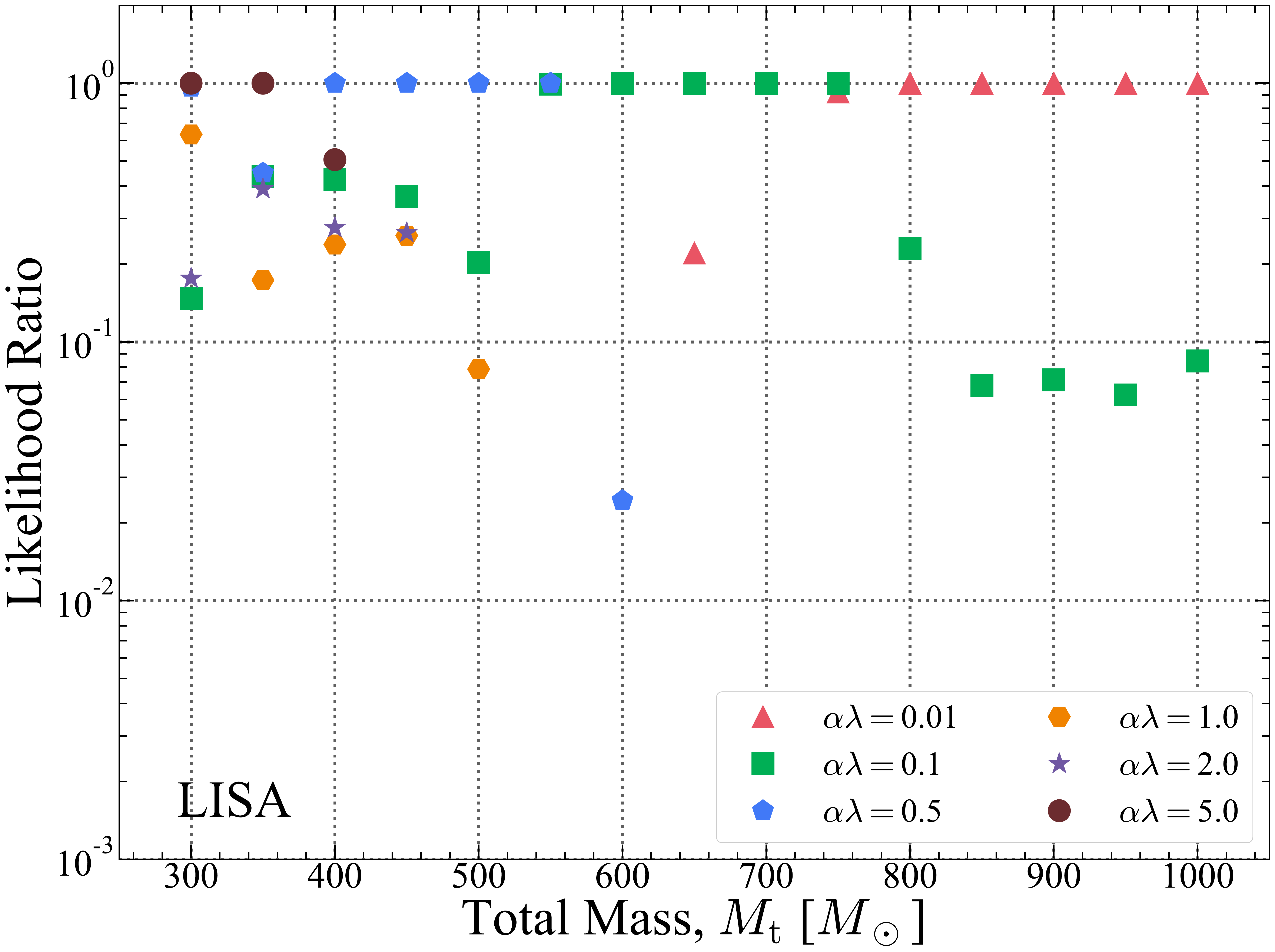}
         \subcaption{LISA}
      \end{minipage}\\
   \caption{The likelihood ratio of $\alpha\lambda$ for each detector.}
   \label{fig: lr al}
\end{figure*}

\begin{figure*}
      \begin{minipage}[t]{0.49\linewidth}
         \centering
         \includegraphics[width=\linewidth]{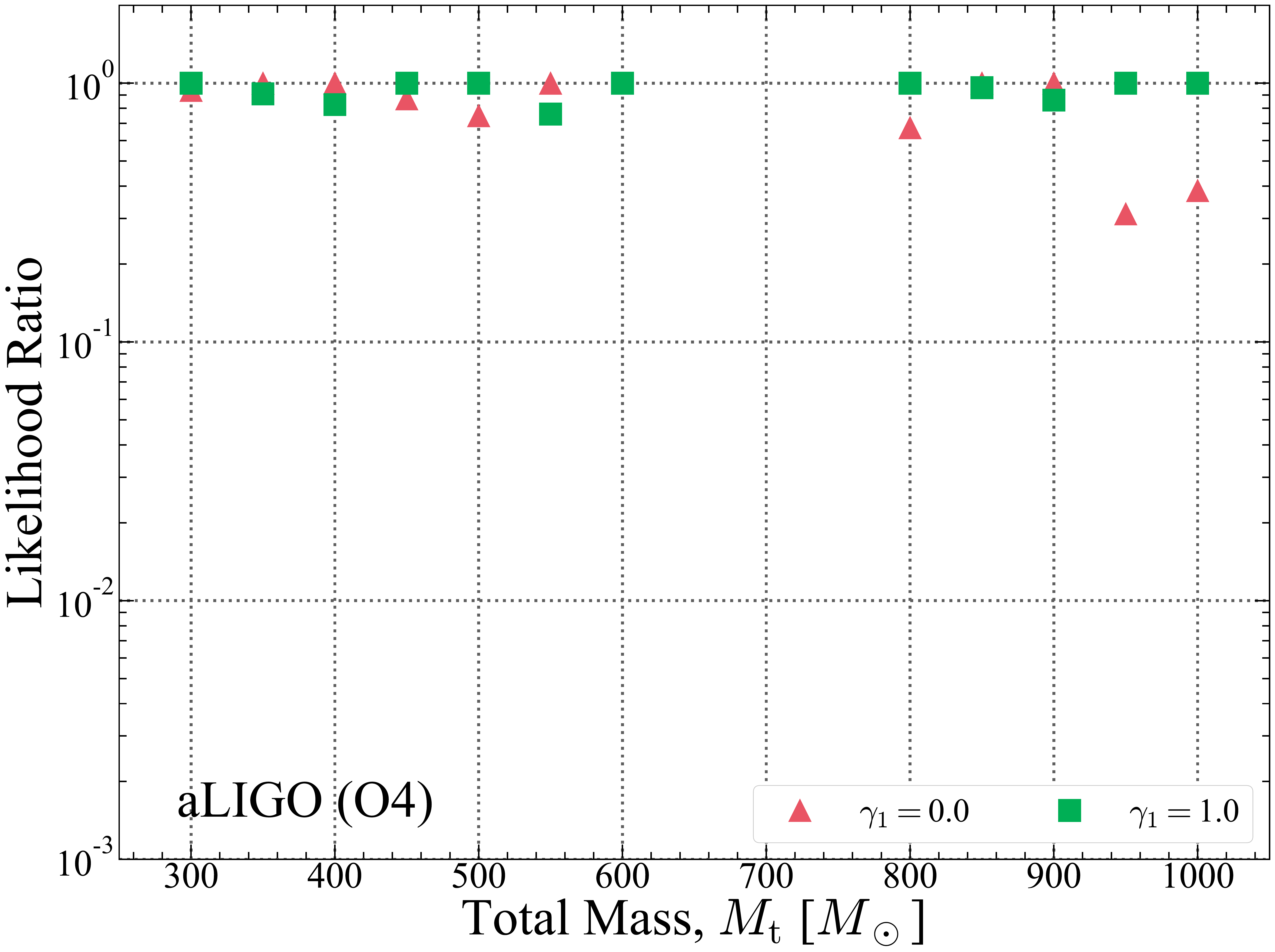}
         \subcaption{advanced LIGO (O4)}
      \end{minipage}
      \begin{minipage}[t]{0.49\linewidth}
         \centering
         \includegraphics[width=\linewidth]{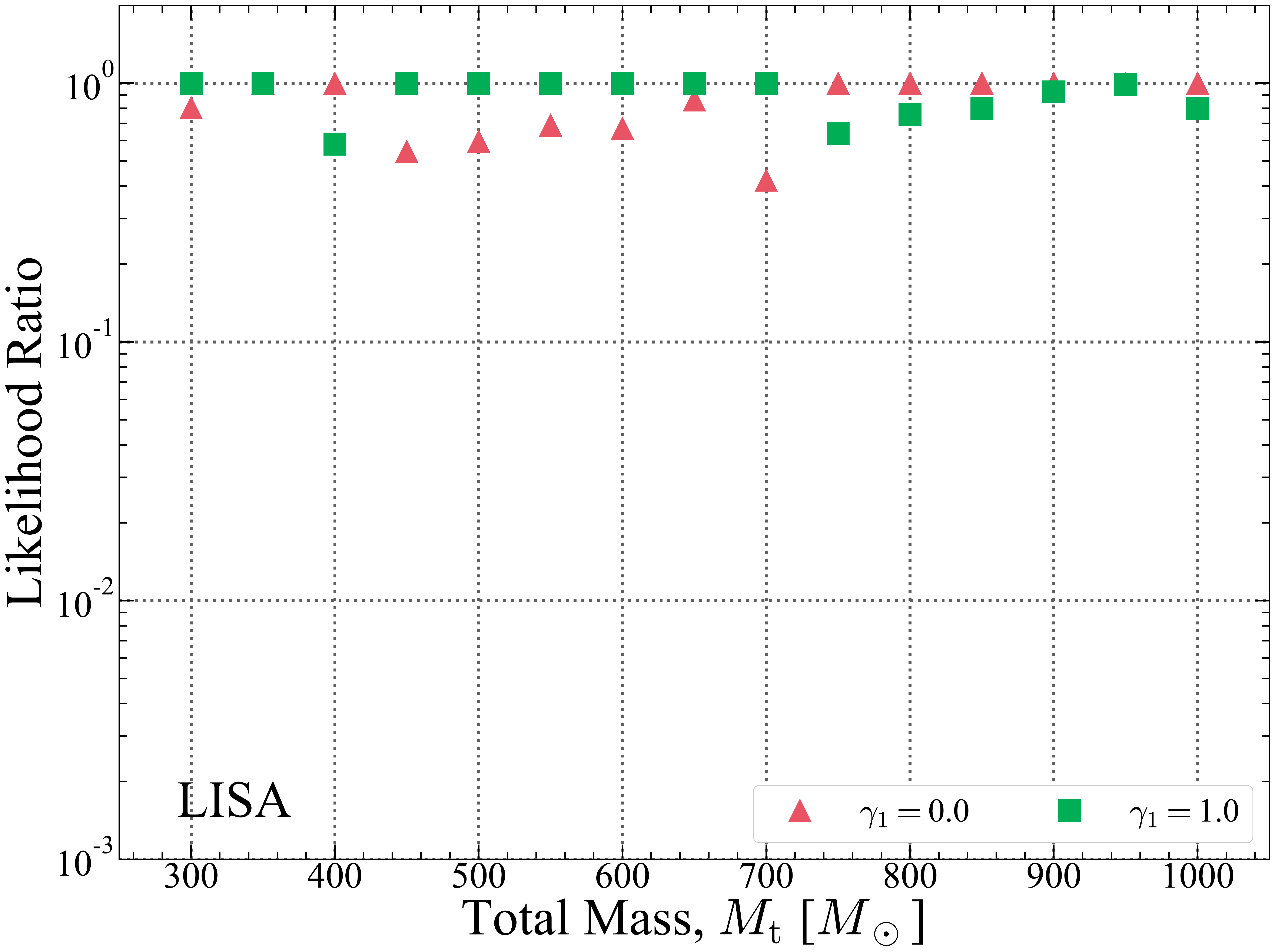}
         \subcaption{LISA}
      \end{minipage}
   \caption{The likelihood ratio of $\gamma_1$ for each detector.}
   \label{fig: lr g1}
\end{figure*}

\begin{figure*}
      \begin{minipage}[t]{0.49\linewidth}
         \centering
         \includegraphics[width=\linewidth]{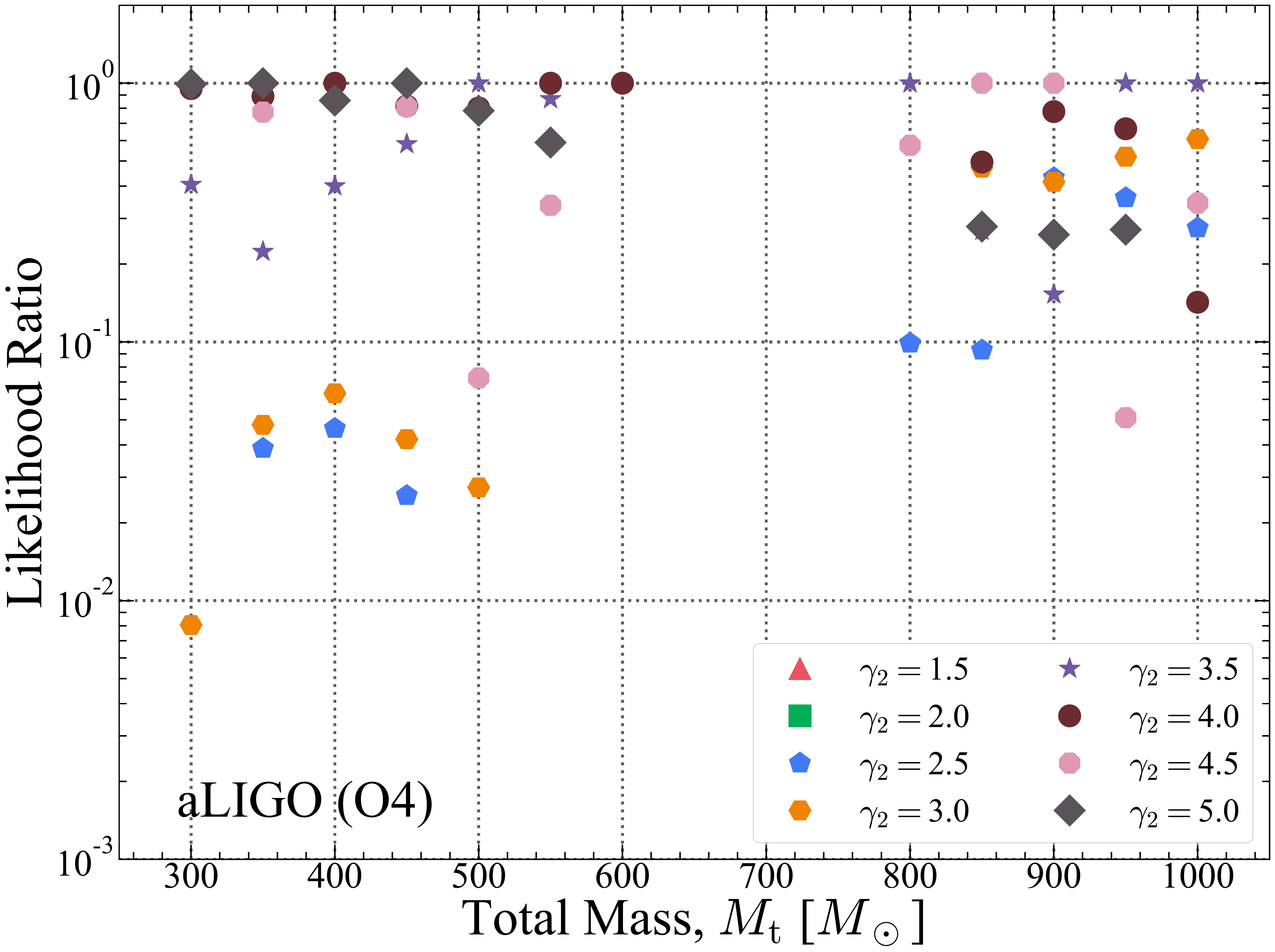}
         \subcaption{advanced LIGO (O4)}
      \end{minipage}
      \begin{minipage}[t]{0.49\linewidth}
         \centering
         \includegraphics[width=\linewidth]{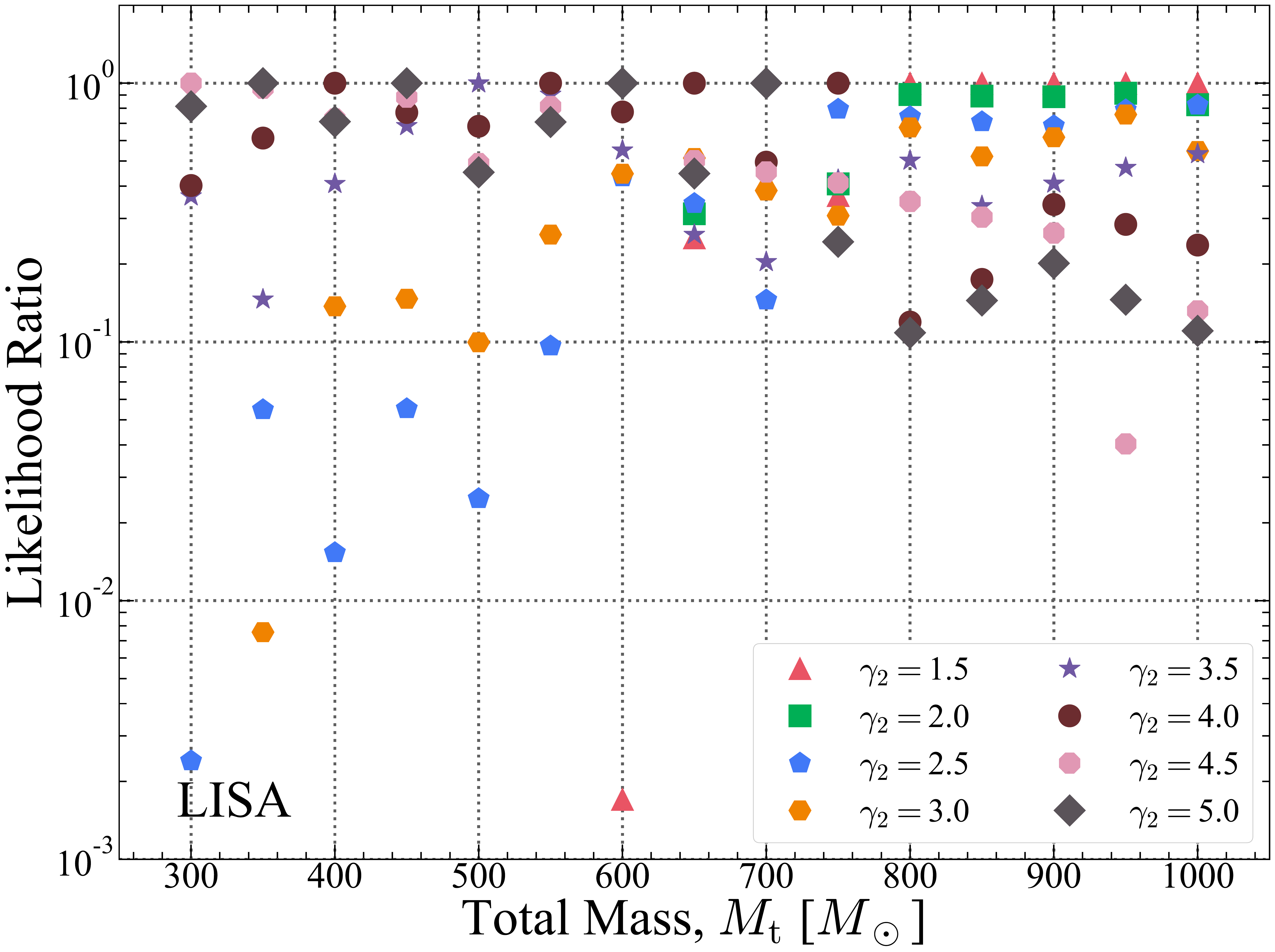}
         \subcaption{LISA}
      \end{minipage}
   \caption{The likelihood ratio of $\gamma_2$ for each detector.}
   \label{fig: lr g2}
\end{figure*}

\begin{figure*}
      \begin{minipage}[t]{0.49\linewidth}
         \centering
         \includegraphics[width=\linewidth]{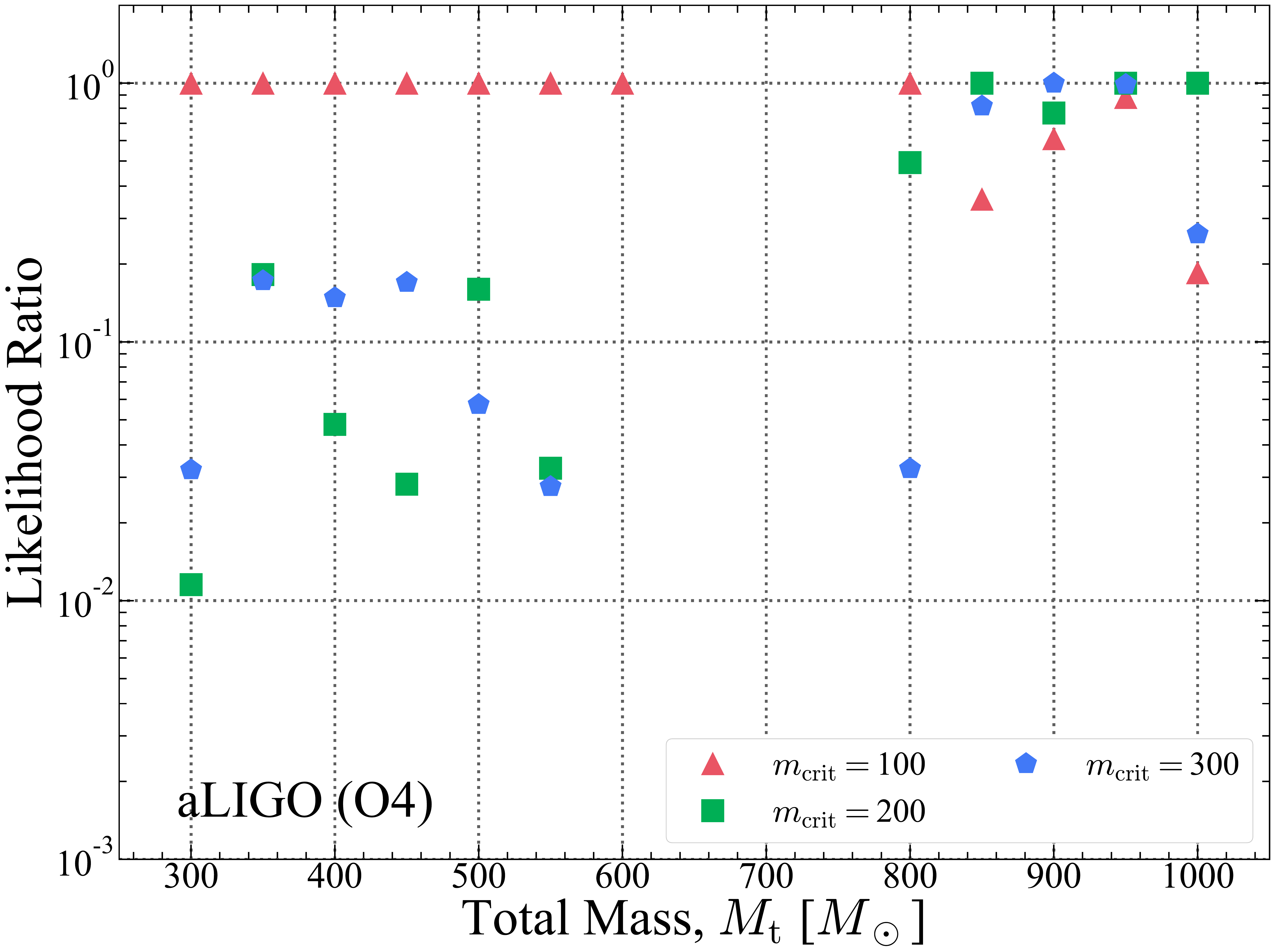}
         \subcaption{advanced LIGO (O4)}
      \end{minipage}
      \begin{minipage}[t]{0.49\linewidth}
         \centering
         \includegraphics[width=\linewidth]{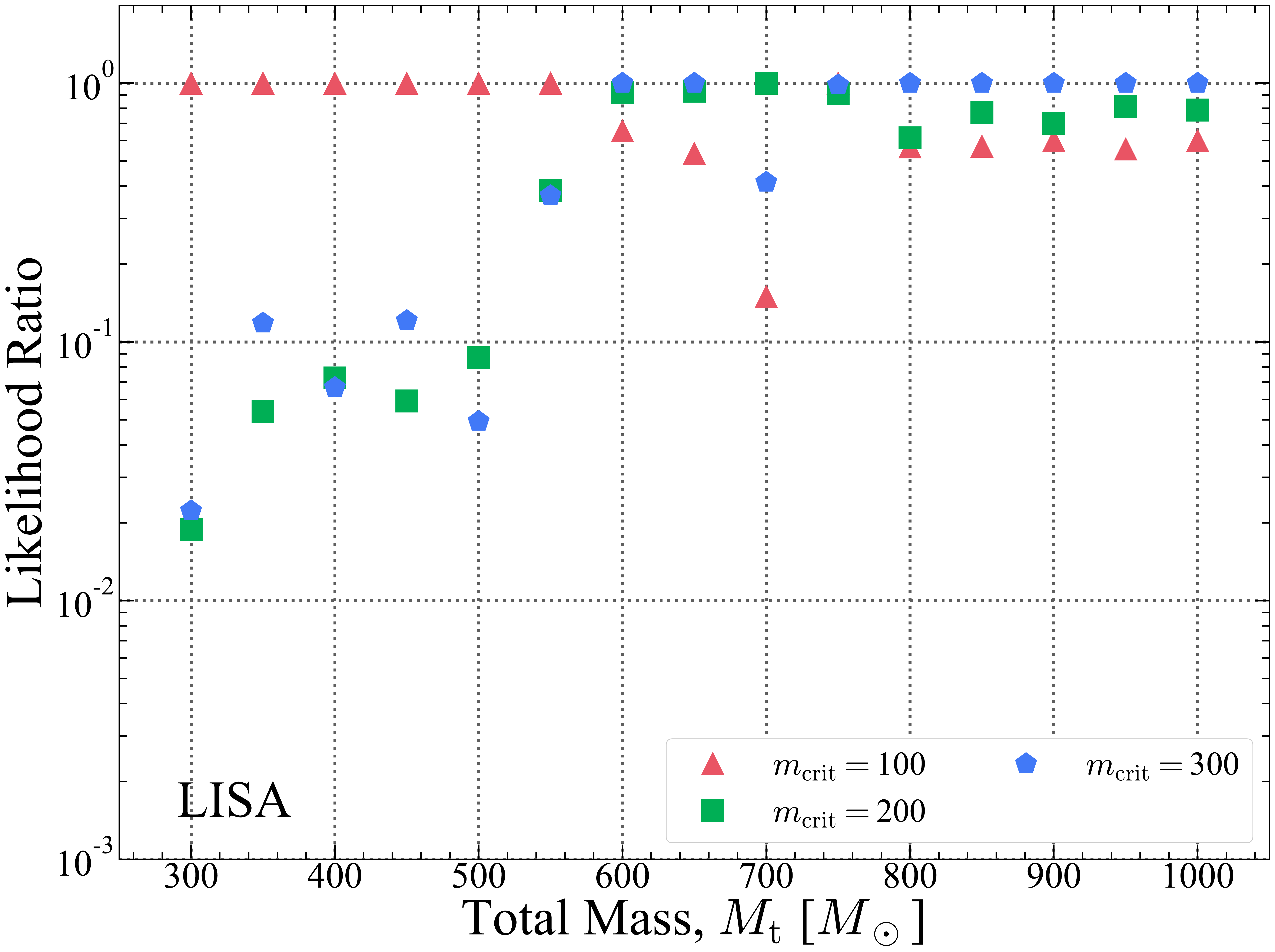}
         \subcaption{LISA}
      \end{minipage}
   \caption{The likelihood ratio of $m_\mathrm{crit}$ for each detector.}
   \label{fig: lr mc}
\end{figure*}

From Table \ref{table: detectionrate_001}--\ref{table: detectionrate_50},
Einstein telescope will be able to detect several tens to thousands BBHs with IMBHs in the future.
Therefore, we may be able to impose restrictions on models by using mass distribution.
On the other hand, advanced LIGO (O4) and LISA may only see one to a few BBHs.
Thus, in this section, in order to impose restrictions on the model when only one BBH with IMBHs is detected,
we compute and show the likelihood ratio 
for each parameter ($\alpha\lambda$, $\gamma_1$, $\gamma_2$ and $m_\mathrm{crit}$)
assuming that one BBH with IMBHs with a certain total mass is observed by a certain detector.
We simply assume that IMF models occurs with equal probability and $\alpha\lambda$
have a probability density, $p(\log(\alpha\lambda))\propto\const$.

Figure \ref{fig: lr al} shows the likelihood ratio to the maximum likelihood for $\alpha\lambda$.
For example, from Figure \ref{fig: lr al} (a), if a BBH with total mass $\sim500\msun$ are
detected by advanced LIGO (O4), we can find that most likely model is $\alpha\lambda=0.5$ and
about three times more likely than the second most likely model, $\alpha\lambda=0.1$.
As described in Section \ref{subsubsec: primary mass hh}, the smaller $\alpha\lambda$ value is, the larger
the maximum primary BH mass.
Therefore, as the detected BBH total mass increases, $\alpha\lambda$ value with the maximum likelihood decreases.
We can impose restrictions on $\alpha\lambda$ to some degree
if a BBH with total mass $\gtrsim500\msun$ are detected by advanced LIGO (O4) or LISA.

Figure \ref{fig: lr g1}--\ref{fig: lr mc} show the likelihood ratio of IMF parameters.
It is hard to distinguish between $\gamma_1=0.0$ and 1.0 (see Figure \ref{fig: lr g1}).
If more massive BBH is detected, smaller $\gamma_2$, i.e., shallower IMF, and 
larger $m_\mathrm{crit}$ are preferred.
However, the difference between maximum likelihood and that of the other models is not so large
that it is difficult to distinguish models.

In Figure \ref{fig: lr al}(a)--\ref{fig: lr mc}(a), the likelihood 
in the range $M_\mathrm{t}=$600--800 is not plotted.
This is because, as can be seen from Figure \ref{fig: detection aligo}, 
BBHs in this mass range do not merge within the horizon of advanced LIGO.
When $\alpha\lambda=0.1$, the total mass of BBH mergers with short delay time
is $\sim$ 250--1500 $\msun$.
On the other hand, BBHs with long delay time have two sub-populations 
with slightly different formation channels, and total mass is $\lesssim600\msun$ and
$\gtrsim800\msun$.
Since the horizon of advanced LIGO is $z\sim1$ (see Figure \ref{fig: horizon}), only BBHs with long delay time
can be observed by advanced LIGO.
Therefore, when $\alpha\lambda=0.1$, BBHs with $M_\mathrm{t}=$ 600--800 $\msun$
cannot be detected by advanced LIGO.
On the other hand, since LISA can detect BBHs with short delay time, likelihood ratios
are plotted in the range $M_\mathrm{t}=$ 600--800 $\msun$.

\section{Summary} \label{sec: summary}
In this paper, we perform a binary population synthesis calculation for 
very massive ($\sim$ 100--1000 $\msun$) Pop. III stars and derive the various property
of BBH mergers with IMBHs.
We adopt wide range of common envelope parameter $\alpha\lambda$ from 0.01 to 5
and investigate the dependence of $\alpha\lambda$ on the primary
BH mass, mass ratio, effective spin and delay time distribution of BBH mergers.
We find that 
if $\alpha\lambda$ is smaller, the maximum mass of primary BH mass is larger (Section \ref{sec: results}).

We derive the redshift evolution of merger rate density and compare our results with 
the observation.
So far, no BBH with IMBHs have been discovered by gravitational wave detector, and thus
only upper limit on merger rate density have been obtained.
In this study, we adopt various IMF models and impose restrictions on IMF models
by using this upper limit obtained from non-detection (Section \ref{subsec: comparison}).

We also compute the detection rate by advanced LIGO (O4), Einstein telescope and LISA,
and obtain the mass distribution of detected BBHs with IMBHs (Section \ref{subsubsec: detection rate}).
We find that Einstein telescope is a promising detector for the Pop.III BBH mergers with IMBHs, 
and the detection rate is from $\sim$ 10 $\mathrm{yr}^{-1}$ to 1000 $\mathrm{yr}^{-1}$.

Finally, we calculate the likelihood ratio 
for model parameter ($\alpha\lambda$, $\gamma_1$, $\gamma_2$ and $m_\mathrm{crit}$)
assuming that one BBH with IMBHs with a certain total mass is observed by certain detector.
We find that we can impose restrictions on $\alpha\lambda$ to some degree
if a BBH with total mass $\gtrsim500\msun$ are detected by advanced LIGO (O4) or LISA.

\section*{Acknowledgements}
This work was supported by JSPS KAKENHI grant No. 17H06360 and 19K03907
(A. T.), 21K13915, 22K03630(T. K), 20H05249 (T. Y.), and 17H01130, 17K05380 and 21H01123
(H. U.), and by the University of Tokyo Young Excellent Researcher 
program (T. K.).
We use the Python packages, {\tt numpy} \citep{numpy} and 
{\tt matplotlib} \citep{matplotlib} for our analysis.


\section*{Data Availability}
Results will be shared on reasonable request to authors.



\bibliographystyle{mnras}
\bibliography{main.bib} 






\bsp	
\label{lastpage}
\end{document}